\documentclass[longauth]{aa}

\usepackage{graphicx}
\usepackage{txfonts}
\usepackage{comment}
\usepackage{lscape}
\usepackage{longtable}

\usepackage{natbib,twoopt}
\bibpunct{(}{)}{;}{a}{}{,}             

\def\ls{\vskip 12.045pt} 

\usepackage[usenames]{color}                                   

\newcommand\pycasso{{\sc p}y{\sc casso}}          	
\newcommand\starlight{{\sc starlight}}          	

\newcommand\ageL{$\langle {\rm log}\,age\rangle _L$}
\newcommand\logZM{$\langle {\rm log}\,Z_\star\rangle _M$}
\newcommand\ageM{$\langle {\rm log}\,age\rangle _M$}

\begin{document}

\title{The CALIFA survey across the Hubble sequence:} 
\subtitle{Spatially resolved stellar population properties in galaxies}

\authorrunning{IAA et al.}
\titlerunning{Metallicity, age, and stellar mass density across the Hubble sequence}

\author{
R. M. Gonz\'alez Delgado\inst{1},
R. Garc\'{\i}a-Benito\inst{1}, 
E. P\'erez\inst{1},
R. Cid Fernandes\inst{2},
A. L.\ de Amorim\inst{2},
C. Cortijo-Ferrero\inst{1},
E. A. D. Lacerda\inst{1,2},
R. L\'opez Fern\'andez\inst{1},
N. Vale-Asari\inst{2},
S. F. S\'anchez\inst{3},
M. Moll\'a\inst{4},
T. Ruiz-Lara\inst{5},
P.  S\'anchez-Bl\'azquez\inst{6},
C. J. Walcher\inst{7},
J. Alves\inst{8},
J. A. L. Aguerri\inst{9},
S. Bekerait\'e\inst{7},
J. Bland-Hawthorn\inst{10},
L. Galbany\inst{11,12},
A. Gallazzi\inst{13},
B. Husemann\inst{14},
J. Iglesias-P\'aramo\inst{1,15},
V. Kalinova\inst{16},
A. R. L\'opez-S\'anchez\inst{17},
R. A. Marino\inst{18},
I. M\'arquez\inst{1},
J. Masegosa\inst{1},
D. Mast\inst{19},
J. M\'endez-Abreu\inst{20},
A. Mendoza\inst{1},
A. del Olmo\inst{1},
I. P\'erez\inst{5},
A. Quirrenbach\inst{21},
S. Zibetti\inst{12},
CALIFA collaboration\inst{22}
}

\institute{
Instituto de Astrof\'{\i}sica de Andaluc\'{\i}a (CSIC), P.O. Box 3004, 18080 Granada, Spain. ( \email{rosa@iaa.es})
\and
Departamento de F\'{\i}sica, Universidade Federal de Santa Catarina, P.O. Box 476, 88040-900, Florian\'opolis, SC, Brazil
\and
Instituto de Astronom\'\i a,Universidad Nacional Auton\'oma
de Mexico, A.P. 70-264, 04510, M\'exico,D.F.
\and
Departamento de Investigaci\'on B\'asica, CIEMAT, Avda.\ Complutense 40, E-28040 Madrid, Spain
\and
Departamento de F\'\i sica Te\'orica y del Cosmos, University of Granada, Facultad de Ciencias (Edificio Mecenas), E-18071 Granada, Spain
\and
Depto. de F\'{\i}sica Te\'orica, Universidad Aut\'onoma de Madrid, 28049 Madrid, Spain
\and
Leibniz-Institut f\"ur Astrophysik Potsdam (AIP), An der Sternwarte 16, D-14482 Potsdam, Germany
\and
University of Vienna, T\"urkenschanzstrasse 17, 1180, Vienna, Austria
\and
Instituto de Astrof\'\i sica de Canarias (IAC), E-38205 La Laguna, Tenerife, Spain
\and
Sydney Institute for Astronomy, The University of Sydney, NSW 2006, Australia
\and
Millennium Institute of Astrophysics and Departamento de Astronom\'\i a, Universidad de Chile, Casilla 36-D, Santiago, Chile
\and
Departamento de Astronom\'ia, Universidad de Chile, Casilla 36-D, Santiago, Chile
\and
INAF $-$ Osservatorio Astrofisico di Arcetri, Largo Enrico Fermi 5,
50125 Firenze, Italy
\and
European Southern Observatory, Karl-Schwarzschild-Str. 2, 85748 Garching b. M\"unchen, Germany
\and
Estaci\'{o}n Experimental de Zonas Aridas (CSIC), Ctra. de Sacramento s/n, La Ca\~{n}ada, Almer\'{\i}a, Spain
\and
Department of Physics 4-181 CCIS, University of Alberta, Edmonton AB T6G 2E1, Canada
\and
Australian Astronomical Observatory, PO BOX 296, Epping, 1710
NSW, Australia
\and
CEI Campus Moncloa, UCM-UPM, Departamento de Astrof\'{i}sica y CC. de la Atm\'{o}sfera, Facultad de CC.\ F\'{i}sicas, Universidad Complutense de Madrid, Avda.\,Complutense s/n, 28040 Madrid, Spain
\and
Instituto de Cosmologia, Relatividade e Astrof\'{i}sica - ICRA, Centro Brasileiro de Pesquisas F\'{i}sicas, Rua Dr.Xavier Sigaud 150, CEP 22290-180, Rio de Janeiro, RJ, Brazil
\and
School of Physics and Astronomy, University of St. Andrews, North Haugh, St. Andrews, KY169SS, UK
\and
Landessternwarte, Zentrum fur Astronomie der Universitat Heidelberg, K\"{o}nigstuhl 12, D-69117 Heidelberg, Germany
\and
http://califa.caha.es
}

\date{Feb. 2015}

 
\abstract
{
Various different physical processes contribute to the star formation and stellar mass assembly histories of galaxies. One important approach to understand the significance of these different processes on galaxy evolution is the study of the stellar population content of today's galaxies in a spatially resolved manner. The aim of this paper is to characterize in detail the radial structure of stellar population properties of galaxies in the nearby universe, based on a uniquely large galaxy sample considering the quality and coverage of the data. The sample under study was drawn from the CALIFA survey and contains 300 galaxies observed with integral field spectroscopy. These cover a wide range of Hubble types, from spheroids to spiral galaxies, while stellar masses range from $M_\star \sim 10^9$ to $7 \times 10^{11}$ $M_\odot$. We apply the fossil record method based on spectral synthesis techniques to recover the following physical properties for each spatial resolution element in our target galaxies: the stellar mass surface density ($\mu_\star$), stellar extinction ($A_V$), light-weighted and mass-weighted ages (\ageL, \ageM), and mass-weighted metallicity (\logZM). To study mean trends with overall galaxy properties, the individual radial profiles are stacked in seven bins of  galaxy morphology (E, S0, Sa, Sb, Sbc, Sc and Sd). 
We confirm that more massive galaxies are more compact, older, more metal rich, and less reddened by dust. Additionally, we find that these trends are preserved spatially with the radial distance to the nucleus. Deviations from these relations appear correlated with Hubble type: earlier types are more compact, older, and more metal rich for a given M$_\star$, which evidences that quenching is related to morphology, but not driven by mass. Negative gradients of \ageL\ are consistent with an inside-out growth of galaxies, with the largest \ageL\ gradients in Sb--Sbc galaxies. Further, the mean stellar ages of disks and bulges are correlated, with disks covering a wider range of ages, and late type spirals hosting  younger disks. However, age gradients are only mildly negative or flat beyond $R \sim 2$ HLR, indicating that  star formation is more uniformly distributed or that stellar migration is important at these distances. The gradients in stellar mass surface density depend mostly on stellar mass, in the sense that more massive galaxies are more centrally concentrated. Whatever sets the concentration indices of galaxies obviously depends less on quenching / morphology than on the depth of the potential well. There is a secondary correlation in the sense that at the same $M_\star$ early type galaxies have  steeper gradients. The $\mu_\star$ gradients outside 1 HLR show no dependence on Hubble type. We find mildly negative \logZM\ gradients, shallower than predicted from models of galaxy evolution in isolation. In general, metallicity gradients depend on stellar mass, and less on morphology, hinting that metallicity is affected by the depth of both - potential well and morphology/quenching. Thus, the largest \logZM\ gradients occur in Milky Way-like Sb--Sbc galaxies, and are similar to those measured above the Galactic disk. Sc spirals show flatter \logZM\ gradients, possibly indicating a larger contribution from secular evolution in disks. The galaxies from the sample have decreasing-outwards stellar extinction; all spirals show similar radial profiles, independent from the stellar mass, but redder than E's and S0's.  Overall we conclude that quenching processes act in manners that are independent of mass, while metallicity and galaxy structure are influenced by mass-dependent processes.
}



\keywords{Techniques: Integral Field Spectroscopy; galaxies: evolution; galaxies: stellar content; galaxies: structure; galaxies: fundamental parameters; galaxies: bulges; galaxies: spiral}
\maketitle

\section{Introduction}

\label{sec:Introduction}

Galaxies are a complex mix of stars,  gas, dust, and dark matter, distributed in different components (bulge, disk, and halo) whose present day structure and dynamics  are intimately linked to their assembly and evolution over the history of the Universe. Different observational and theoretical approaches can be followed to learn how galaxies form and evolve. 

Theoretically, the formation of large-scale structures arise through the evolution of cold dark matter. In this picture, small-scale density perturbations in the dark matter collapse and form the first generation of dark matter halos, that subsequently merge to form larger structures such as  clusters and superclusters \citep{springel05,delucia06}. This basic hierarchical picture is able to explain the global evolution of the star formation rate density of the universe, with galaxy peak formation epoch at redshift 2--3 \citep[e.g.][and references therein]{madau14}.  The stellar components formed at earlier epochs likely evolved into elliptical galaxies and bulges through mergers of the primordial star-forming disks \citep{elmegreen07,bournaud07}. However, this framework fails to explain how the galaxy population emerges at $z \sim 1$, and how the present day Hubble sequence of galaxies was assembled.

The growth of galaxies is not related in a simple way to the build up of dark matter; the interplay of energy and matter exchange (between the process of gas accretion and cooling and star formation) is essential to grow the gaseous and stellar components in galaxies. Feedback mechanisms resulting from stellar winds, supernova explosions, and AGN are  relevant to stop the gas collapse and cooling, quenching the star formation and hence galaxy growth \citep{silk98,hopkins11}. Although these processes are difficult to implement in theoretical models, they are essential to explain the masses, structures, morphologies, stellar populations, and chemical compositions of galaxies, and the evolution of these properties with cosmic time.

Recently, a new set of cosmological hydrodynamic simulations have  started to predict how the spatially resolved information of the properties of stellar populations in galaxies can constrain the complex interplay between gas infall, outflows, stellar migration, radial gas flows, and star formation efficiency, in driving the inside-out growth of galactic disks \citep{brook12,gibson13,few12,pilkington12a,minchev14}. Radiative cooling, star formation, feedback from supernovae, and chemical enrichment are also included in 
simulations to predict radial metallicity gradients as a function of merging history.  Shallow metallicity gradients are expected if elliptical galaxies result from major mergers \citep[e.g.][]{kobayashi04}, but a minor merger picture for the formation of ellipticals can successfully explain the strong size evolution of massive galaxies \citep{oser12}. This late-time accretion of low mass and metal poor galaxies (dry mergers) into the already formed massive galaxy  can produce a variation of the age and metallicity radial structure of the galaxy as it increases in size. Galactic stellar winds and metal cooling have also an important effect on these ex-situ star formation models, predicting different behaviour of the mass and metallicity assembly in massive early type galaxies, and in the radial gradient of present stellar populations properties of galaxies \citep{hirschmann13,hirschmann15}.

In summary, these theoretical works show that observational data with spatial information of the mass and metallicity assembly and their cosmic evolution, and the present radial structure of stellar population properties (stellar mass surface density, age, metallicity) contain relevant information to constrain the formation history of galaxies, and the physics of feedback mechanisms involved.

Observationally, a first step is to study what kinds of galaxies are there in the Universe, and which are their physical properties. Attending to their form and structure, galaxies can be grouped into a few categories. Results show that most of the massive nearby galaxies are ellipticals, S0's, or spirals \citep{blanton09} following the Hubble tuning fork diagram. In this scheme, S0's are a transition between spirals and ellipticals \citep{cappellari13}, and the bulge/disk ratio increases from late to early type spirals.  At the same time, galaxy properties such as color, mass, surface brightness, luminosity, and gas fraction are correlated with Hubble type \citep{roberts94}, suggesting that the Hubble sequence somehow reflects possible paths for galaxy formation and evolution. However, the processes structuring galaxies along the Hubble sequence are still poorly understood.

Integral Field Spectroscopy (IFS) enables a leap forward, providing 3D information (2D spatial + 1D spectral) on galaxies. 
Such datacubes allow one to recover two-dimensional maps of stellar mass surface density, stellar ages, metallicities,  extinction and kinematics, as well as a suit of nebular properties such as gas kinematics, metallicity, excitation, and etc.
Until a few years ago IFS was used to target small samples of galaxies.  Detailed programs such as SAURON \citep{bacon01}, VENGA \citep{blanc13}, (U)LIRs at z $\leq$0.26 \citep{arribas10}, PINGS \citep{rosales-Ortega10}, or DiskMass Survey \citep{bershady10}, have been limited to less than a hundred galaxies, but it is more than fair to recognize that to get these amounts of IFU data was a challenge at the time. ATLAS3D \citep{cappellari11} represented a step forward, with the observation of a volume-limited sample of 260 galaxies, but with three important limitations: the sample only includes early-type galaxies, the field of view is limited to 1 effective radius, and the spectral range is restricted from H$\beta$ to [NI]$\lambda$5200.

CALIFA (Calar Alto Legacy Integral Field Area) is our ongoing survey of 600 nearby galaxies at the 3.5m at Calar Alto  \citep{sanchez12}\footnote{\url{http://califa.caha.es}}.
The data set  provided by the survey (see \citealt{husemann13} for DR1; \citealt{garcia-benito14} for DR2) is unique to advance in these issues  not only because of its ability to provide spectral and spatial information, but also because: a) It includes a large homogeneous sample of galaxies across the color-magnitude diagram, covering a large range of masses \citep[$10^9$ to $10^{12}$ $M_\odot$,][]{gonzalezdelgado14a}, and morphologies from Ellipticals (E0-E7), Lenticulars (S0-S0a), to Spirals (Sa to Sm) (see \citet{walcher14} for a general description of the sample). b) It has a large field of view ($74{\tt''} \times 65{\tt''}$) with a final spatial sampling of 1 arcsec, and a resolution of $\sim 2.5$ arcsec, allowing to spatially resolve well the stellar population properties, and to obtain the total integrated properties, such as galaxy stellar mass, and stellar metallicity. c) It covers the whole rest-frame optical wavelength at intermediate spectral resolution, including the most relevant absorption diagnostics for deriving the stellar population properties.

Previous papers in this series have used the first $\sim$100 datacubes of the survey to derive spatially resolved  stellar population properties by means of full spectral fitting techniques.
We have obtained that: 
1) Massive galaxies grow their stellar mass inside-out. The signal of downsizing is shown to be spatially preserved, with both inner and outer regions growing faster for more massive galaxies. The relative growth rate of the spheroidal component (nucleus and inner galaxy), which peaked 5--7 Gyr ago, shows a maximum at a critical stellar mass $M_\star \sim 7 \times 10^{10} M_\odot$ \citep{perez13}. 
2) The inside-out scenario is also supported by the negative radial gradients of the stellar population ages \citep{gonzalezdelgado14a}. 
3) Global and local relations between stellar mass, stellar mass surface density and stellar metallicity relation were investigated, along with their evolution (as derived from our fossil record analysis). In disks, the stellar mass surface density regulates the ages and the metallicity. In spheroids, the galaxy stellar mass dominates the physics of star formation and chemical enrichment \citep{gonzalezdelgado14a,gonzalezdelgado14b}.  
4) In terms of integrated versus spatially resolved properties, the stellar population properties are well represented by their values at 1 HLR \citep{gonzalezdelgado14a, gonzalezdelgado14b}. 
The CALIFA collaboration has also compared the age and metallicity gradients in a subsample of 62 face-on spirals and it was found that there is no difference between the stellar population properties in  barred and unbarred galaxies \citep{sanchez-blazquez14}.

In this paper we extend our study of the spatially resolved star formation history of CALIFA galaxies to derive the radial structure of the stellar population properties as a function of Hubble type, and galaxy stellar mass, $M_\star$. The goals are: 
1) To characterize in detail the radial structure of stellar population properties of galaxies in the local universe.
2) To find out how these properties are correlated with  Hubble type,  and if the Hubble sequence is a scheme to organize galaxies by mass and age, and/or  mass and metallicity. 
3) To establish observational constraints to galaxy formation models via the radial distributions and gradients of  stellar populations for disk and bulge dominated galaxies.

This paper is organized as follows: Section 2 describes the observations and  summarizes the properties of the CALIFA galaxies analyzed here. In Sec.\ 3 we summarize our method for extracting the SFH, based on the fossil record method, and we explain the main differences between the analysis presented here and that in previous works. Sec.\ 4 presents results on the galaxy stellar mass, half light and half mass radii (HLR, HMR, respectively), and galaxy averaged stellar metallicity. Sec.\ 5 deals with the spatially resolved properties of the stellar population: stellar mass surface density, $\mu_\star$;  luminosity weighted mean age, \ageL; mass weighted mean metallicity,  \logZM; and stellar extinction, $A_V$. We discuss the results in Sec.\ 6; and Sec.\ 7  presents the conclusions.

\section{Sample, and Observations, data reduction}

\label{sec:Sample}

\subsection{Sample and morphological classification}

The CALIFA mother sample consists of 939 galaxies selected from SDSS survey in the redshift range $z = 0.005$--0.03, and with $r$-band angular isophotal diameter of 45--80$^{\prime\prime}$. These criteria guarantee that the objects fill well 
the $74^{\prime\prime} \times 64^{\prime\prime}$ FoV. The sample includes a significant number of galaxies in different bins in the color-magnitude diagram (CMD), ensuring that CALIFA spans a wide and representative range of galaxy types. 

The galaxies were morphologically classified as Ellipticals (E0--7), Spirals (S0,  S0a, Sab, Sb, Sbc, Sc, Scd, Sd, Sm), and Irregulars (I). The classification was carried out through visual inspection of the $r$-band images averaging the results (after clipping outliers) from five members of the collaboration. Galaxies are also classified as $B$ for barred, otherwise $A$, or $AB$ if it is unsure, and as $M$ if it shows "merger" or "interaction features"  \citep{walcher14}.

The sample for this paper comprises the 312 CALIFA galaxies observed in both V1200 and V500 setups as of January 2014. The 12 galaxies showing "merger or interacting features" are not discussed here, 
leaving a main sample of 300 objects with a well defined morphology. For this work we have grouped galaxies into 7 morphology bins: E, S0 (including S0 and S0a), Sa (Sa and Sab), Sb, Sbc, Sc (Sc and Scd), and Sd (13 Sd, 1 Sm and 1 Irr).

\begin{figure*}
\includegraphics[width=0.48\textwidth]{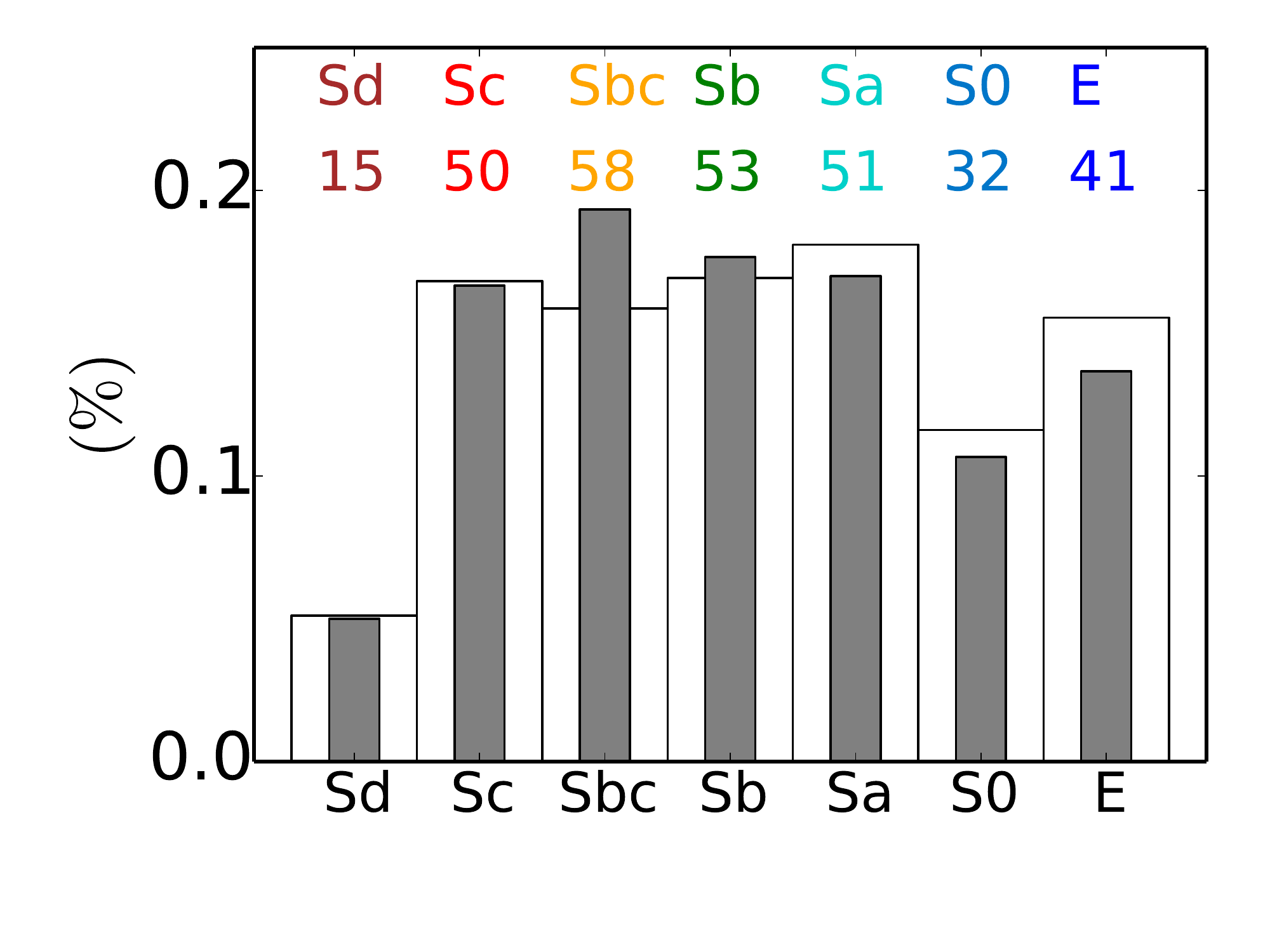}
\includegraphics[width=0.48\textwidth]{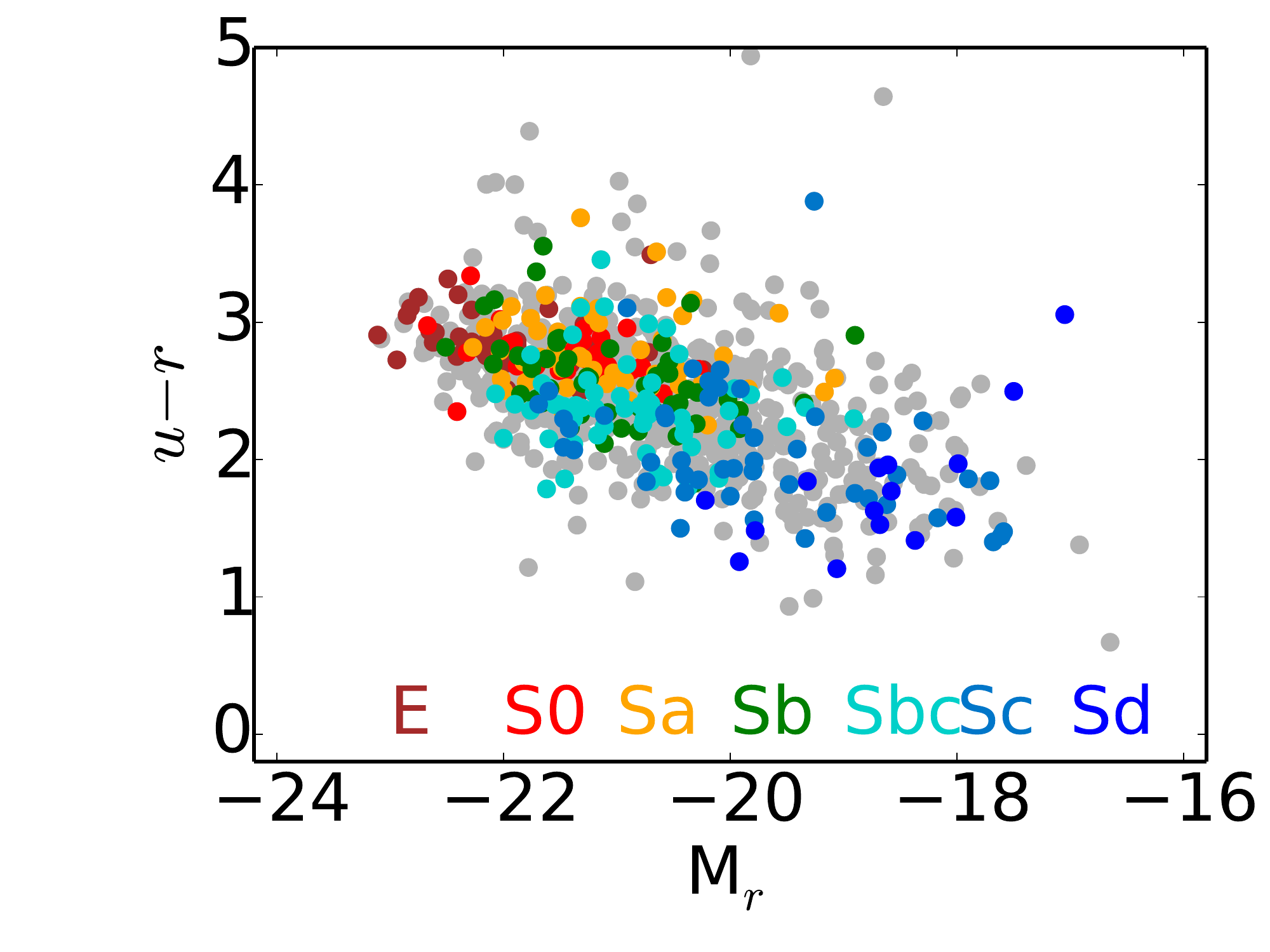}
\caption{{\em Left:} Comparison of the distribution of Hubble types in the CALIFA mother sample (empty bars) and the galaxies analyzed here (filled bars). The number of galaxies in our sample are labeled in colors. The histograms are normalized to form a probability density, i.e., each bar scales with the ratio of the number of galaxies in each bin and the total number of galaxies, such that the two distributions are directly comparable. 
{\em Right:} Color-magnitude diagram. Mother sample galaxies are plotted in grey, while the 300 galaxies analyzed in this work are marked as colored points.}
\label{fig:hist-type}
\end{figure*}

Fig.\ \ref{fig:hist-type} shows that these 300 galaxies provide a fair representation of the CALIFA survey as a whole. The left panel shows scaled histograms of the Hubble type in the mother sample (empty bars) and in our sample (filled bars). The number of objects in each morphology bin for our sample is indicated at the top, with a brown to blue color palette that represents Hubble types from ellipticals to late spirals. This same color scheme is used throughout this paper.
The similarity of the distributions reflects the random sampling strategy of CALIFA, with targets being picked from the mother sample on the basis of visibility criteria alone. The right panel in Fig.\ \ref{fig:hist-type} shows the $u - r$ versus $M_r$ CMD, with grey points representing the mother sample and colored points  the 300 galaxies. As for the Hubble type distribution, a simple visual inspection shows that our subsample is representative of the full CALIFA sample in terms of CMD coverage.

\subsection{Observations and data reduction}

The observations were carried out with the Potsdam Multi-Aperture Spectrometer \citep[PMAS,]{Roth05} in the PPaK mode \citep{verheijen04} at the 3.5m telescope of Calar Alto observatory. PPaK contains 382 fibers of $2.7^{\prime\prime}$  diameter each, and a  $74^{\prime\prime} \times 64^{\prime\prime}$ Field of View \citep[FoV][]{kelz06}. Each galaxy is observed with two spectral settings, V500 and V1200, with spectral resolutions  $\sim 6$ \AA\ (FWHM) and 2.3 \AA, respectively. The V500 grating covers from 3745 to 7300 \AA, while the V1200 covers 3650--4840 \AA. Detailed descriptions of the observational strategy and of the data can be found in \citet{sanchez12}, and \citet{husemann13}.

The datacubes analyzed here have been calibrated with version 1.5 of the reduction pipeline. The main issues addressed by this new version are: {\em (i)} correction of the sensitivity curve for the V500 grating; {\em (ii)} new registering method to determine, for each galaxy, the relative positioning of the 3 pointings of the dithering pattern, and absolute WCS registration; {\em (iii)} a new cube interpolation method. CALIFA pipeline v1.5 improves the flux calibration to an accuracy of 2--3\% and is the current official data release. A detailed account of this new pipeline is presented in the Data Realease 2 article \citep{garcia-benito14}.  

In order to reduce the effects of vignetting on the data, we combine the observations in the V1200 and V500 setups. The combined datacubes were processed as described in \citet{cidfernandes13}. Our analysis requires that spectra have  signal to noise ratio S/N $\geq 20$ in a 90 \AA\ window centered at 5635 \AA\ (rest-frame). When individual spaxels do not meet this S/N threshold, they are coadded into Voronoi zones \citep{cappellari03}. Further pre-processing steps include spatial masking of foreground/background sources, rest-framing  and spectral resampling. The resulting 253418 spectra were then processed through \starlight\ and \pycasso\ (the Python CALIFA \starlight\ Synthesis Organizer), 
producing the stellar population properties discussed here as described in detail in the next section.


\section{Stellar population analysis: Differences with respect to previous work}

Our method to extract stellar population properties from  datacubes has been explained and applied to CALIFA in \citet{perez13}, \citet{cidfernandes13,cidfernandes14}, and \citet{gonzalezdelgado14a,gonzalezdelgado14b}. In short, we analyse the data with the \starlight\ code \citep{cidfernandes05}, which fits an observed spectrum ($O_\lambda$) in terms of a model ($M_\lambda$) built by a non-parametric linear combination of $N_\star$ Simple Stellar Populations (SSPs) from a base spanning different ages ($t$) and metallicities ($Z$). Dust effects are modeled as a foreground screen with a \citet{cardelli89} reddening law with $R_V = 3.1$. Windows around the main optical emission lines and the NaI D absorption doublet (because of its interstellar component) are masked in all spectral fits\footnote{To test the effect of this process in the estimation of ages, we have compared the results for 60 galaxies in common with \citet{sanchez-blazquez14}. This work uses Steckmap  \citep{ocvirk06} and the H$\beta$ line (previously corrected for emission). Statistically, we find that there is no difference in ages (mean = -0.04, std = 0.15 dex) if the same SSP models are used in the two methods.}. 
Bad pixels (identified by the reduction pipeline) are also masked. Results for each spectrum are then packed and organized with the \pycasso\ pipeline.

This working scheme is preserved here, but with three new developments:

\begin{enumerate}

\item The datacubes used in this paper come from the version 1.5  of the reduction pipeline \citep{garcia-benito14}.

\item Larger and more complete SSP bases are employed.

\item A somewhat different definition of mean stellar metallicity is adopted \citep[see][]{gonzalezdelgado14b}.

\end{enumerate}

This section describes the novelties related with the stellar population synthesis. Improvements resulting from the new reduction pipeline are described in Appendix A.

\subsection{SSP spectral bases}

\label{sec:SSP_spectral_bases}

SSP models are a central ingredient in our analysis, linking the results of the spectral decomposition to physical properties of the stellar populations. Our previous applications of \starlight\ to CALIFA explored spectral bases built from three sets of SSP models, labeled as {\it GM}, {\it CB} and {\it BC} in \citet{cidfernandes14}. The first two are again used in this study, but extended to a wider range of metallicities, producing what we will denote as bases {\it GMe} and {\it CBe}.

Base {\it GMe} is a combination of the SSP spectra provided by \citet{vazdekis10} for populations older than $t = 63$ Myr and the \citet{gonzalezdelgado05} models for younger ages. The evolutionary tracks are those of \citet{girardi00}, except for the youngest ages (1 and 3 Myr), which are based on the Geneva tracks \citep{schaller92,schaerer93,charbonnel93}. The IMF is Salpeter. In our previous studies of the first 100 CALIFA galaxies we defined base {\it GM} as a regular $(t,Z)$  grid of these models, with 39 ages spanning $t = 0.001$--14 Gyr and four metallicities from 0.2 to 1.5 $Z_\odot$. 
We now extend the $Z$ range to  use of all seven metallicites provided by \citet{vazdekis10} models: $\log Z/Z_\odot = -2.3$, $-1.7$, $-1.3$, $-0.7$, $-0.4$, 0, and $+0.22$. Because these models lack ages below 63 Myr, these young ages are only covered by the four largest metallicities, such that our extended {\it GM} base is no longer regular in $t$ and $Z$. Base {\it GMe} contains $N_\star = 235$ elements.

Base {\it CBe} is built from an update of the \citet{bruzual03} models (Charlot \& Bruzual 2007, private communication), replacing  STELIB \citep{leborgne03} by a combination of the MILES \citep{sanchez-blazquez06,falcon-barroso11} and {\sc granada} \citep{martins05} spectral libraries (the same ones used in base {\it GMe}). The evolutionary tracks are those collectively known as Padova 1994 \citep{alongi93,bressan93,fagotto94a,fagotto94b,girardi96}. The IMF is that of \citet{chabrier03}. 
Whereas in previous works we limited the $Z$ range to $\ge 0.2$ solar, we now extend this base to six metalicities: $\log Z/Z_\odot = -2.3$, $-1.7$, $-0.7$, $-0.4$, 0, and $+0.4$. Base {\it CBe} contains $N_\star = 246$ elements (41 ages from 0.001 to 14 Gyr and the 6 metallicities above).

The main similarities and differences between bases {\it GMe} and {\it CBe} are the same as between the original 
{\it GM} and {\it CB} bases, thoroughly discussed in \citet{cidfernandes14}. Throughout the main body of this paper we focus on results obtained with base {\it GMe}, but we anticipate that our overall qualitative findings remain valid for base {\it CBe}. The role of base {\it CBe} in this paper is to allow a rough assessment of the uncertainties associated to model choice. 

A minor technical difference with respect to our previous analysis is that we now smooth the spectral bases to 6 \AA\ FWHM effective resolution prior to the fits. This is because the kinematical filter implemented in \starlight\ operates in velocity-space, whereas both CALIFA and the SSP model spectra have a constant spectral resolution in $\lambda$-space, so that effects of the instrumental broadening can only be mimicked approximately by \starlight. We have verified that this modification does not affect the stellar population properties used in this paper.

Appendix \ref{app:BaseExperiments} presents some comparisons of the results obtained with these two bases. Experiments were also performed with bases which extend the age range to 18 Gyr, and configuring  \starlight\ to allow negative values of $A_V$. These tests are also discussed in Appendix \ref{app:BaseExperiments}, which adds to the collection of  ``sanity checks" on the results of our analysis.


\section{Galaxy mass, metric, and stellar metallicity}

This section addresses three relatively unrelated aspects, which are all important to better understand the results presented in the next section, where we examine how the spatial distribution of stellar population properties relates to a galaxy's stellar mass and morphology. First, \S\ref{sec:Mass_x_Morphology} reviews the relation between stellar mass and morphological type for our sample. This strong relation is imprinted on virtually all results discussed in \S\ref{sec:Results}.
Secondly, \S\ref{sec:HMR2HLR} compares our measurements of the Half Light (HLR) and Half Mass Radii (HMR). As discussed by \citet{gonzalezdelgado14a}, these two natural metrics for distances are not identical due to the inside-out growth of galaxies. Here we inspect how the HMR/HLR ratio varies as a function of Hubble type and stellar mass in our sample.
Finally, \S\ref{sec:metallicity} presents our definition of mean stellar metallicity. \citet{gonzalezdelgado14a} showed that stellar mass surface densities, mean ages, and extinction values defined from the integrated spectrum, from galaxy-wide spatial averages, and measured at $R = 1$ HLR all agree very well with each other. Here we extend this test to stellar metallicities. Throughout this section, results for the two SSP models discussed in \S\ref{sec:SSP_spectral_bases} are presented.

\subsection{Stellar masses}
\label{sec:Mass_x_Morphology}
\begin{table}
\caption{Number of galaxies for each Hubble type and M$_\star$ interval ($GMe$)}
\begin{tabular}{lccccccc}
\hline\hline
$\log \ M_\star$ (M$_\odot$) bin &   E & S0 & Sa & Sb & Sbc & Sc & Sd    \\      
\hline
$\leq$9.1   & - & - & - & -    & -     & 2   & 2 \\
9.1-9.6       & - & - & - & -    & -     & 9   & 8 \\
9.6-10.1     & - & - & -  & -   & 2    & 10 & 5 \\
10.1-10.6   & - & -    & 7 & 11   & 16  & 21 & - \\
10.6-10.9   & 3 & 8  & 9 & 14   &21 & 4 & - \\
10.9-11.2   & 8 & 14 & 22 & 17 & 16 & 3 & - \\
11.2-11.5    & 17 & 8 & 13 & 10 & 3 &  1 & - \\
11.5-11.8    & 12 & 2 & - & 1 & - &   & - \\
$\geq$11.8 & 1 & - & - & - & - &   & - \\
\hline
 total          & 40 & 32 & 51 & 53 & 58 & 50 & 15 \\
\hline\hline
\label{tab:Massdistribution}
\end{tabular}
\end{table}

To obtain the total stellar mass of a galaxy we add the mass in each zone, thus taking into account spatial variations of the stellar extinction and $M/L$ ratio. Masked spaxels (e.g., foreground stars) are accounted for using the $\mu_\star$ radial profile as explained in \citet{gonzalezdelgado14a}.

\begin{figure}
\includegraphics[width=0.5\textwidth]{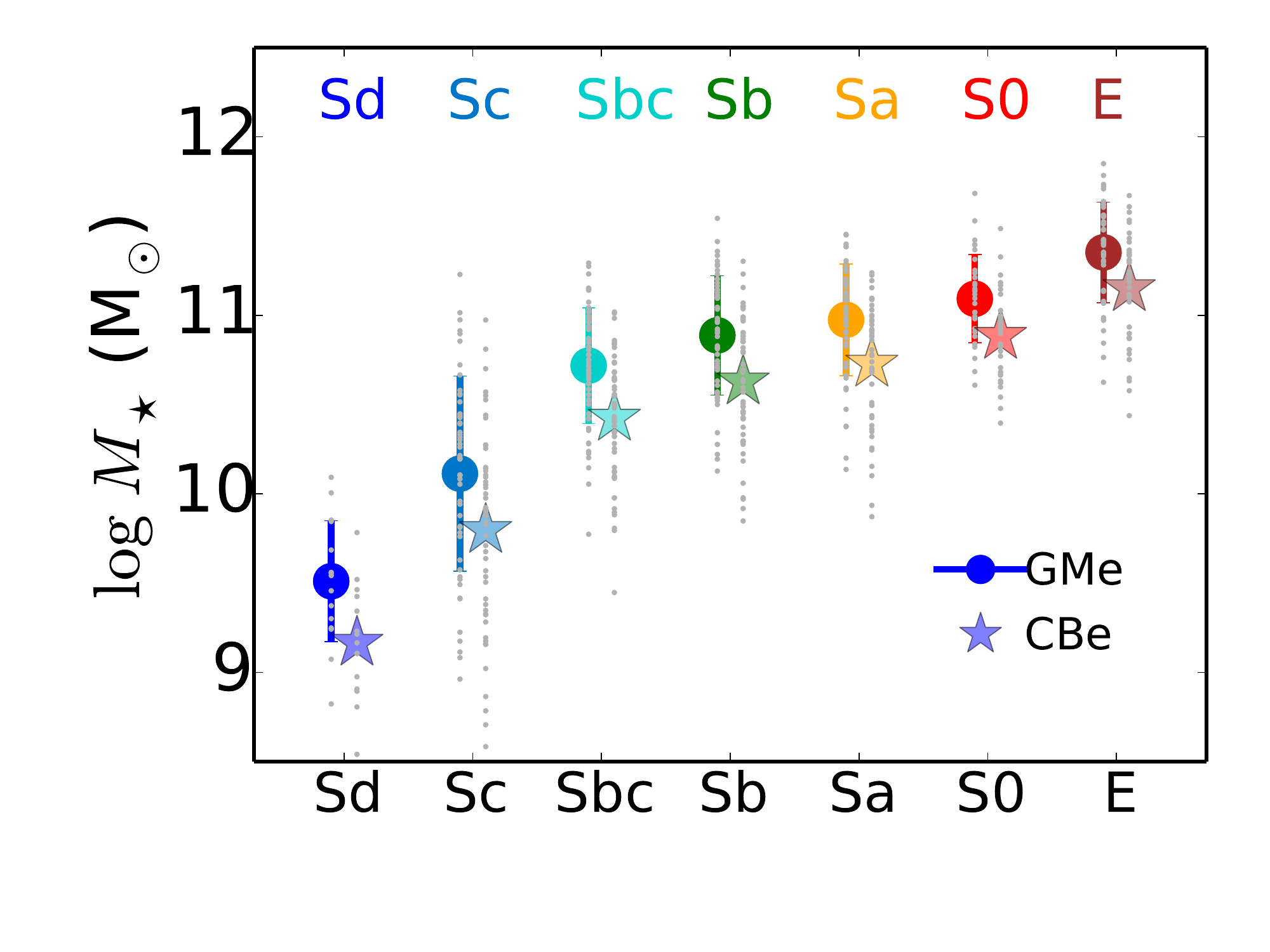}
\caption{ 
Distribution of the stellar masses obtained from the spatially resolved spectral fits of each galaxy for each Hubble type (grey small points). The colored dots (stars) are the mean galaxy stellar mass in each Hubble type obtained with the {\it GMe} ({\it CBe}) SSP models. The bars show the dispersion in mass.} 
\label{fig:mass-distribution}
\end{figure}

Fig.\ \ref{fig:mass-distribution} shows the distribution of $M_\star$ as a function of  Hubble type. Table 1 shows the distribution of galaxies by Hubble type in several bins of $M_\star$.
The masses range from $7\times 10^8$ to $7\times 10^{11} M_\odot$ for fits with {\it GMe} (Salpeter IMF). {\it CBe}-based masses (Chabrier IMF) are on average smaller by a factor 1.84. As for the general galaxy population, mass is well correlated with  Hubble type, decreasing from early to late types. High  bulge-to-disk ratios (E, S0, Sa) are the most massive ones ($\geq 10^{11} M_\odot$), while galaxies with small bulges (Sc--Sd) have  $M_\star \leq  10^{10}  M_\odot$. 
The average $\log M_\star (M_\odot)$ is 11.4, 11.1, 11.0, 10.9, 10.7, 10.1, and 9.5 for E, S0, Sa, Sb, Sbc, Sc, and Sd, respectively. The dispersion is  typicaly 0.3 dex, except for Sc galaxies, that have a dispersion of $\sim 0.5$ dex.

Because CALIFA is not complete for $M_r \geq  -19.5$, this distribution in mass is not completely representative of the local Universe. In particular, it is important to remember that dwarf ellipticals are not included, so $M_\star$ or any other property discussed here for E's are restricted to massive ellipticals.

\subsection{The HMR/HLR}
\label{sec:HMR2HLR}

\begin{figure*}
\includegraphics[width=0.48\textwidth]{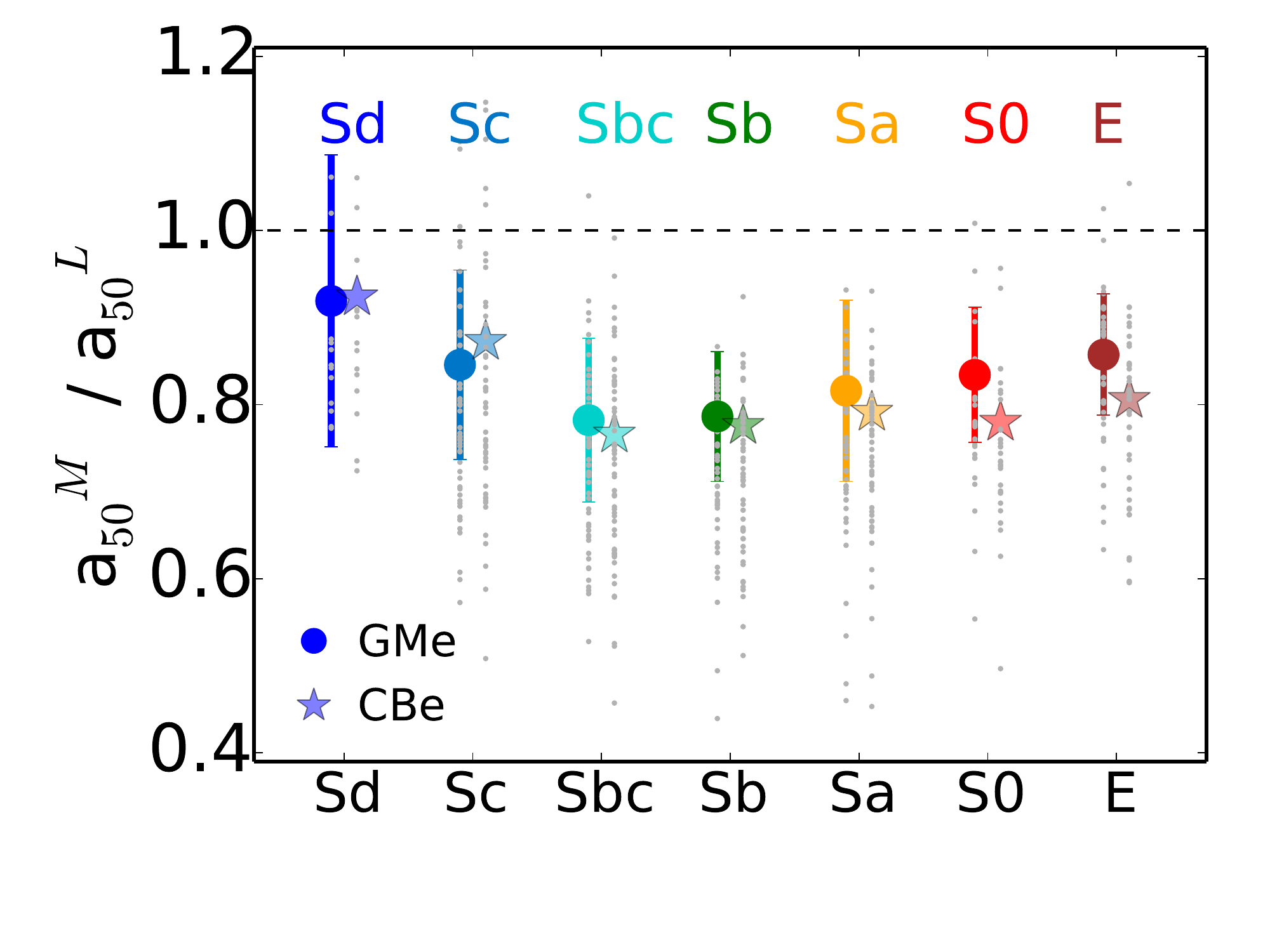}
\includegraphics[width=0.48\textwidth]{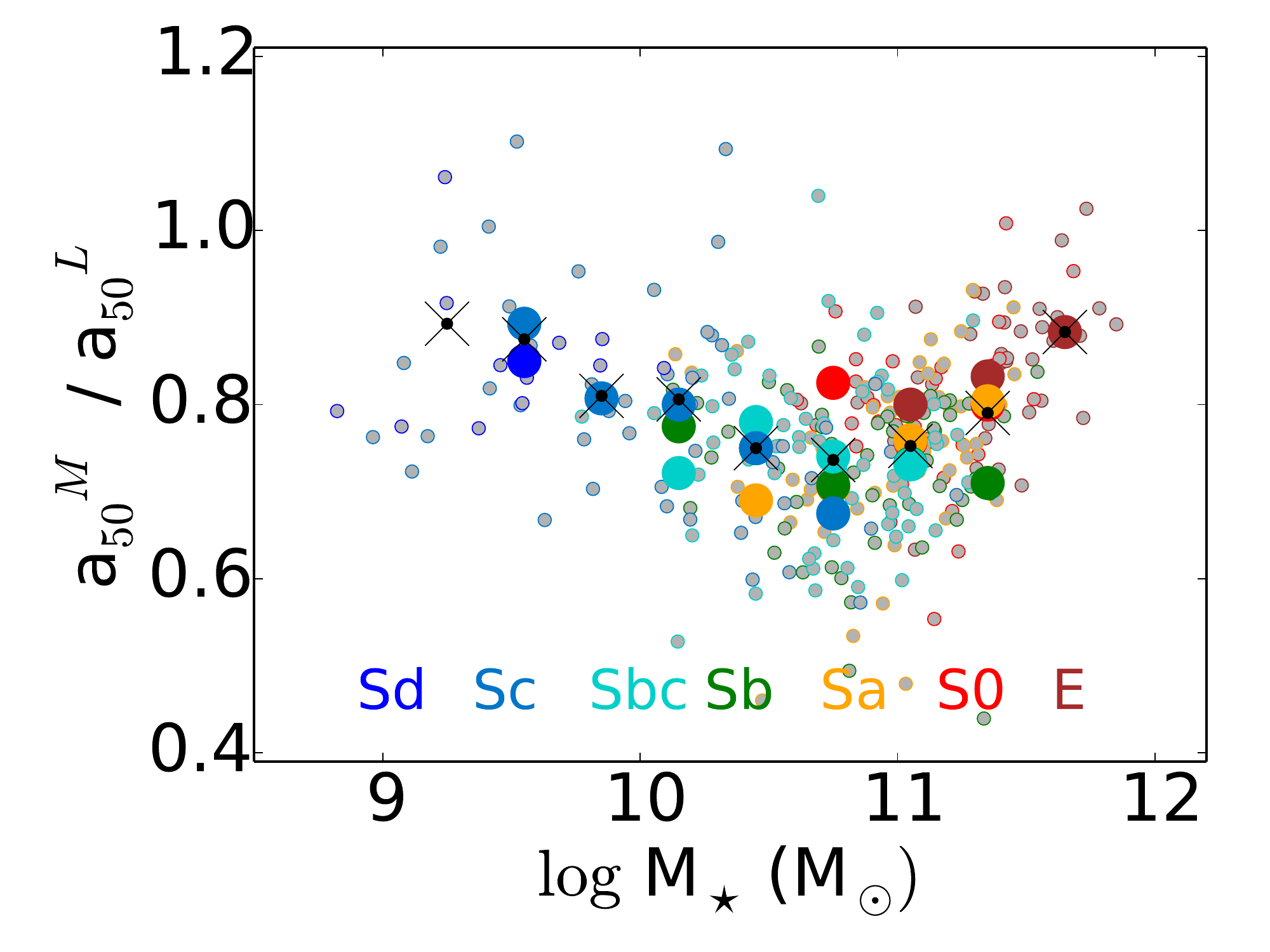}
\caption{Left: The ratio between half mass and half light radius (a$^M_{50}$/ a$^L_{50}$) with the Hubble type (left). Big colored dots represent the averaged a$^M_{50}$/ a$^L_{50}$ in each Hubble type bin, and the lines the dispersion. Stars and big circles show the results obtained with the $GMe$ and $CBe$ bases, respectively. Right: a$^M_{50}$/ a$^L_{50}$ as a function of the galaxy stellar mass. The black circles show the averaged correlation independently of the morphological type.  Large circles represent the averaged relation in mass intervals of 0.25 dex for each color-coded morphological type. } 
\label{fig:HMR}
\end{figure*}

As explained in \citet{cidfernandes13}, we define the HLR as the semi-major axis length of the elliptical aperture  that contains half of the total light of the galaxy at the rest-frame wavelength 5635 \AA. Similarly, the HMR is derived from the 2D distribution of the stellar mass, as the elliptical aperture at which the mass curve of growth reaches 50\% of its asymptote. The ratio between the HMR and the HLR ($a_{50}^M$/$a_{50}^L$) reflects the spatial variation of the star formation history in a galaxy. This ratio is  lower than 1 in almost all cases \citep{gonzalezdelgado14a},  a signpost of the inside-out growth found by \citet{perez13}. 

Fig. \ref{fig:HMR} shows the relation between $a_{50}^M$/$a_{50}^L$ and  Hubble type (left panel), and galaxy stellar mass (right panel). These plots confirm our earlier finding that galaxies are more compact in mass than in light.
If the gradient in stellar extinction is taken into account, the average $a_{50}^M$/$a_{50}^{L_{intrin}} = 0.82 \,(0.80) \pm 0.10 \,(0.13)$ for base {\it GMe} ({\it CBe}).  Fig. \ref{fig:HMR} shows that the ratio decreases from late to early type spirals; while lenticulars and ellipticals have similar $a_{50}^M$/$a_{50}^L$. 

These results are also in agreement with our previous result that $a_{50}^M$/$a_{50}^L$ shows a dual dependence with galaxy stellar mass: It decreases with increasing mass for disk galaxies but it is almost constant in spheroidal galaxies, as confirmed in the right panel of Fig. \ref{fig:HMR}.
Sb-Sbc  galaxies are the ones with the lowest $a_{50}^M$/$a_{50}^L$.

\subsection{Stellar metallicity}

\label{sec:metallicity}

Metallicity is  one of the most difficult stellar population properties to estimate. Reasons for this include: {\it (i)} the coarse metallicity grid of the SSP bases; {\it (ii)} the limitation of the stellar libraries to the solar neighborhood; and {\it (iii)}  inherent degeneracies like the dependence of the continuum shape on extinction, age, and metallicity, whose effects are hard to disentangle. Notwithstanding these difficulties, meaningful estimates of $Z_\star$ can be extracted from observed spectra, particularly by means of full spectral synthesis methods (\citet{sanchez-blazquez11}).

\starlight-based estimates of $Z_\star$ for the same CALIFA sample used in this paper were previously used  by \citet{gonzalezdelgado14b} to study global and local relations of $Z_\star$ with the stellar mass and stellar mass surface density. We have shown there that: {\it (i)}  our sample follows a well defined stellar mass-metallicity relation (MZR), {\it (ii)}  this relation is steeper than the one obtained from O/H measurements in HII regions, but that considering only young stellar populations the two MZR's are similar, and {\it (iii)}  $Z_\star$ is strongly related to $\mu_\star$ in galaxy disks and to $M_\star$ in spheroids. All these results lend confidence to our $Z_\star$ estimates.

Here we review our definition of the mean stellar metallicity, and test whether its value at 1 HLR matches the galaxy wide average value as well as the one obtained from the spatially collapsed data cube.

\subsubsection{Mean stellar metallicity}

The main properties analyzed in this paper are the stellar mass surface density ($\mu_*$), stellar extinction ($A_V$), mean age (\ageL), and metallicity of the stellar population, whose spatial distributions are studied as a function of Hubble type and total stellar mass ($M_\star$). These properties were defined in previous articles in this series. For instance, the mean light weighted log stellar age is defined as 

\begin{equation} 
\label{eq:at_flux}
\langle \log\ age \rangle_L = \sum_{t,Z} x_{tZ} \times \log t
\end{equation}

\noindent \citep[eq.\ 9 of][]{cidfernandes13}, where $x_{tZ}$ is the fraction of flux at the normalization wavelength (5635 \AA) attributed to the base element with age $t$ and metallicity $Z$. The mass weighted version of this index, $\langle \log age \rangle_M$, is obtained replacing $x_{tZ}$ by its corresponding mass fraction $m_{tZ}$. 


While \citet{cidfernandes13} average the base metallicities linearly (their eq.\ 10), in this paper, as in \citet{gonzalezdelgado14b}, we employ a logarithmic average:

\begin{equation} 
\label{eq:alogZ_mass}
\langle \log Z_\star \rangle_M = \sum_{t,Z} m_{tZ} \times \log\ Z
\end{equation}

\noindent for the mass weighted mean $\log Z_\star$ and

\begin{equation} 
\label{eq:alogZ_flux}
\langle \log Z_\star \rangle_L = \sum_{t,Z} x_{tZ} \times \log\ Z
\end{equation}

\noindent for the luminosity weighted mean $\log Z_\star$.
The motivation to use this definition is that the extended SSP bases used in this study span a much wider dynamical range in $Z_\star$ (nearly three orders of magnitude, compared to barely one in our previous papers), which is better handled with a geometric mean (implicit in the use of the logarithm). This is the same reasoning behind the use of 
$\langle \log age \rangle$ instead of $\log \langle age \rangle$. 

To some degree, the definition of mean $Z$ is largely a matter of taste (albeit one with mathematical consequences because of the inequality of the arithmetic and geometric means, $\langle \log Z \rangle\leq \log \langle Z \rangle$), so much so that one finds both types of averaging in the literature. For instance, in  \citet{gallazzi05} metallicities are averaged logarithmically, whereas \citet{asari07} work with arithmetic averages.

As shown in \citet{gonzalezdelgado14b} (see also Fig.\ \ref{fig:metalgalaxy}), our metallicities span about 1 dex for galaxy masses ranging from $10^9$ to $10^{12} M_\odot$, with an MZR which matches well the stellar metallicities of both Milky Way and LMC-like galaxies.

\subsubsection{Galaxy averaged stellar metallicity}

\label{sec:galaxyaveragedmetallicity}

\citet{gonzalezdelgado14a} obtained the important result that galaxy-averaged stellar ages, mass surface density, and extinction are well matched by the corresponding values of these properties at $R = 1$ HLR and also with the values obtained from the analysis of the integrated spectrum (i.e, the one obtained by collapsing the datacube to a single spectrum). The general pattern therefore is that galaxy averaged properties match both the values at 1 HLR and those obtained from integrated spectra. Do our stellar metallicities comply with this rule?

To answer this question we first define the galaxy-wide average stellar metallicity following  eq.\ 2 in \citet{gonzalezdelgado14b}. which gives the mass weighted mean value of $\langle \log Z_{\star, xy} \rangle_M$ as

\begin{equation}
\label{eq:logZmass_Galaxy}
\langle \log Z_\star \rangle_M^{galaxy}  =
\frac{ \sum_{xy} M_{\star,xy} \langle \log Z_\star \rangle_{M,xy} }{ \sum_{xy} M_{\star,xy} }
\end{equation}

\noindent where $M_{\star ,xy}$ is the stellar mass in spaxel $xy$. 

\begin{figure}
\includegraphics[width=0.48\textwidth]{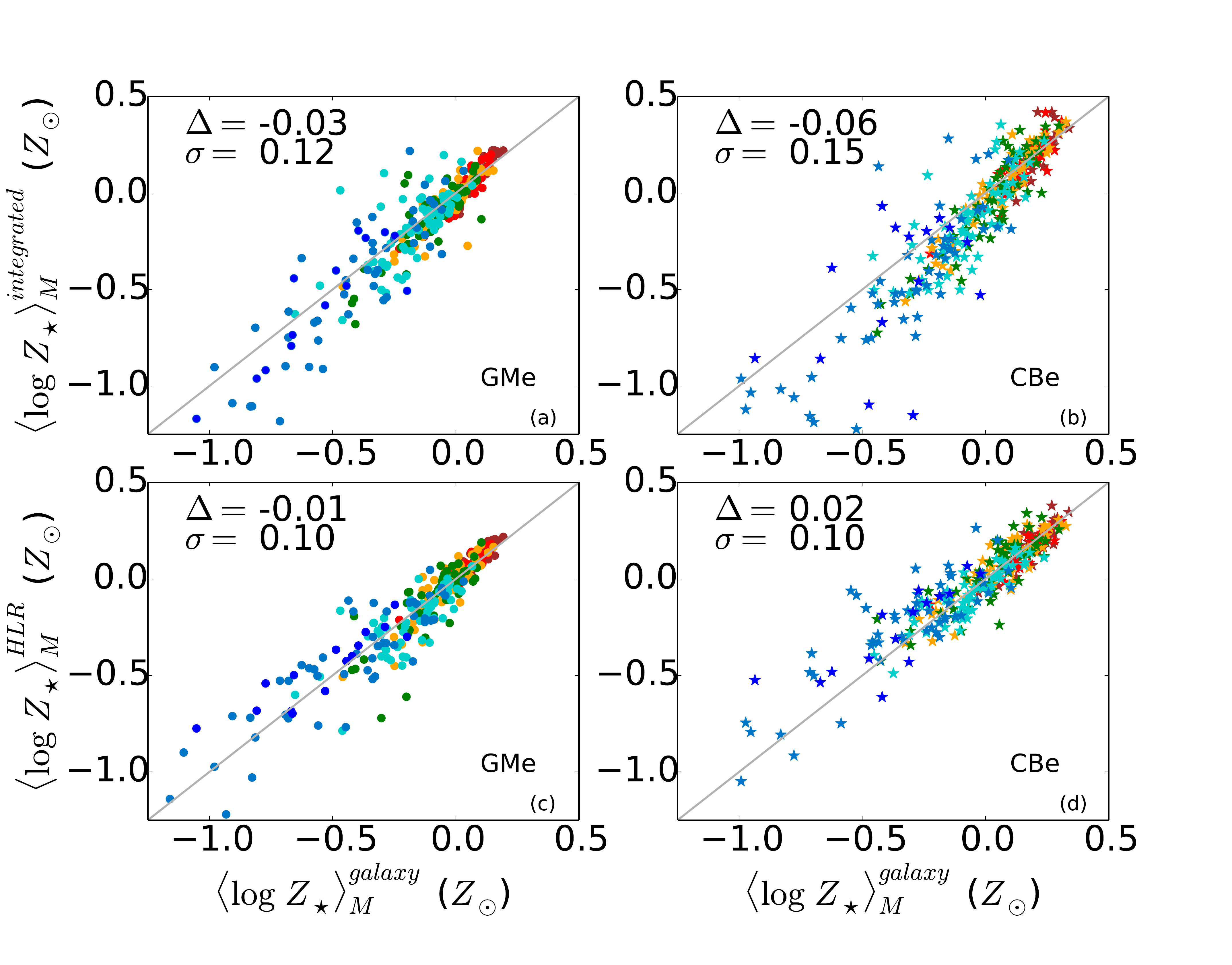}
\caption{Upper panels: Comparison of the galaxy-wide average stellar metallicity (weighted in mass) derived from the spatially resolved spectral analysis ($\langle \log Z_\star \rangle_M^{galaxy}$) and the integrated metallicity derived from fitting the total (integrated) galaxy spectrum ($\langle \log Z_\star \rangle_M^{integrated}$). Lower panel: Comparison of $\langle \log Z_\star \rangle_M^{galaxy}$ with the value measured $R = 1$ HLR ($\langle \log Z_\star \rangle_M^{HLR}$). Left and right panels show results obtained with base {\it GMe} and {\it CBe} SSPs, respectively.  All panels include 300  galaxies. The difference between the y-axis and x-axis is labeled in each panel as $\Delta$, and the dispersion as $\sigma$.}
\label{fig:metalgalaxy}
\end{figure}

Fig.\ \ref{fig:metalgalaxy} compares our results for 
$\langle \log Z_\star \rangle_M^{galaxy}$ with the mass weighted mean $\langle \log Z_\star \rangle_M$ values obtained at $R = 1$ HLR ($\langle \log Z_\star \rangle_M^{HLR}$, bottom panels) and those derived from the integrated spectrum ($\langle \log Z_\star \rangle_M^{integrated}$, top panels), analyzed in the exact same way as the individual zone spectra. Results are shown for both base {\it GMe} (left panels) and {\it CBe} (right).

The agreement is remarkable. The galaxy averaged metallicity and the one at 1 HLR are the same to within a dispersion of 0.1 dex. The integrated metallicity also matches the galaxy averaged value, with only slightly larger dispersions. The largest deviations occur for low metallicity systems. Similar conclusions are reached if the comparison in Fig.\ \ref{fig:metalgalaxy} is done using the light weighted version of eq.\ (\ref{eq:logZmass_Galaxy}),

\begin{equation}
\label{eq:logZlight_Galaxy}
\langle \log Z_\star \rangle_L^{galaxy} = 
\frac{ \sum_{xy} L_{\star, xy} \langle \log Z_\star \rangle_{L,xy} }{ \sum_{xy} L_{\star,xy} }
\end{equation}

\noindent where $L_{\star, xy}$ is the luminosity (corrected by stellar extinction) in each spaxel evaluated at a reference wavelength  (5635 \AA\ in our case).

The stellar metallicities behave as expected, in the sense that, like other properties, their galaxy-wide averages match the values at $R = 1$ HLR, and also the  values derived from integrated spatially unresolved spectroscopy \citep{gonzalezdelgado14a}.  We thus conclude that  galaxy-wide spatially averaged stellar population properties (stellar mass, mass surface density, age, metallicity, and extinction) match those obtained from the integrated spectrum, and that
these spatially averaged properties match those at R= 1 HLR, proving that  effective radii are really effective \citep{gonzalezdelgado14c}.

\section{Spatially resolved stellar population properties as a function of morphology and mass}

\label{sec:Results}

This section presents a series of results derived from our spatially resolved spectral synthesis analysis of CALIFA galaxies. We focus on the following four stellar populations properties: mass surface density ($\mu_*$, \S\ref{sec:subsec_McorSD}), mean ages (\ageL, \S\ref{sec:subsec_ages}), metallicities (\logZM, \S\ref{sec:subsec_Z}), and extinction ($A_V$, \S\ref{sec:subsec_AV}).
Each of these properties is studied by means of {\em (i)} 2D maps of the individual galaxies, {\em (ii)} radial profiles, and {\em (iii)} radial gradients. 
Throughout the section, the emphasis is on evaluating and comparing the roles of morphology and total stellar mass in shaping the observed behavior of these four properties.

Before  discussing the results, we briefly explain how these quantities are obtained and how they are presented.

\ls\noindent {\it 2D maps in the CMD:} Using \pycasso\ we obtain, for each galaxy, 2D maps of each of the four properties. 
The results for all the galaxies are presented in the framework of the color-magnitude diagram, where each map is placed at the galaxy's coordinates in the $u-r$ vs.\ $M_r$ CMD.
Because absolute magnitude is related to $M_\star$ and  redder galaxies are (usually)  older and more metal rich, 
these plots  show the correlations $M_\star$-$\mu_\star$,  $M_\star$--age,  and $M_\star$--metallicity in a 2D fashion. Because in our sample the galaxy Hubble type is correlated with color and luminosity, these plots not only show how the galaxy averaged properties and their radial structure change with the galaxy stellar mass, but also with the morphological type. These maps are shown in the Appendix \ref{ap:2dmap} (Figs.\ \ref{fig:cmd_mu}--\ref{fig:cmd_AV}).

\ls\noindent  {\it Radial profiles:} Each 2D map is azimuthally averaged in order to study the radial variations of each of the four stellar population properties. 
Elliptical apertures 0.1 HLR in width are used to extract the radial profiles, with ellipticity and position angle  obtained from the moments of the 5635 \AA\ flux image.
Expressing radial distances in units of HLR allows the profiles of individual galaxies to be compared on a common metric, as well as averaging (``stacking'') them as a function of  Hubble type or stellar mass.  Radial profiles expressed in units of the HMR were also analyzed and lead to similar shapes, so all profiles presented below use HLR as the unit for radius.

\ls\noindent  {\it Radial gradients:} Inner and outer gradients are defined as differences between the values at $R = 1$ and 0 ($\bigtriangledown_{in}$), and $R = 2$ and 1 ($\bigtriangledown_{out}$), respectively. For instance,

\begin{equation}
\label{eq:innergradient}
\bigtriangledown_{in} \log \mu_\star = 
\log \mu_\star({\rm 1\, HLR}) - \log \mu_\star(0)
\end{equation}

\begin{equation}
\label{eq:outergradient}
\bigtriangledown_{out} \log \mu_\star = 
\log \mu_\star({\rm 2\, HLR}) - \log \mu_\star({\rm 1\, HLR})
\end{equation}

\ls\noindent for $\log \mu_\star$, and similarly for \ageL,  \logZM\ and $A_V$. 
Defined in this way, the gradients have units of dex/HLR (mag/HLR for $\bigtriangledown A_V$). Since the stellar population properties of galaxies at 1 HLR represent very well the galaxy-wide average, $\bigtriangledown_{in}$ ($\bigtriangledown_{out}$) effectively measures how the bulge (disk) properties change with respect to those of the galaxy as a whole.\footnote{Based on the exponential fit analysis developed by \citep{sanchez13} and \citep{sanchez-blazquez14}, we conclude that the regions between 1 and 2 HLR are dominated by the disk component; thus,  $\bigtriangledown_{out}$ measures the disk gradient. However,  $\bigtriangledown_{in}$ is not measuring the bulge gradient. The reason is that the effective radius (R$_e$) of the spheroidal component can be smaller than 1 HLR, and it shows a dependence with the morphological type. Thus, R$_e\sim1$ HLR for E, but is significantly smaller in late type spirals.}

\ls Unless otherwise noted, all results reported below are for the {\it GMe} base, although the whole analysis was carried out with properties derived with both sets of SSP models discussed in \ref{sec:SSP_spectral_bases}. Differences between {\it GMe} and {\it CBe} SSPs go in the following way: 
{\em (a)} The stellar mass surface density is lower with {\it CBe} than with {\it GMe} by 0.27 dex on average, mostly due to the different IMFs (Salpeter in {\it GMe} versus Chabrier in {\it CBe}).
{\em (b)} Variations in stellar extinction are negligible. 
{\em (c)} {\it CBe} yields somewhat younger ages and higher metallicities than {\it GMe}, by an average of 0.14 dex in \ageL\ and 0.12 dex in \logZM. These shifts reflect the age-metallicity degeneracy, and are mainly a consequence of the different sets of metallicities available in these bases. However, radial gradients are not affected by this degeneracy. A detailed comparison of properties derived with the two bases is given in  Appendix \ref{app:BaseExperiments}.


\subsection{Stellar mass surface density }

\label{sec:subsec_McorSD}

2D maps of the stellar mass surface density for the 300 individual  galaxies of our sample are presented in the Appendix \ref{ap:2dmap} (Fig. \ref{fig:cmd_mu}). Here we discuss the radial structure of $\log \mu_\star$ as a function of Hubble type and $M_\star$. 

\subsubsection{$\mu_\star$--morphology and $\mu_\star$--mass relations}

\begin{figure*}
\center{\includegraphics[width=0.85\textwidth]{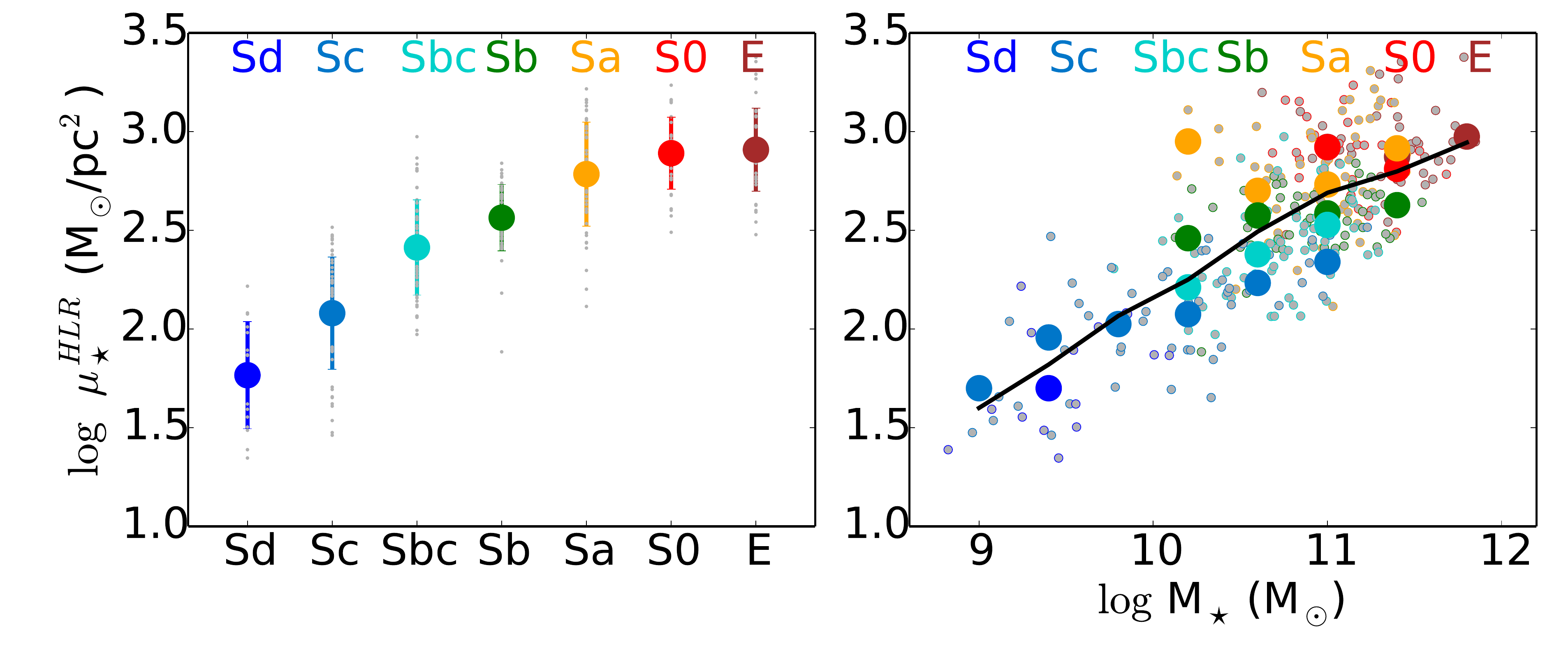}}
\caption{Left panel: stellar mass surface density measured at 1 HLR as a function of Hubble type. Small dots represent $\log \mu_\star$ for each galaxy; the colored circles are the average $\log \mu_\star$ for each Hubble type, and the error bars are the dispersion in $\log \mu_\star$ for each morphological type. Right panel: $\log \mu_\star$-$\log M_\star$ relation. Individual galaxies are represented by small dots colored by their morphological type. The black line is the average $\log \mu_\star$ in galaxy stellar mass bins of 0.4 dex. Large colored circles are the average $\log \mu_\star$ in each bin of mass for each Hubble type.}
\label{fig:mu_tipo_mass}
\end{figure*}

Fig.\ \ref{fig:mu_tipo_mass} shows how $\mu_\star$ measured at 1 HLR changes with  Hubble type (left panel), and with the galaxy stellar mass (right). Recall from \citet{gonzalezdelgado14a} that properties measured at 1 HLR match very well the corresponding galaxy-wide average value, so these plots ultimately show how the global $\mu_\star$ depends on the morphology and on $M_\star$.

The plot shows $\langle\log \mu_\star^{HLR}\rangle$ increasing from late spirals to spheroids, with average and dispersion values of  $3.1 \pm 0.2$, $3.10 \pm 0.18$, $3.05 \pm 0.25$, $2.70 \pm 0.17$, $2.65 \pm 0.24$, $2.40 \pm 0.28$, $2.04 \pm 0.27$, for E, S0, Sa, Sb, Sbc, Sc and Sd, respectively. Note that E and S0 are remarkably similar.

Surface densities also increase with $M_\star$, as seen in the right panel of Fig.\ \ref{fig:mu_tipo_mass}. 
The overall  $\mu^{HLR}_\star$-$M_\star$ relation is relatively smooth, with no evidence of an abrupt change of behavior as that discussed by \citet{kauffmann03} for SDSS galaxies. 
Fig.\ \ref{fig:mu_tipo_mass}, however, reveals that morphology is also behind the dispersion in the $\mu_\star$-$M_\star$ relation.
The black line shows the relation for the full sample, obtained by averaging  $\log \mu_\star$  in 0.4 dex-wide bins in mass, while the big circles break this general relation into different (color-coded) morphological types for the same mass bins. Despite the reduced statistics, it is evident that: (a) for the same stellar mass, early type galaxies are denser than late type ones, and (b) Sa and earlier type galaxies exhibit a much flatter $\mu_\star$-$M_\star$ relation than later types. The overall impression from these results is that morphology, and not only stellar mass, plays a fundamental role in defining stellar surface densities, and it is responsible for the change of slope in the SDSS $\mu_\star$-$M_\star$ relation.

\subsubsection{Radial Profiles}

\begin{figure*}
\includegraphics[width=0.48\textwidth]{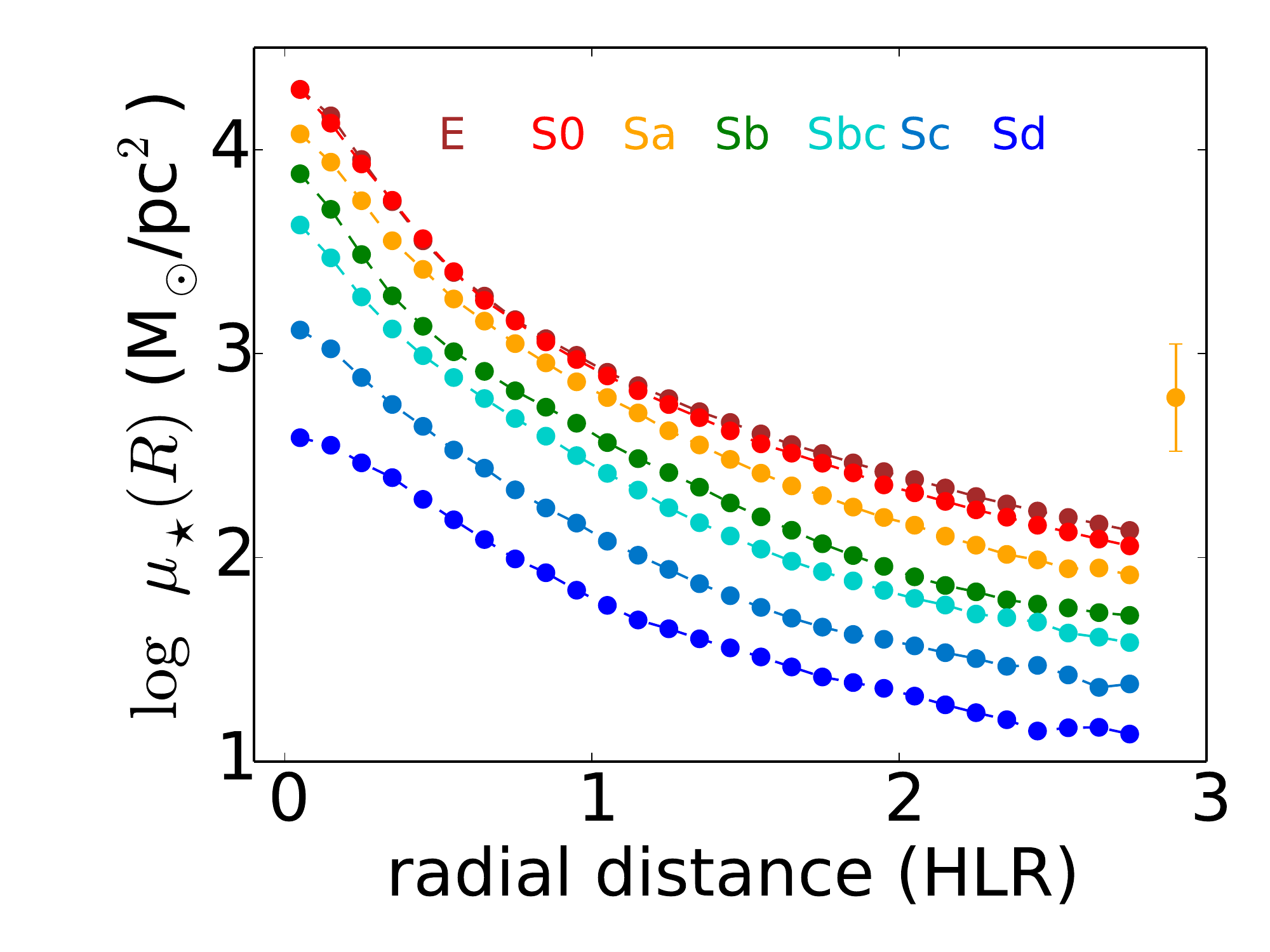}
\includegraphics[width=0.48\textwidth]{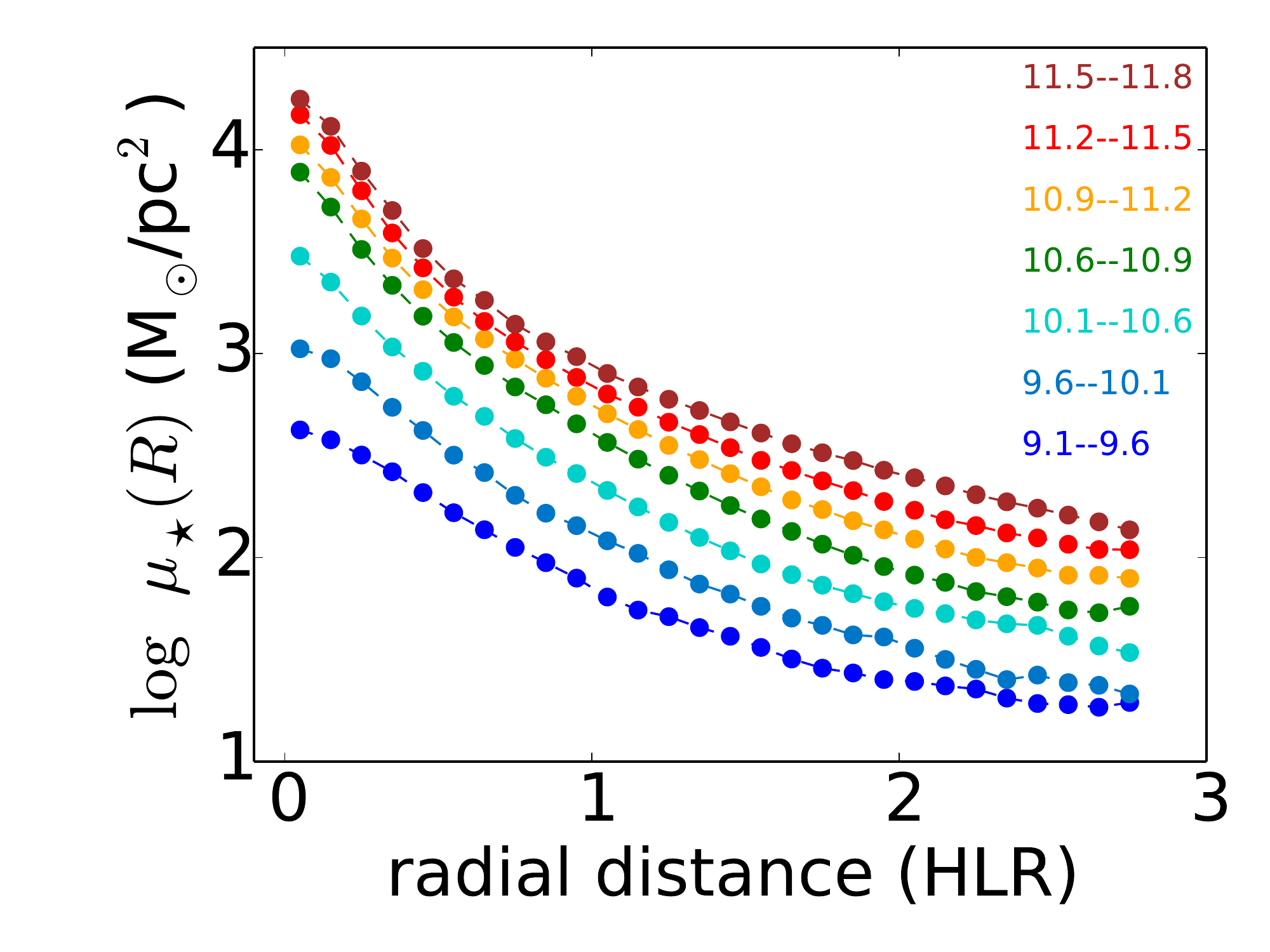}
\caption{(left) Radial profiles (in units of HLR) of the stellar mass surface density obtained with base $GMe$. The results are stacked in seven morphology bins. The error bar in the panel indicate the dispersion at one HLR distance in the galaxies of the Sa bin. It is similar for other Hubble types and radial distances. (right) Radial profiles stacked in seven bins of galaxy stellar mass, $\log M_\star (M_\odot)$: 9.1$-$9.6, 9.6$-$10.1, 10.1$-$10.6, 10.6$-$10.9, 10.9$-$11.2, 11.2$-$11.5, 11.5$-$11.8.}
\label{fig:mu_radialprofiles}
\end{figure*}

Azimuthally averaged radial profiles of $\log \mu_\star$ are shown in Fig.\ \ref{fig:mu_radialprofiles}. Results are stacked by Hubble type (left panel) and mass (right).
In the left panel galaxies are grouped in our seven morphological classes. 
The typical dispersion within these bins is illustrated by the error bar, which shows the standard deviation in $\log \mu_\star \ (R=1\, HLR)$ for galaxies of the Sa class.

A clear trend with Hubble type is seen: The $\mu_\star(R)$ profiles scale with Hubble type from late to early spirals, and this modulation with morphology is preserved at any given distance. E and S0 have remarkably similar profiles, with core and extended envelope equally dense at any given distance, suggesting that the disk of S0 galaxies and the extended envelope of ellipticals have grown their mass by similar processes. 

The right panel of Fig.\ \ref{fig:mu_radialprofiles} shows the radial profiles grouped in seven bins of stellar mass spanning the $\log M_\star (M_\odot) =$ 9.1--11.8 range. These also show that the average of $\log \mu_\star (R)$ is modulated by  $M_\star$. However, this $\mu_\star (R)$-$M_\star$ modulation breaks for early type galaxies (concentration index C ($r_{90}/r_{50}$) $\geq$2.8; see also Fig.13 in \citet{gonzalezdelgado14a}), that in our sample are populated mainly by E and S0, and some Sa. On the other hand these early types are all massive, with $M_\star \geq$ 10$^{11}$ M$_\odot$.

\subsubsection{Radial gradients}

\begin{figure*}
\includegraphics[width=0.48\textwidth]{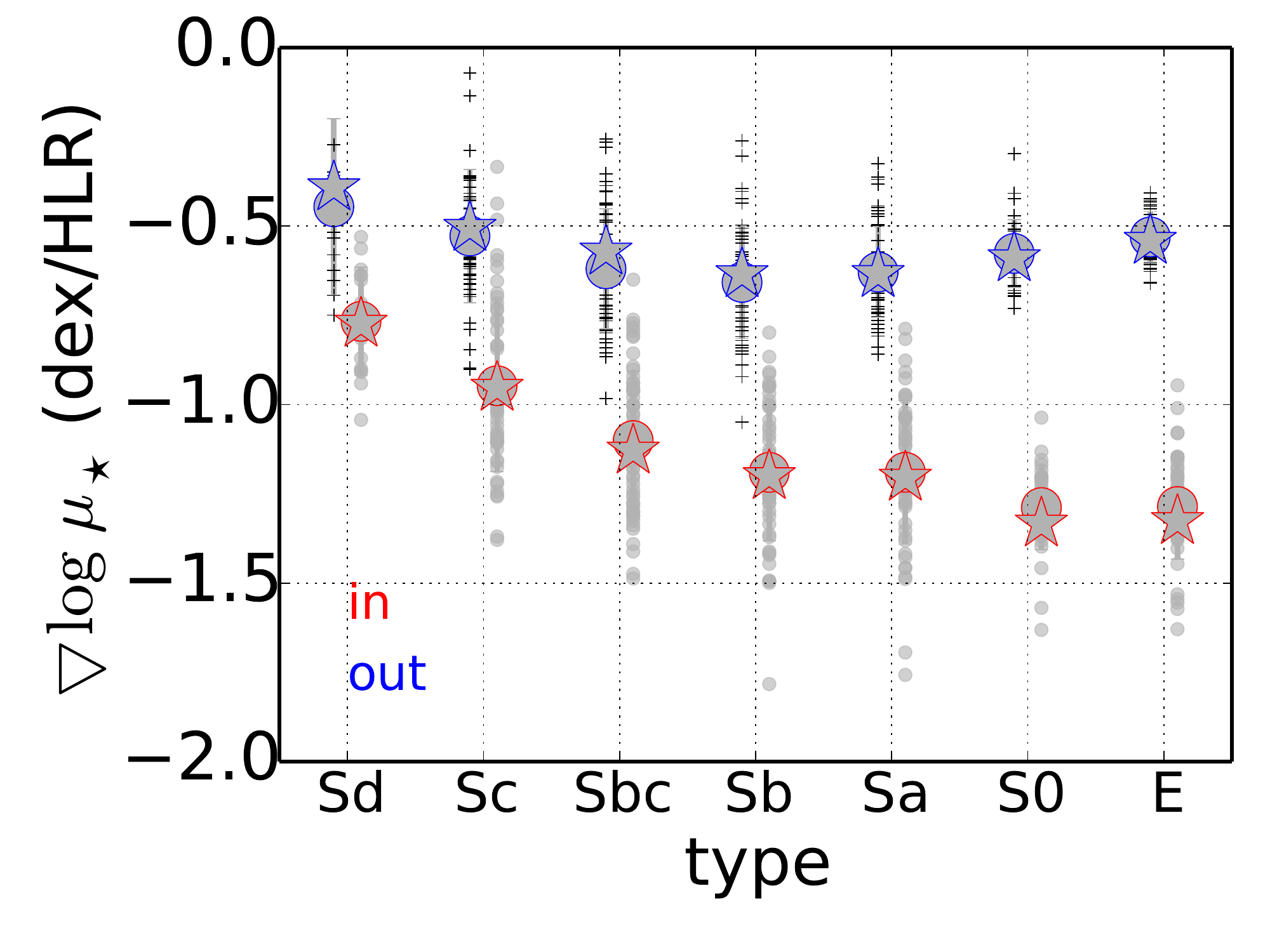}
\includegraphics[width=0.48\textwidth]{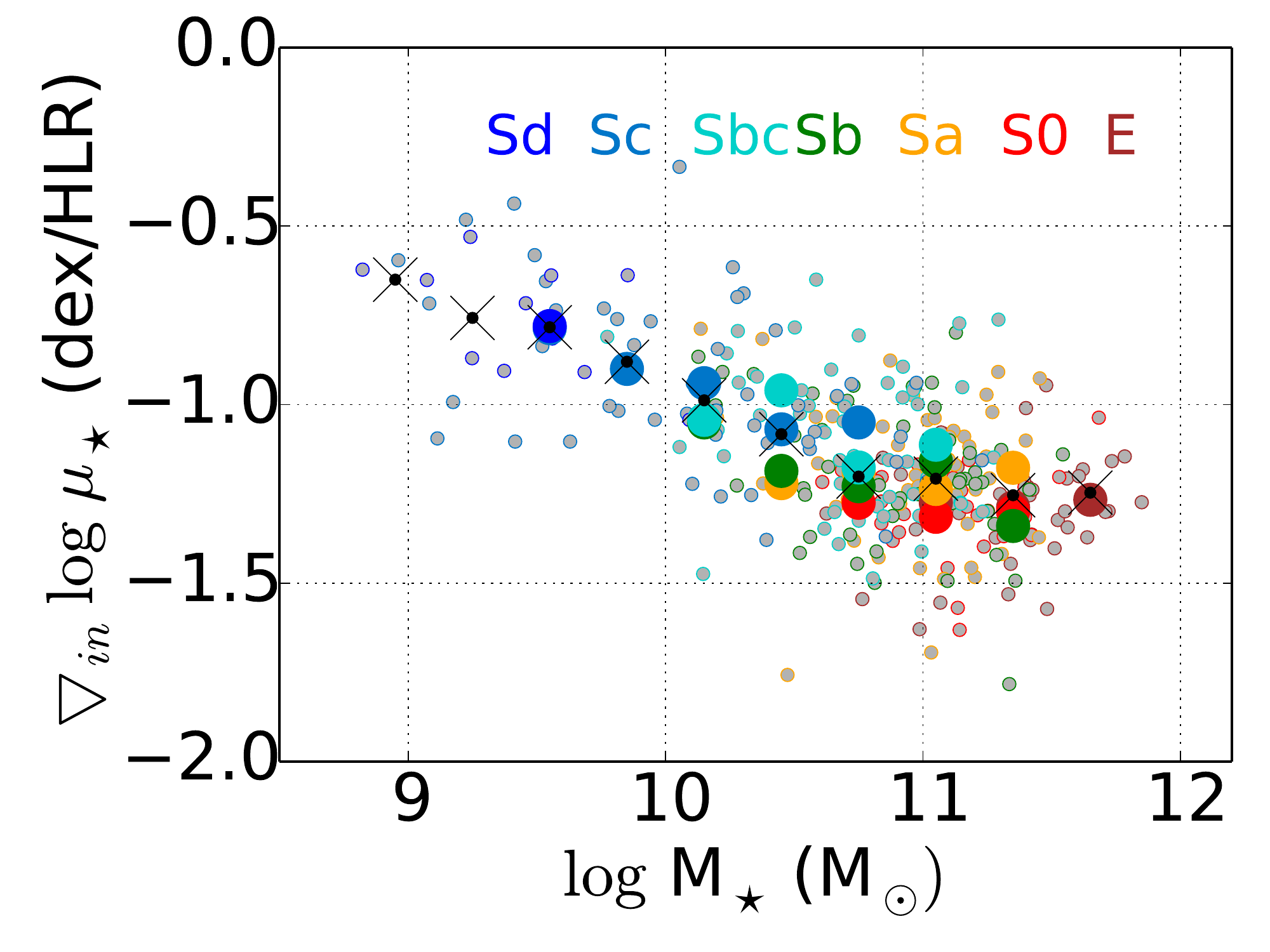}
\caption{(left)  Correlation between the inner (grey-red) and outer (grey-blue) gradient of $\log\ \mu_\star$  and the morphological type. The results are shown for  the $GMe$ (stars) and $CBe$ (circles) SSP models. The inner gradient is calculated between the galaxy nucleus and 1 HLR, and the outer gradient between 1 HLR and 2 HLR. (right) Correlation between the inner gradient of $\log\ \mu_\star$ and the galaxy stellar mass. Small dots represent the results for each galaxy, and black crosses the average for each 0.3 dex mass bin. Large circles represent the averaged inner gradient in mass intervals of 0.3 dex for each color-coded morphological type. Black crosses show the average correlation between the inner gradient of $\log \mu_\star$ and galaxy mass independently of the morphological type.
}
\label{fig:mu_gradient}
\end{figure*}

Inner (0--1 HLR) and outer (1--2 HLR) gradients in $\log \mu_\star$, as defined by equations (\ref{eq:innergradient}) and (\ref{eq:outergradient}), are plotted as a function of morphology and stellar mass in Fig.\ \ref{fig:mu_gradient}.
$\bigtriangledown_{in} \log \mu_\star$ values (corresponding to the core region) are plotted in grey-red, while
$\bigtriangledown_{out} \log \mu_\star$ (which trace the disks of spirals and S0 and the extended envelope of ellipticals) are plotted in grey-blue. Circles and stars show results for bases {\it GMe} and {\it CBe} respectively, illustrating that even though these bases yield different absolute values of $\mu_\star$ the resulting gradients are nearly identical.

A clear correlation exists between $\bigtriangledown_{in} \log \mu_\star$ and Hubble type. The gradient in the inner HLR increases (in absolute values) significantly from late to early spirals, converging to a constant value for E and S0. This relation reflects the variation of the bulge to disk ratio in spirals, and the dominance of the bulge component in spheroids (S0 and E). The outer gradient is weaker (smaller in absolute value) than  the inner one, as expected if a disk component dominates the mass outwards of 1 HLR. 
 
The right panel of Fig.\ \ref{fig:mu_gradient} shows the relation between the inner gradient and the  stellar mass. There is a clear increase (in absolute values) of $\bigtriangledown_{in}\log \mu_\star$ with $M_\star$, with the more massive galaxies having a steeper increase of the central density. The dispersion with respect to the average values (black cross) within $M_\star$-bins  is significant. To check the effect of morphology on this dispersion we have averaged $\bigtriangledown_{in}\log \mu_\star$  in mass intervals for each Hubble type and plotted the resulting averages (large colored circles). 
The general trend that emerges is that, for galaxies of the same mass, early type galaxies tend to be overall  centrally denser  than later  types, in agreement with  Fig.\ \ref{fig:mu_tipo_mass};  
although, there are a few  intervals of stellar mass (e.g. $\log M_\star$ = 11.4 M$_\odot$), in which the variations in $\bigtriangledown_{in}\log \mu_\star$ with  Hubble type are not significant.

It is also worth mentioning that  $\bigtriangledown_{in}\log \mu_\star$ in Sa and Sb is very close to that in S0 and E, and in this sense it would be easy to fade early type spirals into S0's.


\subsection{Ages of the stellar populations}

\label{sec:subsec_ages}

2D maps of the  luminosity weighted mean log stellar ages (eq.\  \ref{eq:at_flux}) for the 300 galaxies are presented in Fig.\ \ref{fig:cmd_ageL}. Here we discuss the radial structure of \ageL\ and its relation to Hubble type and $M_\star$.
The presentation follows the same script used in the presentation of $\mu_\star$-related results in \S\ref{sec:subsec_McorSD}.

\subsubsection{Age--morphology and age--mass relations}

\begin{figure*}
\center{
\includegraphics[width=0.85\textwidth]{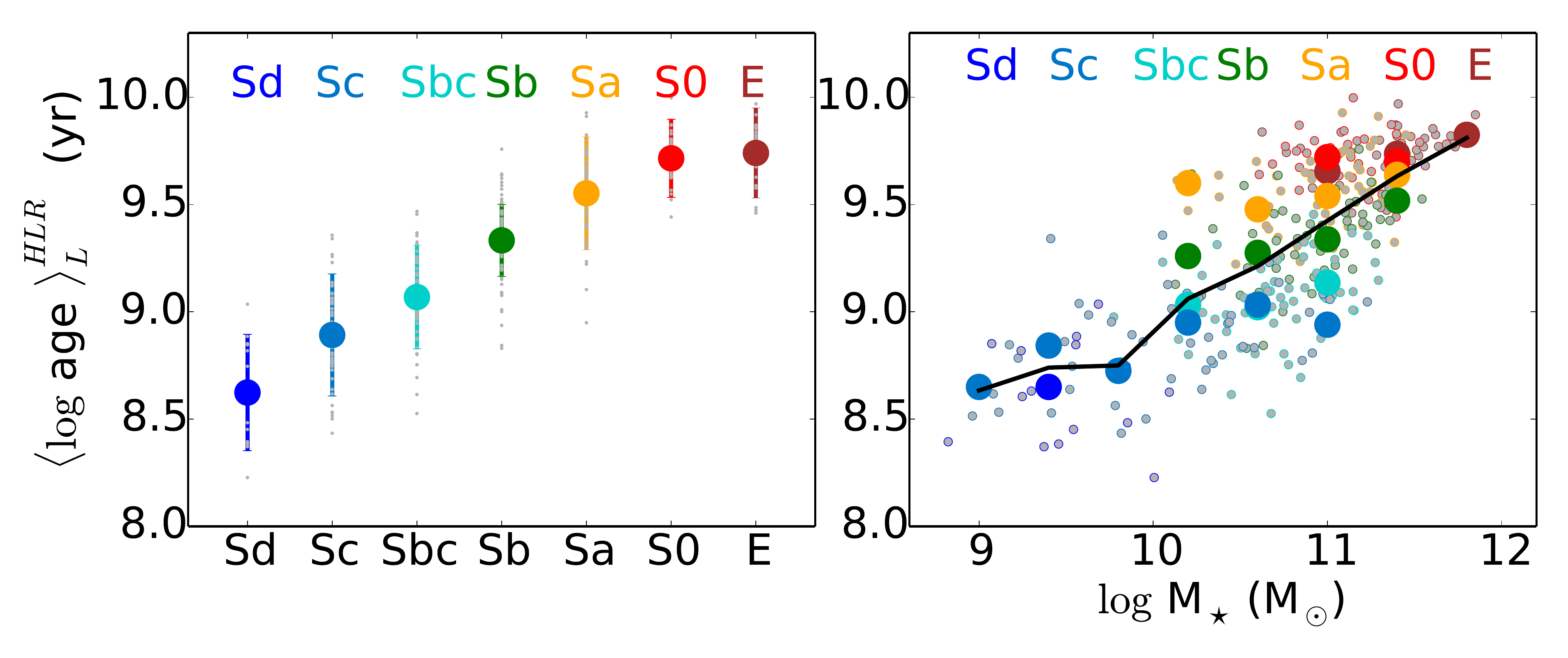}}
\caption{\ageL\ measured at 1 HLR as a function of  Hubble type (left) or galaxy stellar mass (right). 
Symbols and  colors  are as in Fig. \ref{fig:mu_tipo_mass}. The black line is the average \ageL\ in galaxy stellar mass bins of 0.4 dex.}
\label{fig:ageL_tipo}
\end{figure*}

Fig.\ \ref{fig:ageL_tipo} shows how the mean age of the stellar populations at 1 HLR changes along the Hubble sequence (left panel), and with the galaxy stellar mass (right).  Similarly to $\log \mu_\star^{HLR}$, \ageL$^{HLR}$ represents well the galaxy-wide averaged stellar population age (\ageL$^{galaxy}$, \cite{gonzalezdelgado14a}). 

Clearly,  \ageL$^{HLR}$  scales with Hubble type,  increasing steadily from Sd to Sa. S0 and ellipticals have stellar populations of similar mean age, and older than  spirals.  The average and dispersion values of \ageL$^{HLR}$ (yr) are $8.62 \pm 0.22$, $8.89 \pm 0.22$, $9.07 \pm 0.19$, $9.33 \pm 0.21$, $9.55 \pm 0.19$, $9.71 \pm 0.11$, and $9.74 \pm 0.11$, for Sd, Sc, Sbc, Sb, Sa, S0 and E, respectively. 

Mean ages also increase with the galaxy mass  (right panel of Fig. \ref{fig:ageL_tipo}), a ``downsizing"  behavior that has been  confirmed with widely different samples and methods. For instance, our age-mass relation is similar to that derived for SDSS galaxies by \citet{gallazzi05} (their figure 8). They found that there is a transition at $M_\star \sim 3 \times 10^{10} M_\odot$\footnote{Equivalent to $\sim 5.5 \times 10^{10} M_\odot$ for our IMF.}, below which galaxies are typically young and above which they are old. This is the same mass at which \citet{kauffmann03} find the $\mu_\star$-$M_\star$ relation to flatten. 

Unlike in these SDSS-based works, we do not see sharp transitions as a function of $M_\star$ in neither $\mu_\star$ nor \ageL, although differences in sample selection and statistics prevent a proper comparison. We do, however, note a common behavior in the right panels of Figs.\ \ref{fig:mu_tipo_mass} and \ref{fig:ageL_tipo}, in the sense that the dispersion above $\sim 10^{10} M_\odot$ is strongly related to morphology. 

Like in Fig.\ \ref{fig:mu_tipo_mass} (right panel), the black line in the right panel of Fig.\  \ref{fig:ageL_tipo} shows the age-mass relation for the whole sample, obtained by averaging \ageL$^{HLR}$ values in $M_\star$ bins. Small dots show individual galaxies, while the large colored circles represent the mass-binned average \ageL$^{HLR}$ for each Hubble type. 
As with the $\mu_\star$-$M_\star$ relation, breaking the age-mass relation into morphological types reveals clean trends. In this case, we see that, for a fixed $M_\star$, earlier type galaxies are older than later types. The corollary is that mass is not the sole property controlling the SFH of a galaxy. In fact, given the $\sim$ flat age-mass relations for Sa, S0 and E, morphology seems to a more relevant factor, at least in these cases.

\subsubsection{Radial profiles}

\begin{figure*}
\includegraphics[width=0.48\textwidth]{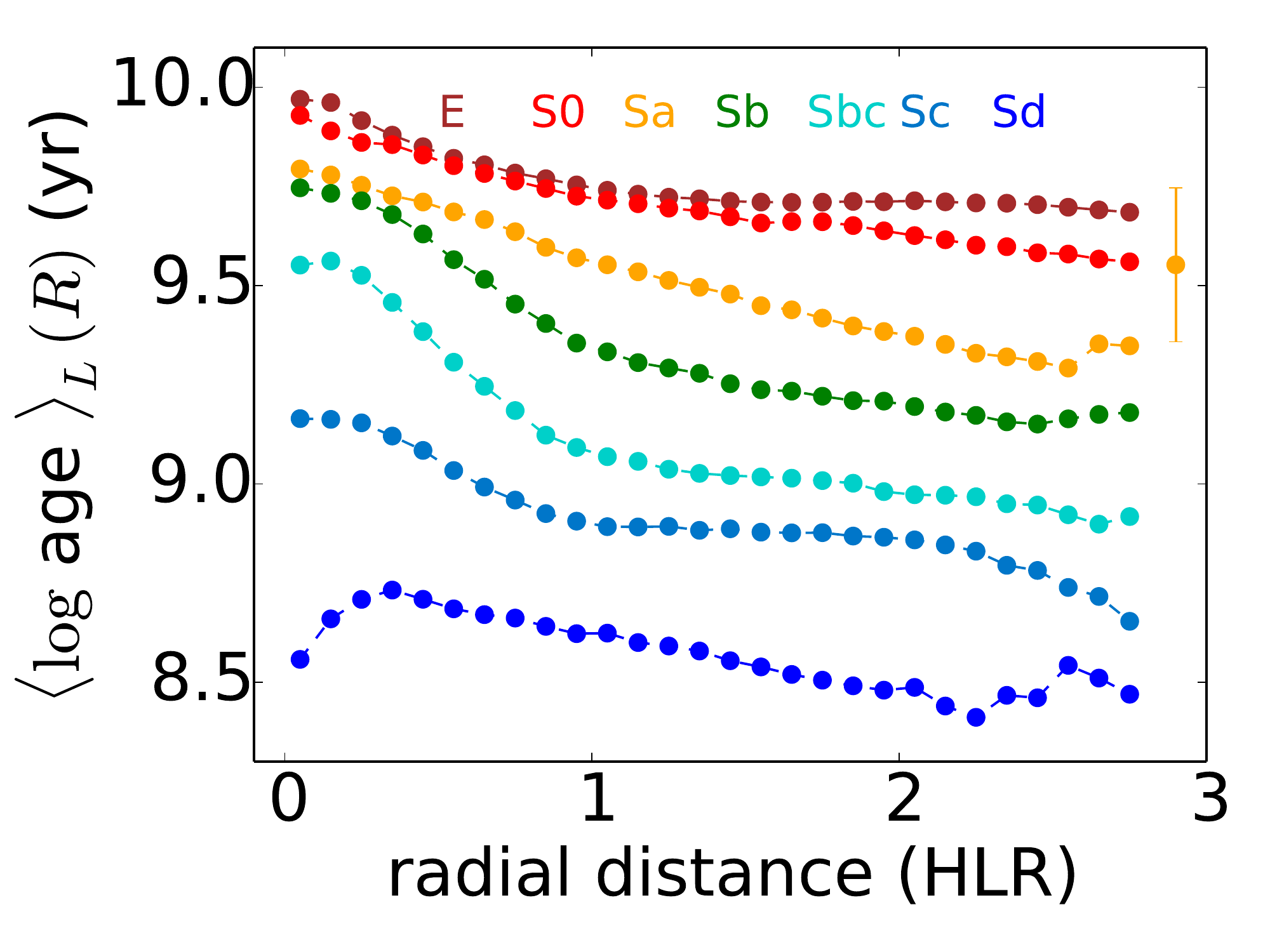}
\includegraphics[width=0.48\textwidth]{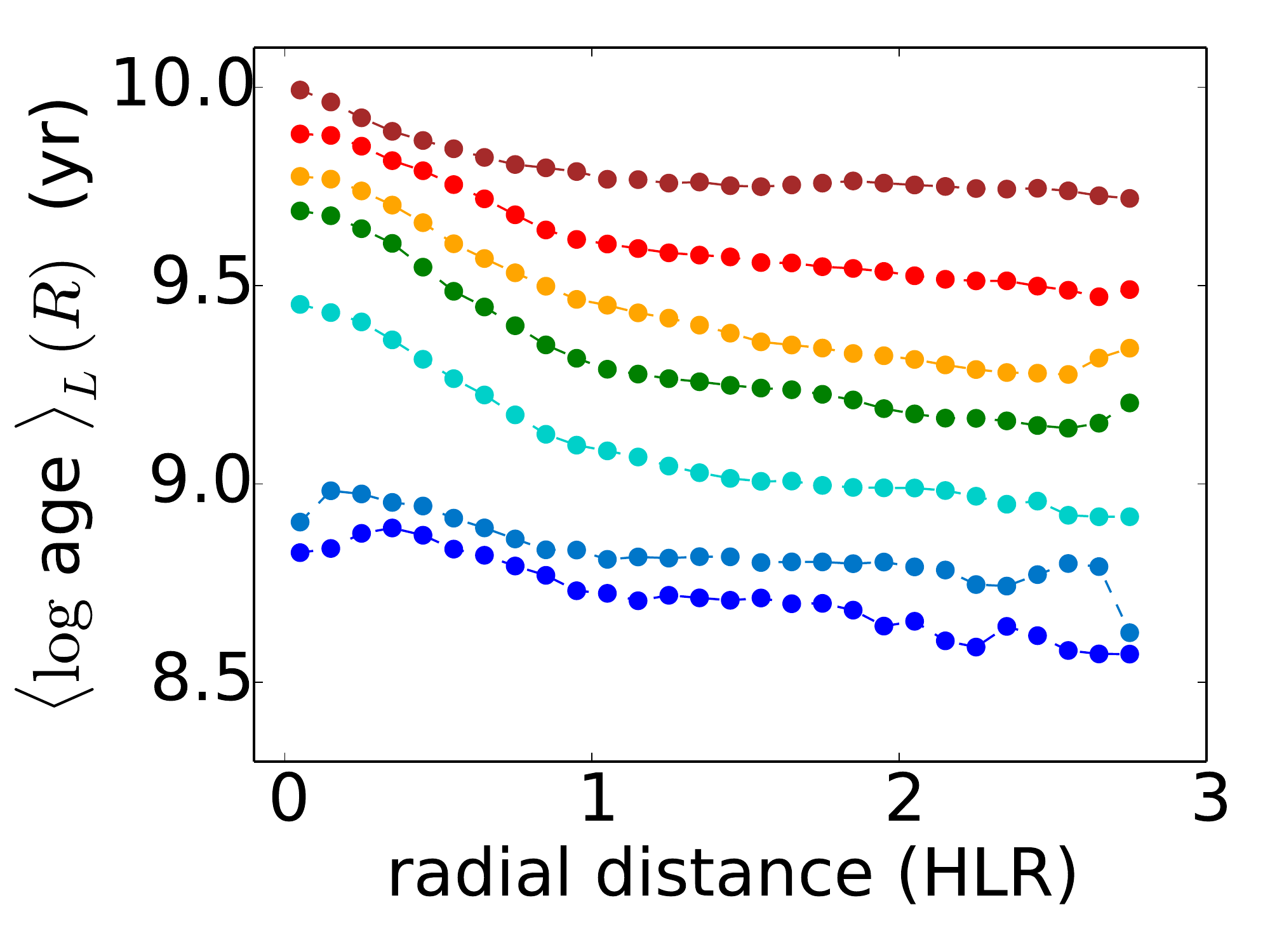}
\caption{ 
Radial profiles of \ageL\ as a function of  Hubble type (left) and in seven bins of galaxy stellar mass (right). These bins 
are $\log M_\star (M_\odot)$ =  9.1$-$9.6, 9.6$-$10.1, 10.1$-$10.6, 10.6$-$10.9, 10.9$-$11.2, 11.2$-$11.5, 11.5$-$11.8. Symbols and colors are as in Fig. \ref{fig:mu_radialprofiles}. 
These results are obtained with base $GMe$. 
}
\label{fig:ageL_radialprofiles}
\end{figure*}

Fig.\ \ref{fig:ageL_radialprofiles} shows the age radial profiles obtained by stacking galaxies as a function of  Hubble type and mass. 
The  \ageL$(R)$ profiles scale with Hubble type, but by different amounts at the center than at 1 HLR. 
At any radial distance, however, the early type galaxies are older than later type ones. 
E and S0 are again very similar at all radii. This suggests that E and S0 have similar histories not only on average, but also spatially resolved, at least in the inner 2 HLR.
Negative age gradients are detected in all  galaxies (except perhaps in Sd, whose ages profiles are flatter than in the other spirals\footnote{The small drop of \ageL\ toward the center of Sd galaxies is caused by a couple of galaxies with young nuclear regions. Given that this group is the least populated in our analysis (only 15 galaxies), better statistics is needed to evaluate the reality of this feature.}). 
These negative gradients reflect the inside-out growth of  galaxies. Furthermore, the decrease of \ageL\ with $R$ indicates that quenching happens earlier at the galaxy center; and also earlier in early type galaxies (spheroids and Sa) than in later type spirals (Sbc--Sc).

The radial profiles also show a clear trend with $M_\star$ (Fig.\ \ref{fig:ageL_radialprofiles}, right), with the more massive galaxies being older everywhere, hence  preserving the downsizing pattern at all  radial distances. Comparing the left and right panels in Fig. \ \ref{fig:ageL_radialprofiles}, one sees that grouping galaxies by their stellar mass leads to a reduced vertical stretch in their \ageL$(R)$ profiles than when the averaging is done by morphology. But the profiles expand similar vertical scale if galaxies earlier than Sd and more massive than 10$^{9.6}$ M$_\odot$ are considered; indicating that the effect of morphology and stellar mass are not easily disentangled here. However, in  \S\ref{sec:subsec_quenching},  Fig. \ref{fig:at_flux_RadialProfiles_tipo_Mass} shows that 
the dispersion in the \ageL$(R)$ profiles  between  galaxies of the same $M_\star$ and different Hubble type is significant, and larger than  between the  \ageL$(R)$ profiles of galaxies of different $M_\star$ but the same Hubble type. These results in agreement with  Fig.\ \ref{fig:ageL_tipo} indicate that the age profiles are more related to morphology than to M$_\star$. Since \ageL$(R)$ is essentially a first moment of the spatially resolved SFH, we can conclude that the SFH and its radial variation are modulated primarily by the galaxy morphology, with mass playing a secondary role.

\subsubsection{Radial gradients}

\begin{figure*}
\includegraphics[width=0.48\textwidth]{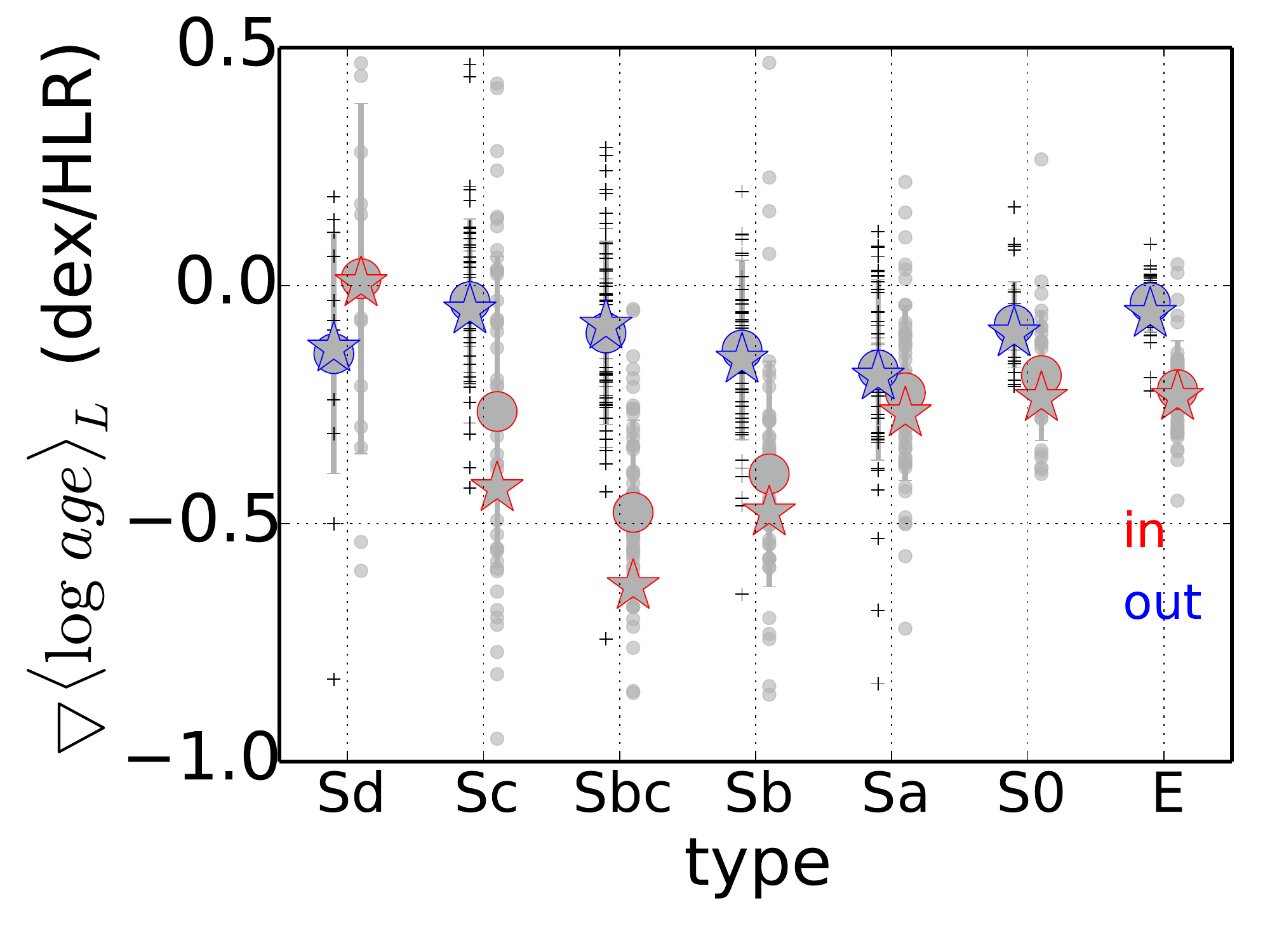}
\includegraphics[width=0.48\textwidth]{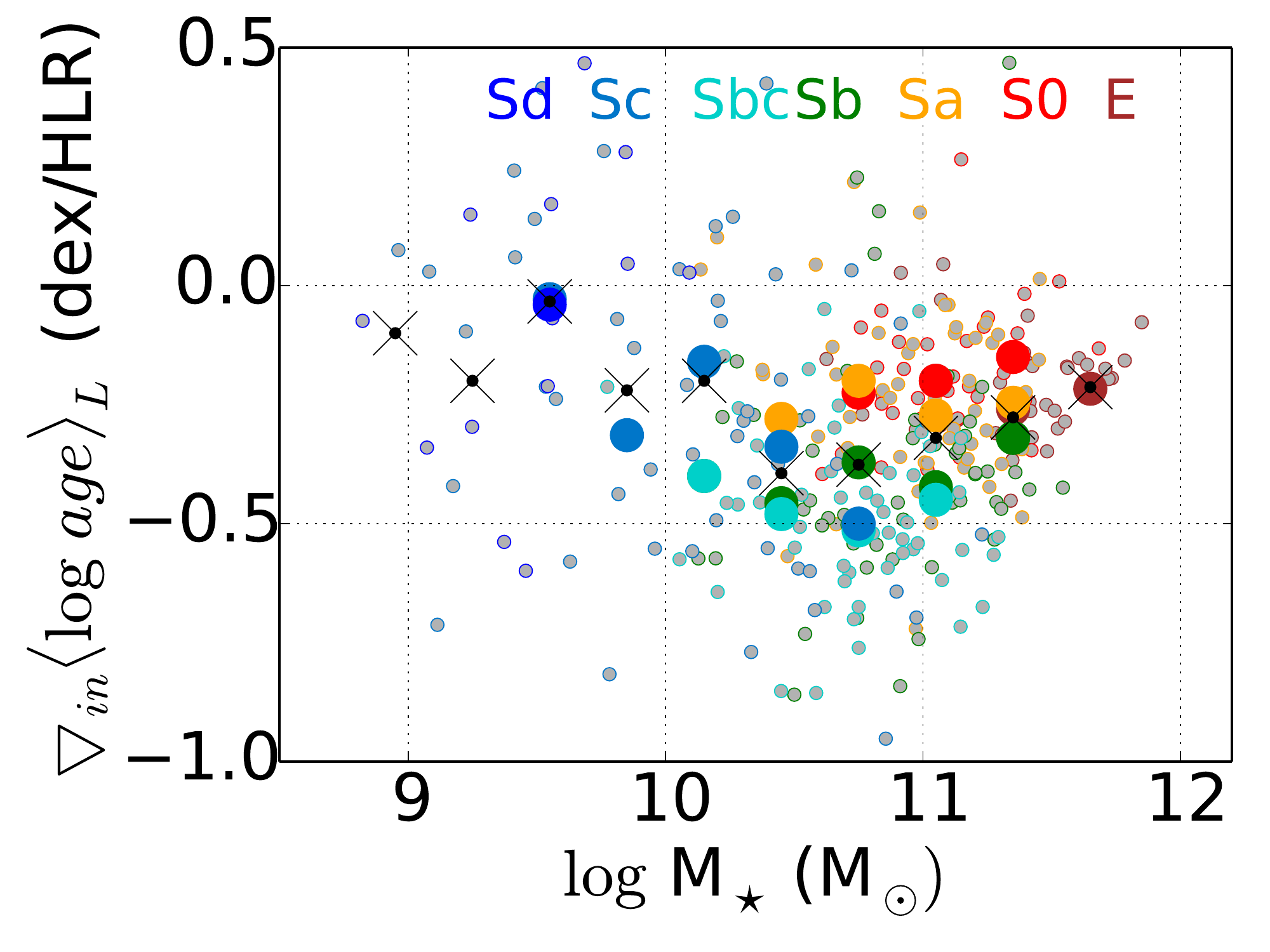}
\caption{(left) As Fig. \ref{fig:mu_gradient} but for \ageL. The inner gradient shows a clear dependence with  Hubble type, that seems to be stronger than with the galaxy mass. Sb-Sbc-Sc galaxies have larger inner  gradients with $CBe$ than with $GMe$, but both sets of models show a similar dependence with Hubble type. (right) The inner gradient of \ageL\ as a function of  galaxy mass. Colors and symbols are as in Fig. \ref{fig:mu_gradient}.}
\label{fig:ageL_gradient}
\end{figure*}

Gradients in \ageL, computed as indicated in eqs.\ (\ref{eq:innergradient}) and (\ref{eq:outergradient}), are plotted in 
Fig.\ \ref{fig:ageL_gradient} against Hubble type (left panel) and stellar mass (right).
The figure layout is exactly as in Fig.\ \ref{fig:mu_gradient}. Whilst in that plot results obtained with bases {\it GMe} and {\it CBe} (circles and stars in the left panel, respectively) could hardly be distinguished, here the results for these two sets of SSPs do not overlap so precisely, although the differences  in $\bigtriangledown\langle \log age \rangle_L$ are clearly very small (see \S\ref{app:Bases_SSP}).

A clear relation exists between $\bigtriangledown_{in} \langle \log age \rangle_L$ and  morphology: The inner age gradient increases from early type galaxies to Sb-Sbc spirals, which are the galaxies with the largest variation between the age of the stellar population at the bulge and the disk. Spirals of later type (Sc and Sd) have flatter radial profiles than Sb-Sbc. The outer (between 1 and 2 HLR) age gradient shows a similar bimodal behavior as  $\bigtriangledown_{in} \ \langle \log age \rangle_L$, but with a smaller amplitude.

The right panel of Fig.\ \ref{fig:ageL_gradient} shows the behavior of $\bigtriangledown_{in}\langle \log age \rangle_L$ with $M_\star$. The gradient tends to increase (become more negative) from low mass galaxies (which have roughly flat  profiles) up to about $10^{11} M_\odot$, at which point the trend reverses and $\bigtriangledown_{in}\langle \log age \rangle_L$ decreases with increasing $M_\star$. This is best seen following the black crosses, that trace the mass-binned mean relation.
The dispersion with respect to this relation is significant and is related to the morphology, as seen through the large colored circles. The tendency is that, at a given mass, S0 and early type spirals have weaker $\bigtriangledown_{in}\langle \log age \rangle_L$ than Sb-Sbc. This dependence of  age gradients with the Hubble type at a fixed $M_\star$ indicates again that the spatial variation of the SFH is mainly driven by the morphology and not by the stellar mass.

However, the morphology (understood as the B/D ratio \citep{graham08}) can not be the only driver of the spatial variation of the SFH along all the Hubble sequence. Fig.\ \ref{fig:ageL_gradient} shows that there is not  a monotonic  relation between the B/D ratio and $\bigtriangledown_{in}\langle \log age \rangle_L$, with galaxies with the smaller B/D ratio having the largest variations in \ageL\ between the central core and the disk. This bimodal behavior seen in Fig.\ \ref{fig:ageL_gradient} suggests that other physical properties are also important in establishing the spatial variation of the SFH, which on the other hand is reflecting the different bulge formation processes along the Hubble sequence.

\subsection{Stellar metallicity}

\label{sec:subsec_Z}

Fig.\ \ref{fig:cmd_logZM} presents the images of the  mass weighted mean (logarithmic) stellar metallicity (cf.\  eq.\ \ref{eq:alogZ_mass}). Here we discuss the radial structure of \logZM\ as a function of  Hubble type and $M_\star$.

\subsubsection{Metallicity-morphology and mass-metallicity relations}

\begin{figure}
\includegraphics[width=0.5\textwidth]{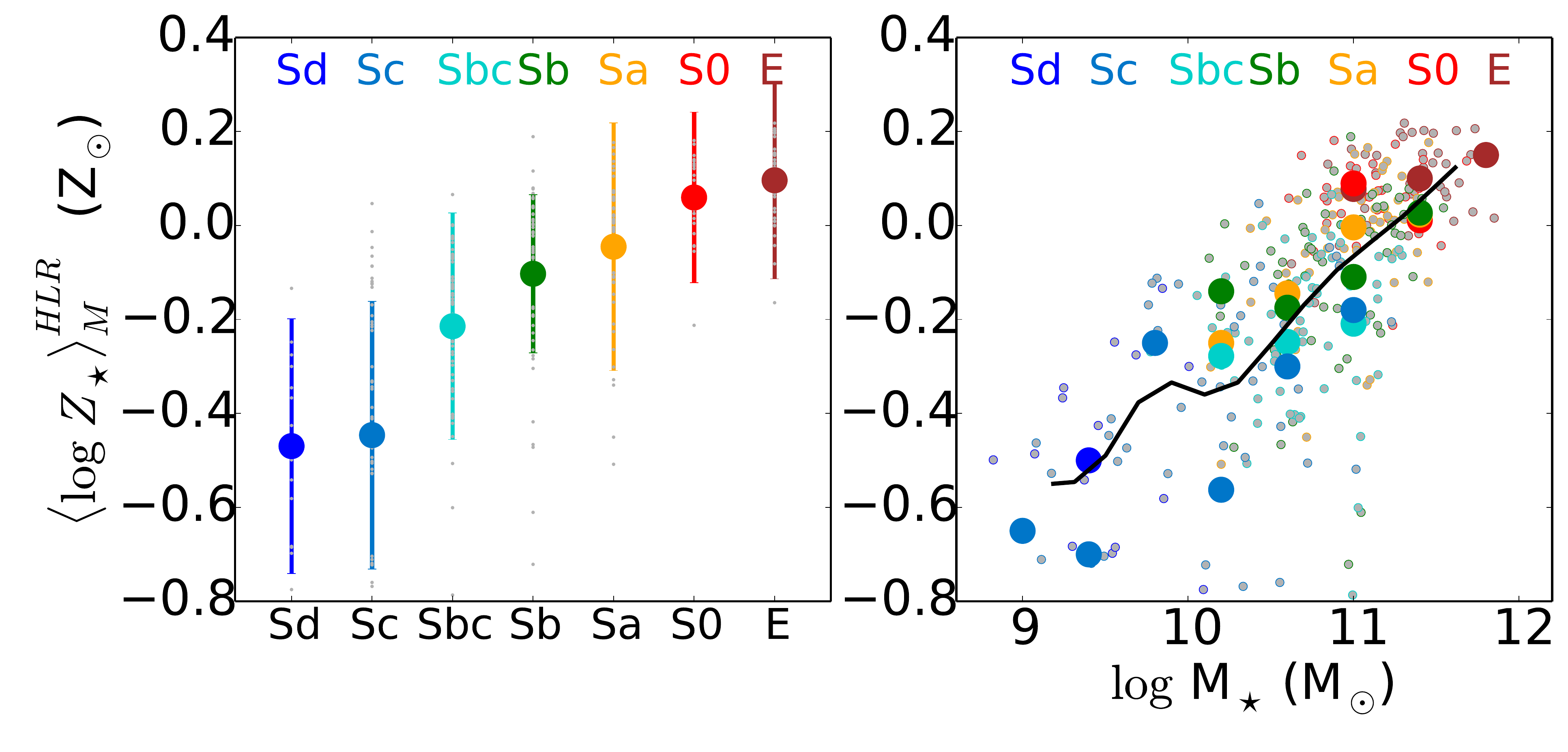}
\caption{\logZM\ measured at 1HLR as function of  Hubble type (left) and galaxy stellar mass (right). 
Symbols and colors are as  in Fig. \ref{fig:mu_tipo_mass}. The black line is the average \logZM\ obtained in 0.4 de bins of $\log M_\star$ .}
\label{fig:alogZM_tipo}
\end{figure}

Fig.\ \ref{fig:alogZM_tipo} shows how the stellar metallicity measured at 1 HLR changes with the Hubble type (left panel) and with the galaxy stellar mass (right). 

Stellar metallicities grow systematically from late to early type galaxies. The statistics within each Hubble type are \logZM$^{HLR} (Z_\odot) = -0.05 \pm 0.13$, $-0.05 \pm 0.33$, $-0.21 \pm 0.16$, $-0.10 \pm 0.18$, $-0.05 \pm 0.15$, $+0.06 \pm 0.08$, and $+0.10 \pm 0.08$ for Sd, Sc, Sbc, Sb, Sa, S0, and E, respectively.

Not surprisingly, metallicities also grow with $M_\star$, as shown in the right panel of Fig.\ \ref{fig:alogZM_tipo}. Since we have shown in \S\ref{sec:galaxyaveragedmetallicity}
that the galaxy-wide average stellar metallicity is well represented by the metallicity at 1 HLR,  this plot is in fact equivalent to the global mass-stellar metallicity relation (MZR). We have previously found that this relation is steeper than the one derived from HII regions, which is similar to the flatter stellar MZR obtained when we consider only young stars \citep{gonzalezdelgado14b}. 
As in Fig.\ \ref{fig:mu_tipo_mass}, the smoothed black curve is obtained by averaging \logZM$^{HLR}$ in 0.4 dex bins of $\log M_\star$. The dispersion in the MZR is significant, and larger than the dispersion produced by the galaxy morphology as shown by the  distribution of large colored circles. These circles are the average \logZM$^{HLR}$ in each mass bin for each Hubble type, and show the tendency of earlier type galaxies to be more metal rich than late type galaxies of the same stellar mass.

\subsubsection{Radial profiles}

\begin{figure*}
\includegraphics[width=0.48\textwidth]{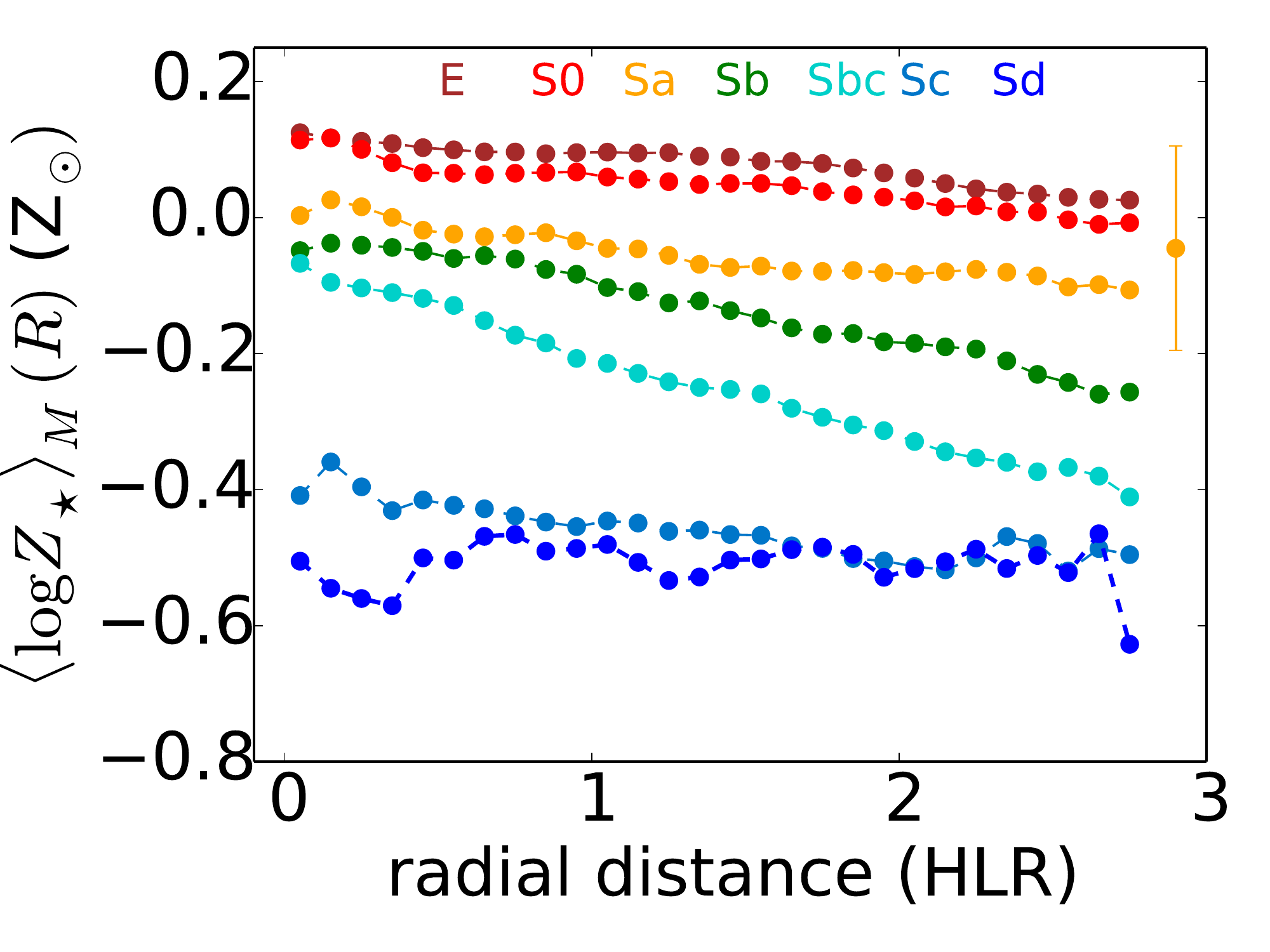}
\includegraphics[width=0.48\textwidth]{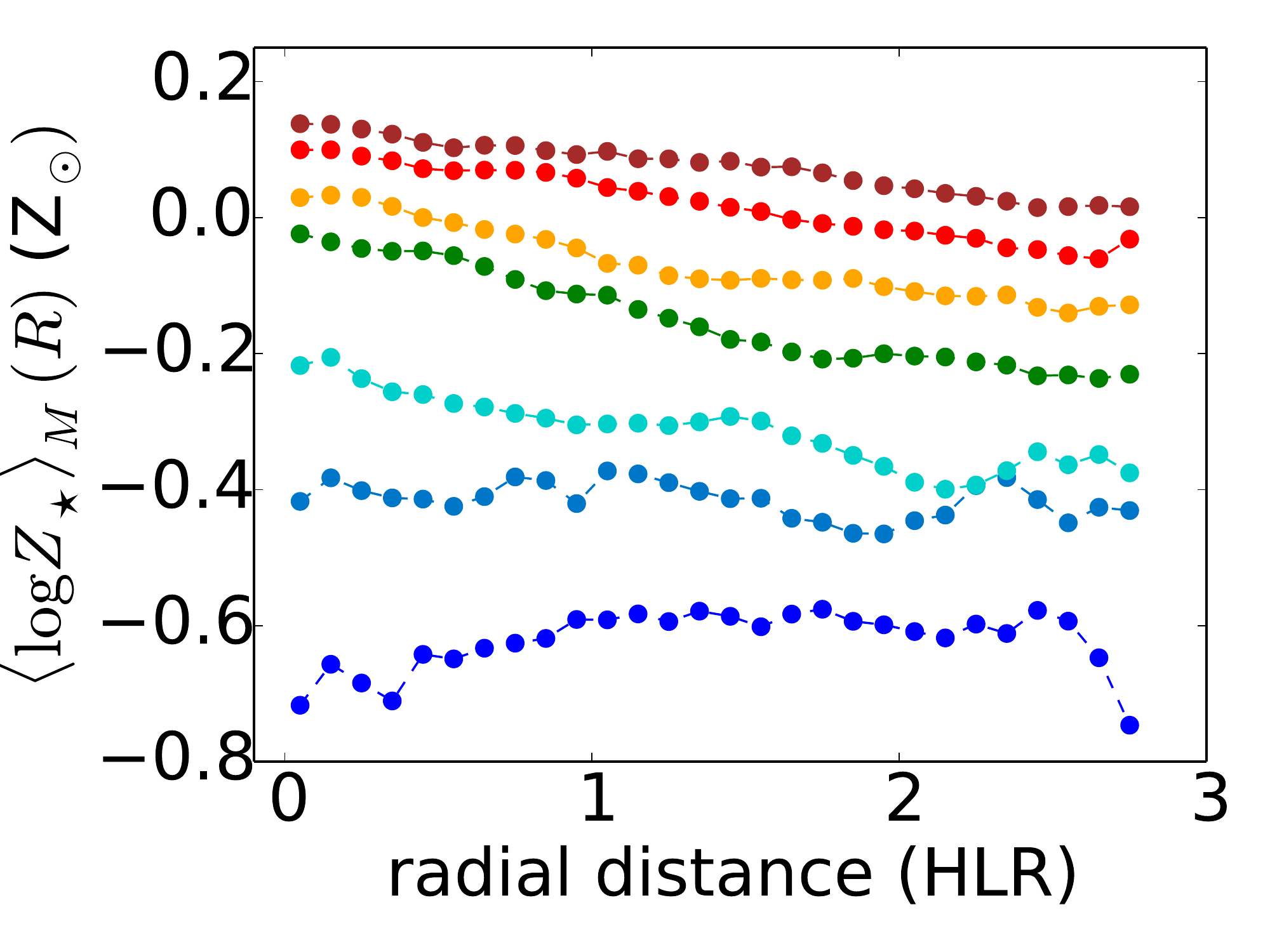}
\caption{Radial profiles of \logZM\ as a function of  Hubble type (left) and  of galaxy stellar mass (right). Mass bins 
are $\log M_\star (M_\odot)$ =  9.1$-$9.6, 9.6$-$10.1, 10.1$-$10.6, 10.6$-$10.9, 10.9$-$11.2, 11.2$-$11.5, 11.5$-$11.8. Symbols and colors are as Fig. \ref{fig:mu_radialprofiles}. 
These results are obtained with base $GMe$.  } 
\label{fig:logZM_radialprofiles}
\end{figure*}

Fig.\ \ref{fig:logZM_radialprofiles} shows the results of stacking the radial profiles of \logZM\ as a function of  Hubble type and $M_\star$. Outwards decreasing \logZM\  is detected for most morphological classes, but  flat profiles are found for Sc-Sd galaxies. Intermediate type spirals (Sb-Sbc) stand out as the ones with the largest variations in stellar metallicity.

The behavior of the radial  variation of the stellar metallicity with $M_\star$ (right panel in Fig. \ref{fig:logZM_radialprofiles})  is similar to the behavior with  morphology. Most  galaxies have \logZM\  that decreases with $R$, except for the two lowest mass bins, which show flat profiles. The largest spatial variations are also found in galaxies in the intermediate mass bins ($10 \leq \log M_\star (M_\odot) \leq 11$). 

These negative radial gradients of the metallicity are also an indicator of the inside-out formation processes in galaxies. The inversion of the gradient in  late type spirals and in low mass spirals may be an indicator of the secular processes or the outside-in formation scenario in these  galaxies \citep{perez13}.

\subsubsection{Radial gradients}

\begin{figure*}
\includegraphics[width=0.48\textwidth]{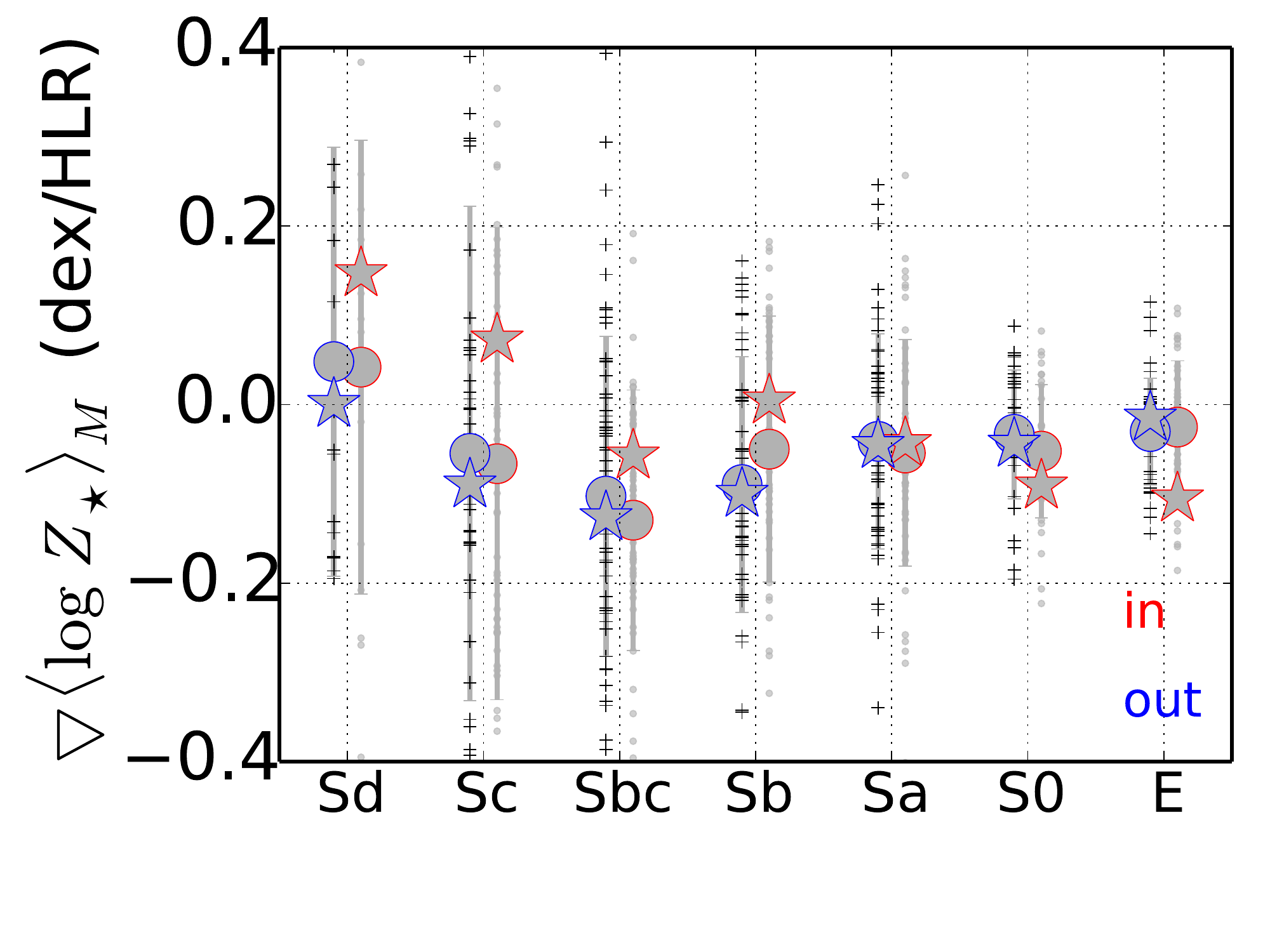}
\includegraphics[width=0.48\textwidth]{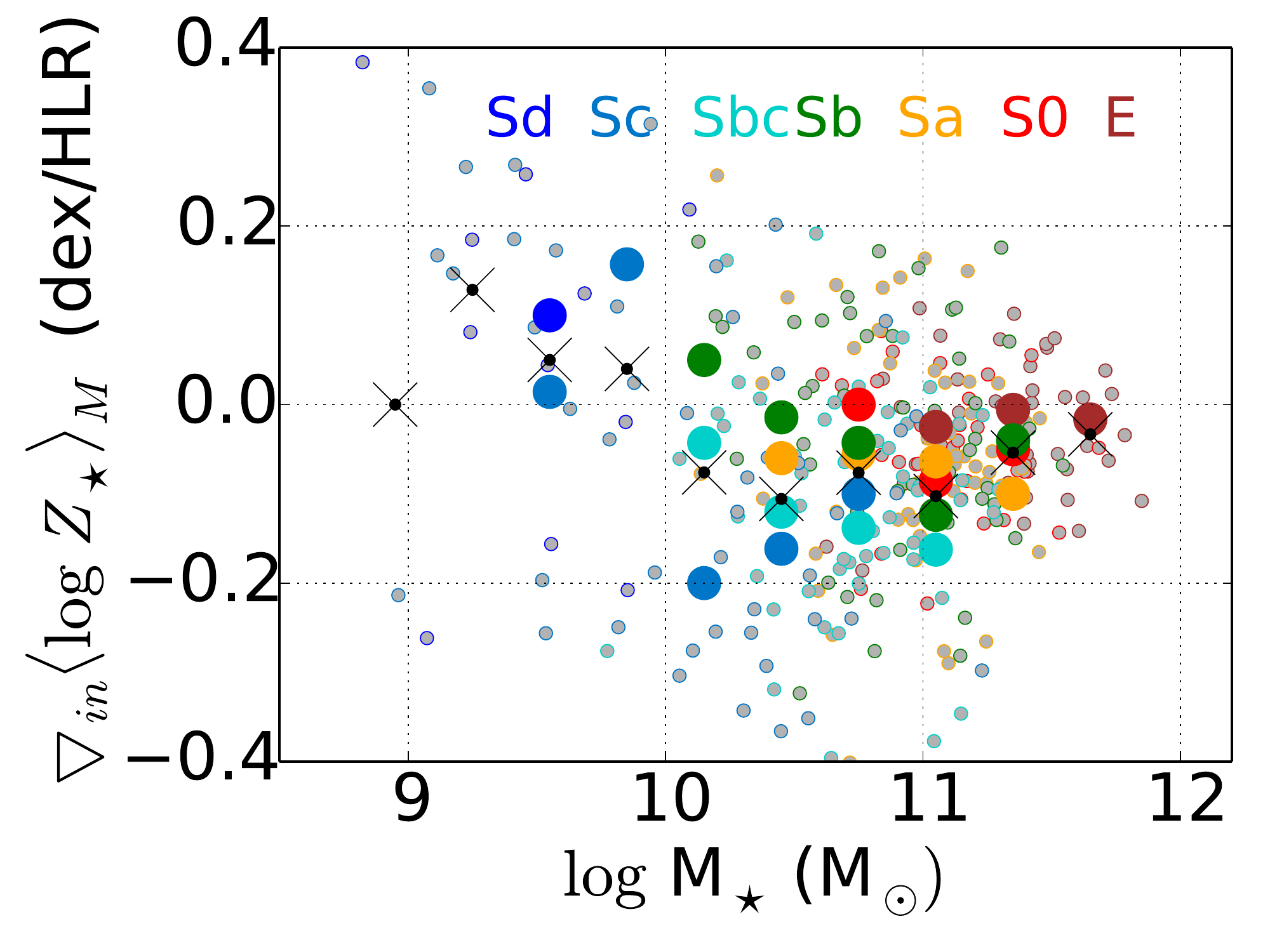}
\caption{Left: As Fig. \ref{fig:mu_gradient} but for \logZM.  Right: The inner gradient as a function of the galaxy stellar mass. Symbols and colors are as in Fig. \ref{fig:mu_gradient}.} 
\label{fig:logZM_gradient}
\end{figure*}

Fig.\ \ref{fig:logZM_gradient} clones Figs.\ \ref{fig:mu_gradient} and \ref{fig:ageL_gradient}  for  \logZM\ gradients. On the left panel, one sees that, as for stellar densities and ages, results for bases {\it GMe} and {\it CBe} are very similar. On average,  galaxies have \logZM\ gradients $\sim-0.1$ dex/HLR, similar to the value obtained from nebular oxygen abundances \citep{sanchez13}. 
Outer and inner gradients are not significantly different.
Despite the large scatter, there is a hint of a bimodal distribution as that found for stellar ages, also with intermediate type spirals in a pivotal position and late type spirals with the flattest gradients, at least in a statistical sense.

The right panel of Fig. \ref{fig:logZM_gradient} shows  $\bigtriangledown_{in}\langle \log Z_\star \rangle_M$ as a function of $M_\star$. The dispersion is significant, but on average there is a tendency to turn flat profiles into negative gradient ones as $M_\star$ increases from $10^9$ to $10^{10} M_\odot$.  The largest gradients are found between $10^{10}$ and $10^{11} M_\odot$. More massive galaxies tend to have weaker stellar metallicity gradients.
The dispersion is significant throughout this relation. A  trend with morphology is seen in the sense that, for a given mass, early types are the ones with weaker gradients.


\subsection{Stellar extinction}

\label{sec:subsec_AV}

\starlight\ models the stellar extinction as a foreground screen, parametrized by $A_V$ and following the Galactic reddening law.  Images showing the spatial distribution of $A_V$ for our 300 galaxies are presented in Fig.\ \ref{fig:cmd_AV}. Here we present  $A_V$ related results as a function of  Hubble type and $M_\star$, following the same script adopted in the discussion of $\mu_\star$, \ageL, and \logZM\ in the previous subsections, thus completing the list of stellar population properties studied in this work.  Unlike masses, ages, and metallicities, extinction is more easily affected by inclination effects, so the results reported below should be interpreted with caution. Section \ref{sec:inclination} explores this issue in depth.

\subsubsection{Extinction--morphology and extinction--mass relations}

\begin{figure}
\includegraphics[width=0.5\textwidth]{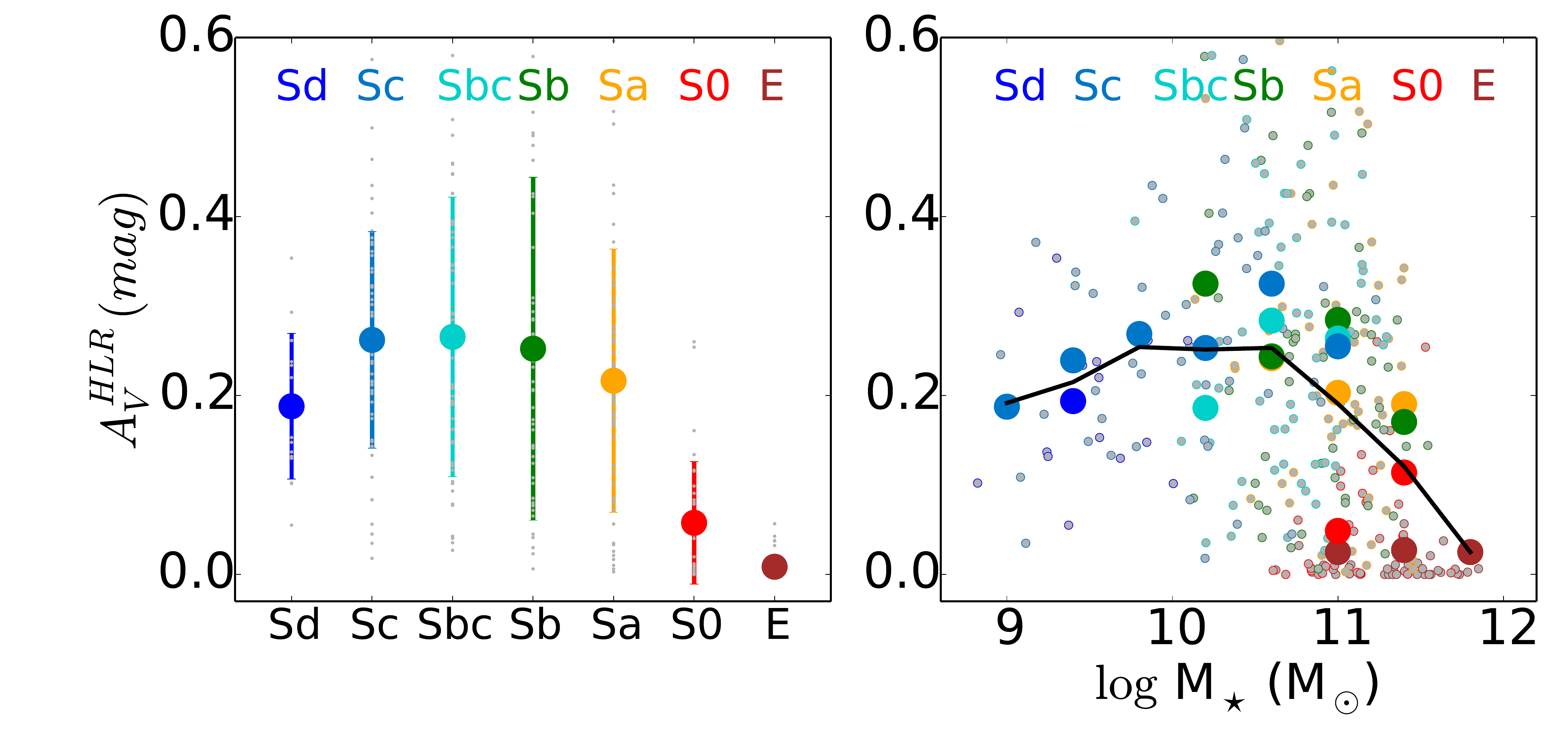}
\caption{A$_V$ measured at 1HLR as function of  Hubble type (left) and galaxy stellar mass (right). 
Symbols and colors are as in  Fig. \ref{fig:mu_tipo_mass}. The black line is the average \logZM\ obtained in 0.4 dex bins of $\log M_\star$.}
\label{fig:AV_tipo}
\end{figure}

\begin{figure*}
\includegraphics[width=0.48\textwidth]{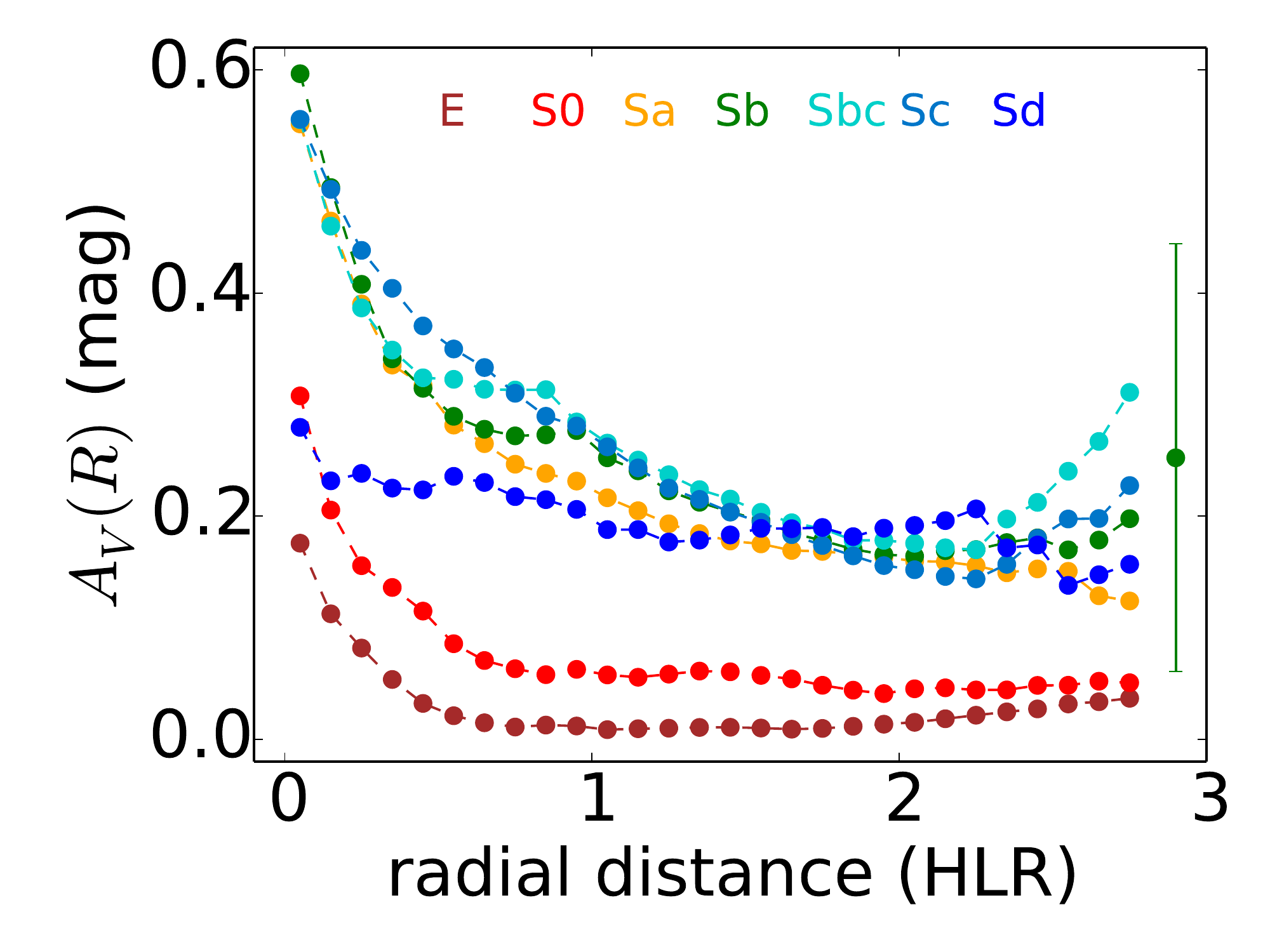}
\includegraphics[width=0.48\textwidth]{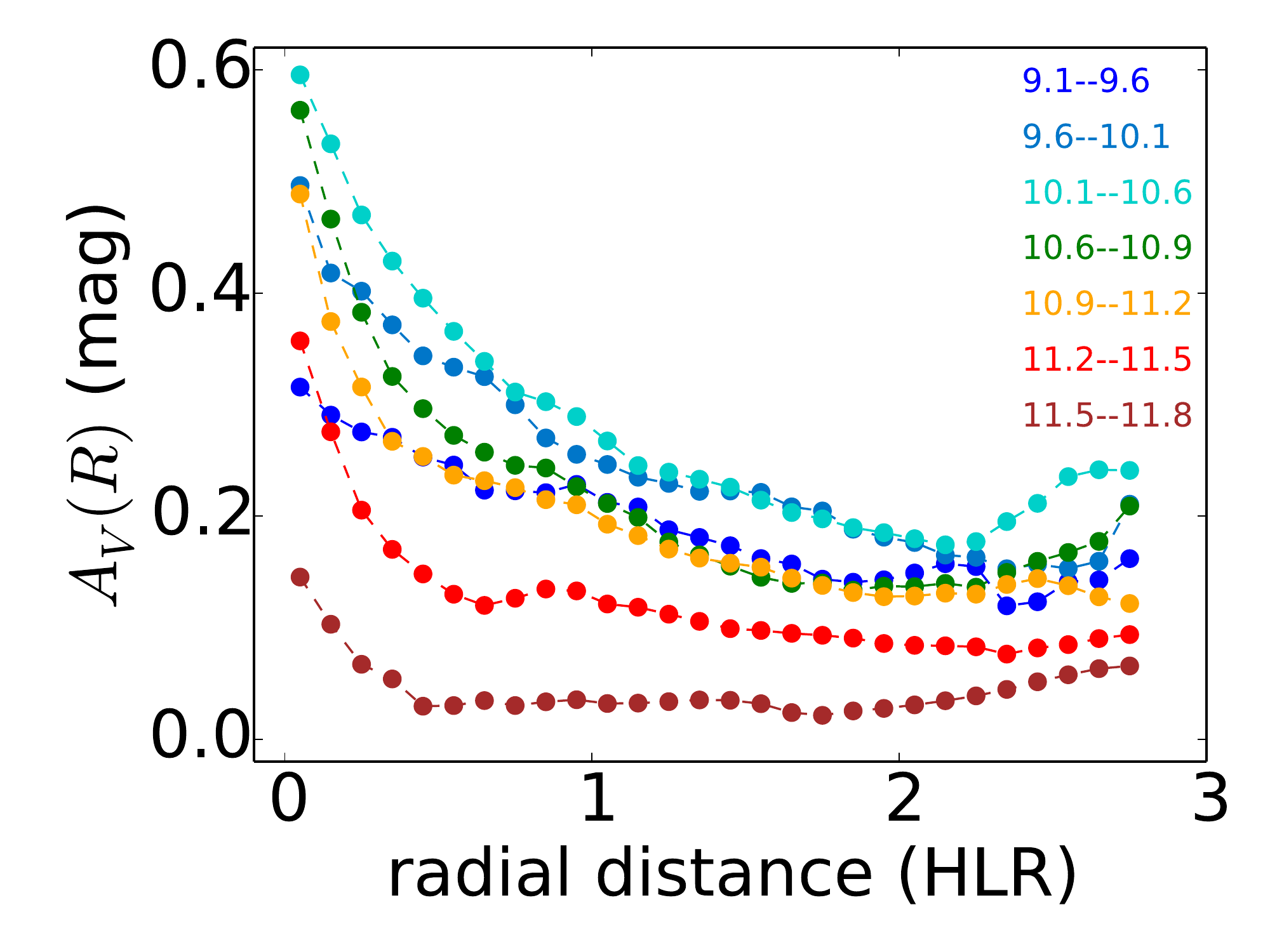}
\caption{Radial profiles of A$_V$ as a function of  Hubble type (left), and in seven bins of galaxy stellar mass (right). These bins 
are $\log M_\star (M_\odot)$ =  9.1$-$9.6, 9.6$-$10.1, 10.1$-$10.6, 10.6$-$10.9, 10.9$-$11.2, 11.2$-$11.5, 11.5$-$11.8. Symbols and colors are as Fig. \ref{fig:mu_radialprofiles}. 
These results are obtained with base $GMe$. } 
\label{fig:AV_radialprofiles}
\end{figure*}

Fig.\ \ref{fig:AV_tipo} shows how the stellar extinction at 1 HLR changes with Hubble type (left panel), and with stellar mass (right panel). As with other properties, $A_V^{HLR}$ represents well the mean extinction of the galaxy\footnote{The galaxy average extinction for each galaxy is  calculated as the mean of all the 20 radial values obtained for each galaxy between the center and  2 HLR.}  as well as the $A_V$ value derived from spectral fits of the integrated spectra.\footnote{The difference  between $A_V^{HLR}$ and $\langle A_V \rangle^{galaxy}$ is $-0.03\pm0.06$, while between A$_V^{HLR}$ and $A_V^{integrated}$ it is $-0.0\pm0.1$.} The left panel in Fig.\ \ref{fig:AV_tipo} shows  $A_V^{HLR}$ as a function of morphology. Ellipticals and S0s  have almost no extinction, with mean $A_V^{HLR} = 0.01\pm0.01$, and $0.06\pm0.07$ mag, respectively. Sa, Sb and Sc galaxies have $A_V^{HLR}$  around 0.25  mag, and somewhat smaller ($0.19\pm0.08$ mag) in Sd's. 

There is no clear behavior of stellar extinction with galaxy stellar mass. In general, galaxies with $M_\star \leq 10^{11} M_\odot$ have  $A_V^{HLR} = 0.2$--0.3 mag. More massive galaxies are less extinguished, and for fixed mass early types tend to have smaller $A_V^{HLR}$, but the dispersion is large.

\subsubsection{Radial profiles}

Fig.\ \ref{fig:AV_radialprofiles} shows $A_V(R)$ profiles stacked  by   Hubble type (left panel), and mass (right). Spirals have $A_V \sim 0.2$ mag in the disk and up to 0.6 mag in the center. Their $A_V$ profiles are similar for all Hubble types except for Sd's, where $A_V$ does not change much from disk to center. 
Ellipticals and S0's also show negative $A_V$ gradients, although at distances larger than 1 HLR they are almost dust-free. 
The radial profiles in different bins of $M_\star$ (right panel) show a similar behavior to that with  morphology. Except for the most massive bins, shifted to lower extinction values, all other mass-binned $A_V(R)$ profiles are similar.

\begin{figure*}
\includegraphics[width=0.48\textwidth]{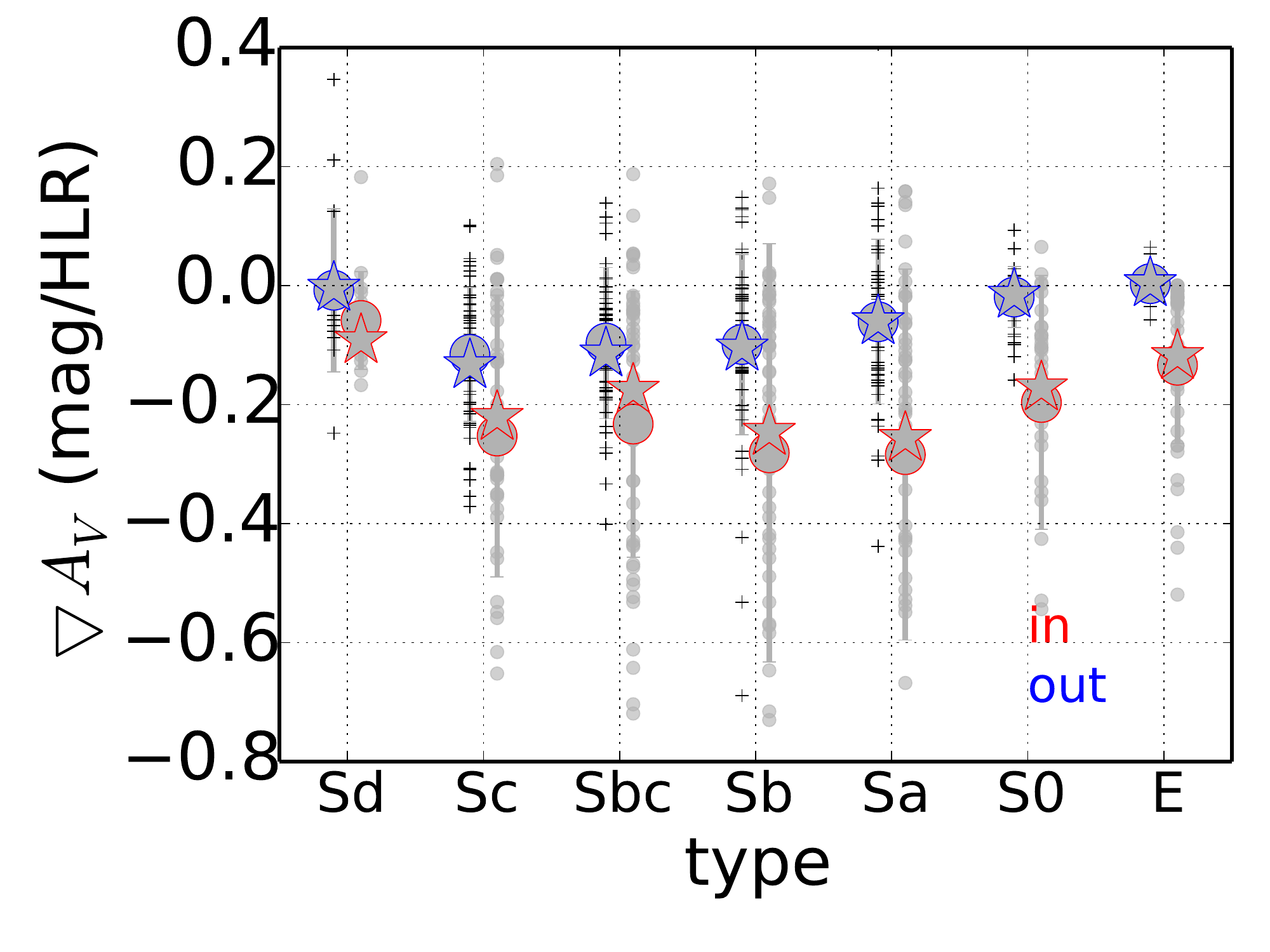}
\includegraphics[width=0.48\textwidth]{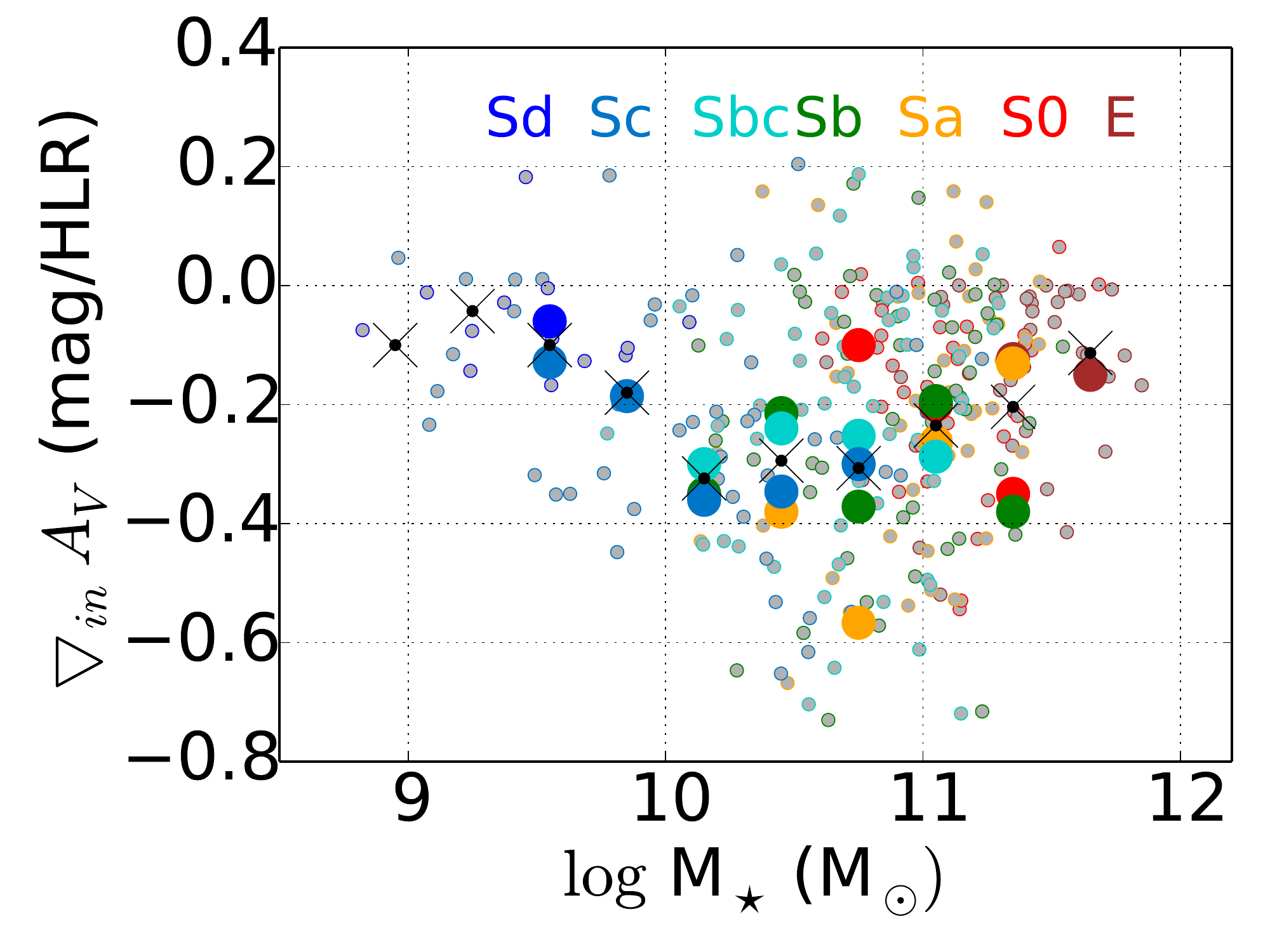}
\caption{Left: As Fig. \ref{fig:mu_gradient} but for A$_V$.  Right: The inner gradient as a function of  galaxy stellar mass. Symbols and colors are as in Fig. \ref{fig:mu_gradient}.
 } 
\label{fig:AV_gradient}
\end{figure*}

\subsubsection{Radial gradients}

$A_V$ gradients  are shown in Fig.\ \ref{fig:AV_gradient}, which is formatted as Figs.\ \ref{fig:mu_gradient}, \ref{fig:ageL_gradient} and \ref{fig:logZM_gradient}. As for the previous properties, results for bases {\it GMe} and {\it CBe} are nearly indistinguishable, as illustrated by the overlap of circles and stars in the left panel. $\bigtriangledown_{in}  A_V$ and $\bigtriangledown_{out} A_V$ show similar behavior with  morphology, although the inner gradient is always higher  than the outer one. In Ellipticals the gradient of $A_V$ exists only in the central region. With the exception of Sd galaxies, spirals have $\bigtriangledown_{in} \, A_V\sim - 0.25$ mag/HLR. 

On average,  $\bigtriangledown_{in} A_V$  gets stronger with increasing $M_\star$ up to $10^{11} M_\odot$ (Fig.\ \ref{fig:AV_gradient}, right) and weakens towards higher mass, spheroid dominated systems. The dispersion with respect to the mass-binned relation (traced by the black crosses)  is large, and not clearly related to  morphology (coded by the colored circles). 

As a whole, and despite the general trends summarized above, of the four properties analyzed in this section, $A_V$ is the one for which tendencies with Hubble type and stellar mass are less clear. A major reason for this is that, unlike for $\mu_\star$, \ageL, and \logZM,  $A_V$ estimates are sensitive to inclination effects. This is explained next.

\subsection{Effect of inclination on the radial profiles}

\label{sec:inclination}

\begin{figure*}[!ht]
\includegraphics[width=\textwidth]{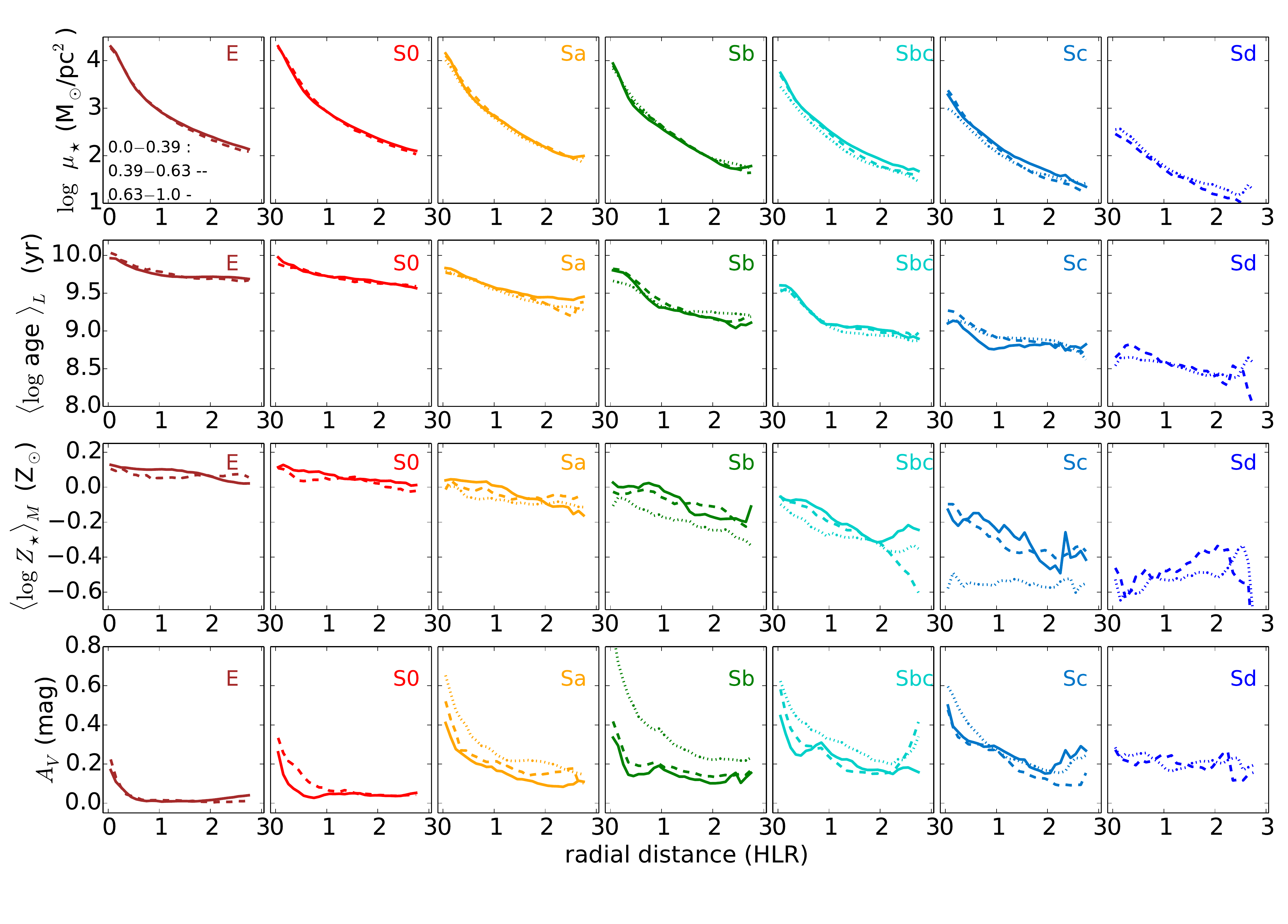}
\caption{Radial profiles of $\log \mu_\star$, \ageL, \logZM, and $A_V$ (from the upper to the bottom panels) for the different Hubble types (from E to Sd from left to right panel) and three different bins of the ratio of minor to major photometric radius of the galaxy: solid line ($0.63 < b/a$, face on),  dashed line (0.39 $\leq$ b/a $\leq$ 0.63), and dotted line ($b/a < 0.39$, edge on). 
}
\label{fig:inclination}
\end{figure*}

An implicit hypothesis throughout the analysis presented so far is that galaxy inclination does not affect our estimates of the stellar population properties and their radial distributions. One expects this assumption to break down in the case of $A_V$, which should increase from face on to edge on  galaxies, although it is not unreasonable to conceive that inclination effects propagate to the spectral synthesis-based estimates of stellar mass surface densities, mean ages, and metallicities.
It is therefore relevant to evaluate if and how inclination affects our results.

In order to do so, we have divided the 300 galaxies in three subsamples on the basis of the $b/a$ ratio (minor to major isophotal axes), as measured in SDSS $R$-band images. The three subsamples, each containing 100 galaxies, cover {\em (i)} $b/a \leq 0.39$, edge on, {\em (ii)} $0.39 < b/a \leq0.63$, and  {\em (iii)}  $b/a > 0.63$, face on. 
Galaxies in each sub-sample were grouped by Hubble type, and their radial profiles of $\log \mu_\star$, \ageL, \logZM, and $A_V$ averaged as previously done for the whole sample in the left panels of Figs.\ \ref{fig:mu_radialprofiles}, \ref{fig:ageL_radialprofiles}, \ref{fig:logZM_radialprofiles}, and \ref{fig:AV_radialprofiles}.

Fig.\ \ref{fig:inclination} shows the resulting stacked profiles of $\log \mu_\star$, \ageL, \logZM, and $A_V$. Solid, dashed and dotted lines show profiles for the ``face-on" ($b/a > 0.63$), intermediate inclination ($0.39 < b/a \leq0.63$), and ``edge-on''  ($b/a \leq 0.39$) samples respectively, and each column is for one of the seven Hubble types used throughout the paper. Average profiles were only computed for morphology-inclination bins containing at least 4 galaxies.

Stellar mass surface density, age, and metallicity profiles show a negligible variation among the $b/a$-based subsamples. This result indicates that inclination does not affect the estimates of these properties in any systematic way. Any difference is at a level not significant insofar as this paper is concerned. The exception is the \logZM\ profiles for ``edge-on'' Sc's, which differ substantially from the profiles of less inclined Sc's. It so happens, however, that the sub-group of $b/a \leq 0.39$ Sc's has a mean stellar mass 0.4 dex lower than other Sc's, which could explain their lower metallicities without implying inclination effects.

The one property which does vary systematically with $b/a$ is $A_V$, and it does so in the expected sense: Spirals with lower $b/a$ have larger extinction. This is particularly evident in Sb's. This dependence hinders the interpretation of the stacking results presented in \S\ref{sec:subsec_AV}, and explains why no clean tendencies of the $A_V$ values and profiles with morphology and stellar mass were identified.


\section{Discussion}

This section is divided into four main parts. First we summarize our results in the context of related studies. 
We then discuss our findings in the context of the growth of galaxies -- theoretical expectations, high redshift observations, and previous results of inside-out growth for CALIFA galaxies. 
In the third part we explore what the results tell us about the quenching of star formation in galaxies.
Finally, we discuss the theoretical predictions for the radial variations of age and metallicity in early types and in spirals from different simulations of galaxy formation. We compare our results for variations of the radial structure in the inner (R$\leq1$ HLR) and outer ($1\leq$R$\leq3$ HLR) parts with other observational results in the literature. 

\subsection{Age and metallicity of bulges and disks}

The analysis of SDSS data has generated a general knowledge of how the age and metallicity of galaxies change with $M_\star$, color, or concentration index  \citep[e.g][]{kauffmann03, gallazzi05, mateus06}. These studies have confirmed that, in general, early type galaxies are old and metal rich, while late type galaxies are younger and more metal poor. 
Numerous (single spectra or longslit) studies have reported also that ellipticals are metal rich, and have a range of ages,  2--10 Gyr, that depend on stellar velocity dispersion \citep[e.g.][]{trager00, thomas05, sanchez-blazquez06, graves09, johansson12}.

Our spatially resolved analysis introduces a significant improvement in the study of the structure of galaxies. For example, we compute ages and metallicities of bulges in disk galaxies and compare them with elliptical galaxies in a systematic way, avoiding problems derived from the lack of spatial resolution that some of the previous studies have. 

\begin{figure*}
\includegraphics[width=0.33\textwidth]{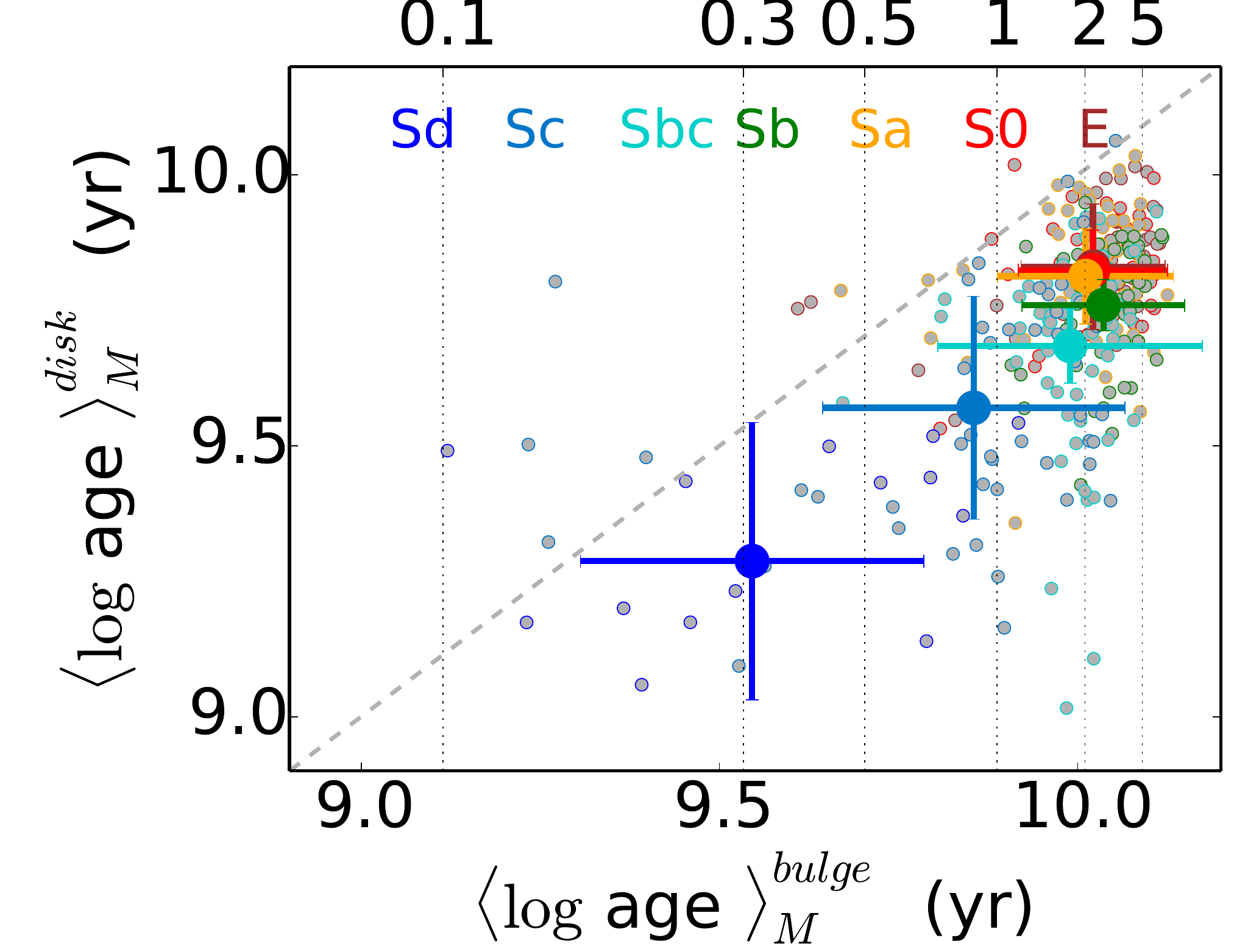}
\includegraphics[width=0.33\textwidth]{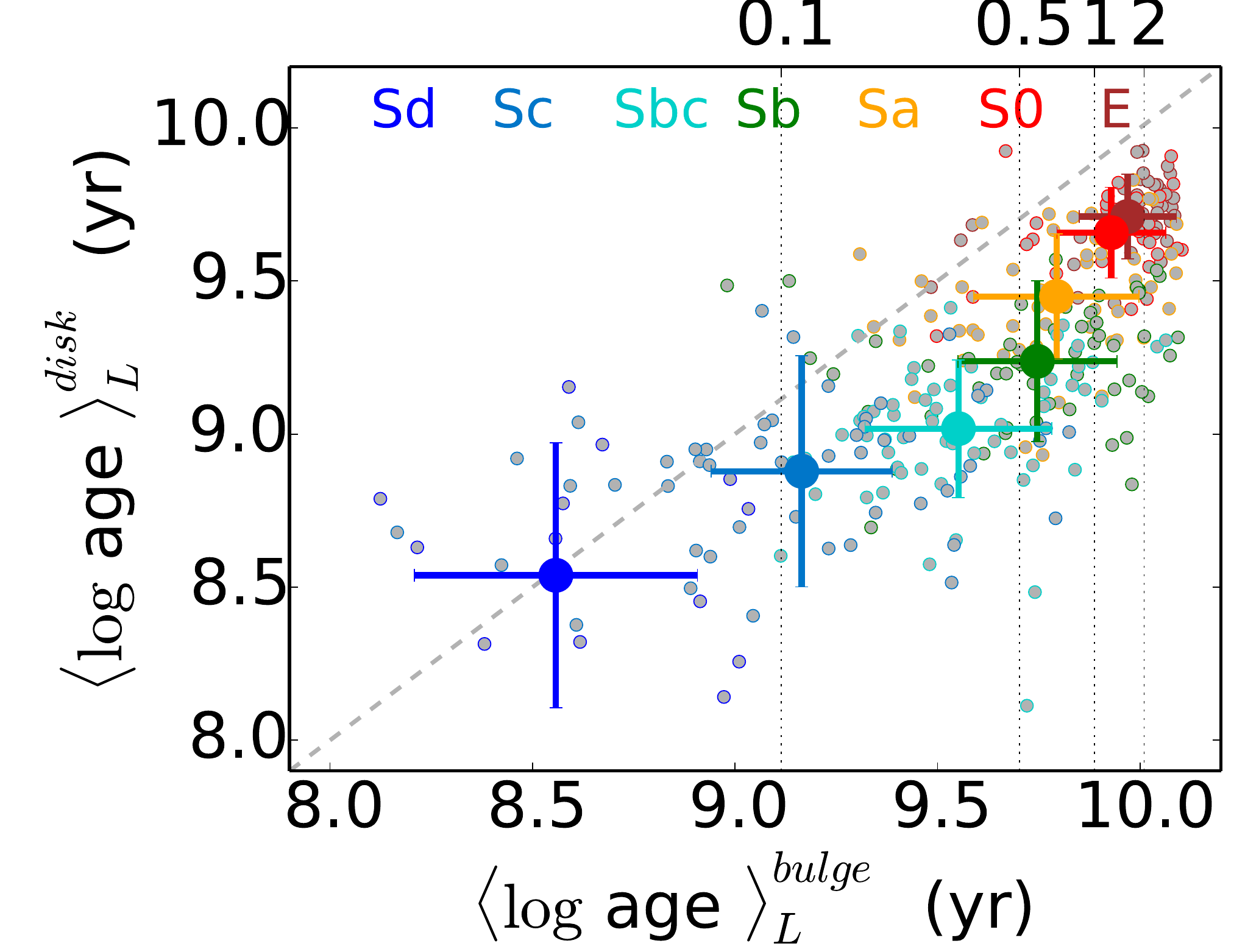}
\includegraphics[width=0.33\textwidth]{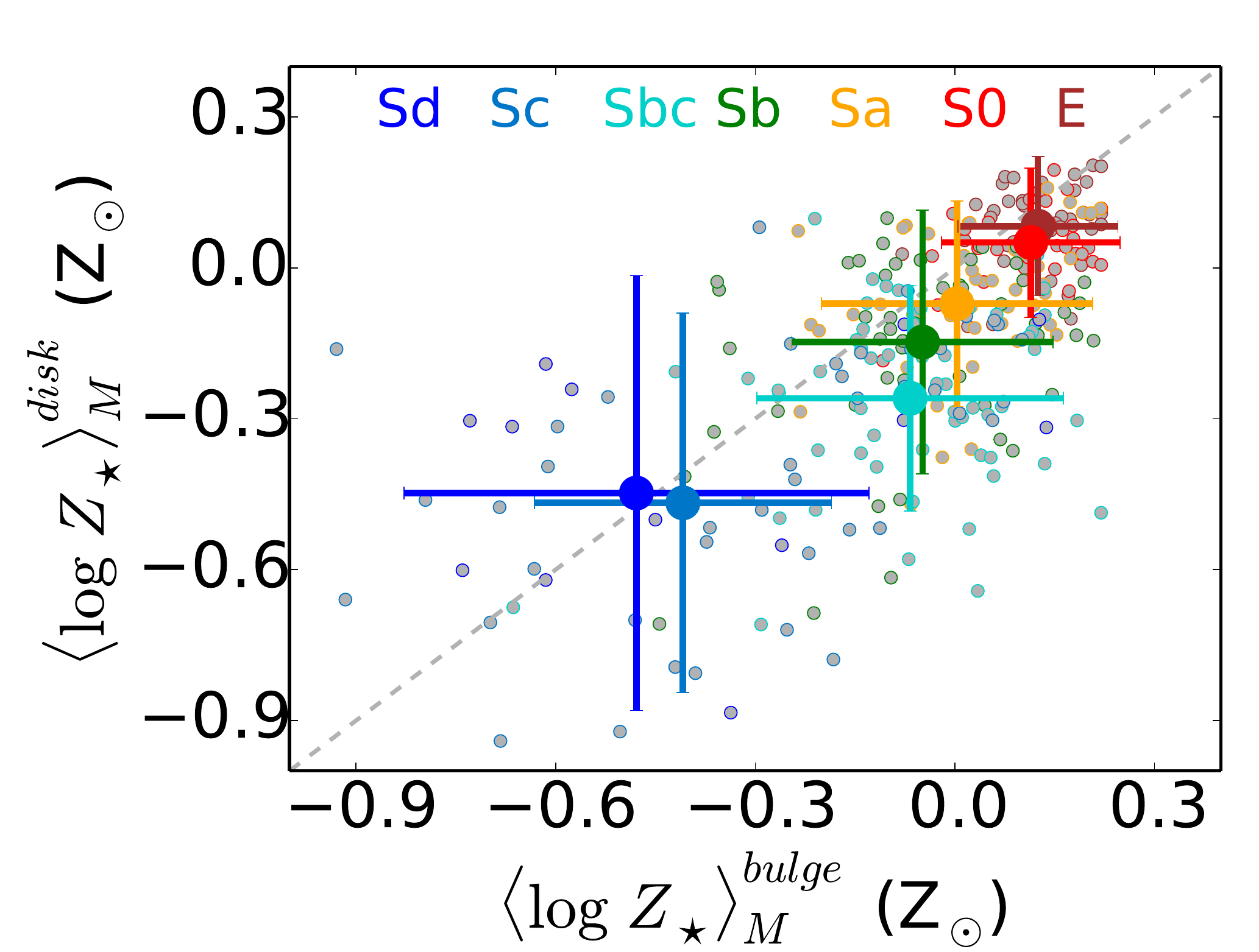}
\caption{Left:  Mass-weighted age at the galaxy center (\ageM$^{bulge}$) and at 1.5 HLR  (\ageM$^{disk}$) for the different Hubble types. Small dots are the individual radial points, while big colored dots represent mean values for each Hubble type, with the error bars indicating the dispersion in the mean. Middle: As in the left panel but for the light-weighted age (\ageL). Right: As in the left panel for \logZM. The top horizontal axes in the left and middle panels show the redshift scale. The diagonal line in the three panels is the bisector.} 
\label{fig:agebulgedisk}
\end{figure*}

We compute the luminosity-weighted and the mass-weighted age and metallicity: (i) in the central part of galaxies (values averaged within 0.25 HLR) as representative of the stellar population properties of bulges and central core of ellipticals; and (ii) in the outer part of galaxies, values averaged in a ring at 1.5$\pm$0.1 HLR, as representative of disks. Fig. \ref{fig:agebulgedisk} plots the individual results as small dots; large dots represent the average values for each (color-coded) Hubble type, and the error bars show the dispersion. While \ageL\ gives information about the globally `averaged' star formation history, \ageM\ informs when most of the stellar mass was formed. 

Fig. \ref{fig:agebulgedisk} shows that the bulges of Sa-Sb and the cores of E-S0 formed at a similar epoch; they are very old ($\sim$10 Gyr)  and metal rich ($\gtrsim$1 Z$_\odot$). Thus, they probably formed by common processes, that occurred rapidly and early on. However, the bulges in Sc-Sd galaxies (shown as the two darkest shade of blue) are younger and have a more extended star formation history (both \ageM\ and \ageL\ are smaller), and have lower stellar metallicities. Thus, Sc-Sd galaxies formed in later times and/or by different processes. 

Many bulges in late type spirals are in fact pseudo-bulges. Unlike true bulges, pseudo-bulges are thought to grow by secular processes, where material from the disk (by the loss of angular momentum) is the main source of star formation in the central parts of these galaxies \citep[e.g.][]{kormendy04}. 
We may see this effect at work in Fig. \ref{fig:ageL_radialprofiles} and Fig. \ref{fig:logZM_radialprofiles}, as a flattening of the radial profiles of \ageL\ and \logZM, and the positive \ageL\ gradient in the core of Sc galaxies. Some effects of the secular processes due to the disk may also be present in the bulges of Sa-Sb. For example, Fig. \ref{fig:agebulgedisk} shows that bulges of Sa-Sb have \ageL$\sim$6 Gyr, younger than the 10 Gyr epoch of formation derived from \ageM; and this may be understood if some disk stars are rearranged into the bulges or if dissipation processes bring  gas triggering new star formation in the center. 

Fig. \ref{fig:agebulgedisk} also shows that disks are younger and more metal poor than bulges. Both \ageL\ and \ageM\ are lower in disks than in their respective bulges, indicating that disks  formed later than bulges, and that star formation continues for a longer time in disks that in bulges, probably as a consequence of a continuing availability of gas in disks  \citep{roberts94}. This indicates a general scenario of inside-out formation.

 \subsection{Inside-out growth of spheroids and spirals}

\subsubsection{Theoretical expectations and recent results from high redshift galaxies}

Models of galaxy formation  predict a  common inside-out view  for the mass assembly in galaxies \citep[e.g.][]{kauffmann93, aumer13}. First, the bulge formed at high redshift; then, the disk was built around the bulge in the case of spirals. In the case of ellipticals, the central core formed at z$\geq$2, and the envelope grew later through minor mergers \citep[e.g]{oser10, hilz13}. 
Observational evidences come from the significant size evolution in early type galaxies (ETG), that grow as size $\propto M_\star^2$  \citep{vandokkum10, patel13}. 

More recently,  \citet{vandokkum14} find evidence against the inside-out formation scenario for spirals. For a sample of MW-like spirals at redshift z=2.5, they estimate the dependence of the radius with $M_\star$, and find that their size--$M_\star$ relation is similar to the size--$M_\star$ of similar galaxies at z=0. They conclude that the mass growth took place in a fairly uniform way, with the galaxies increasing their mass at all  radii, thus, their R$_{eff}$ barely grows. These results seem to be supported by numerical simulations by \citet{elmegreen08}, that find that bulges can be formed by migration of unstable disks. Other observational evidence come from the detection of clumpy star forming disks in galaxies at z$\sim$2 \citep{genzel08, foersterschreiber11a}, that may indicate an early build up of bulges by secular evolution. Thus, studies at high redshift are providing new results that draw a complex landscape of galaxy build up. For example, \citet{wuyts11} also find clumpy disk star formation, but at the same time conclude that there is a Hubble sequence in place at least since z$\sim$2.5. On the other hand, there is other evidence that galaxies rapidly assemble inside-out at z=1 \citep{nelson12, szomoru10, szomoru12}; while \citet{hammer05} find evidence that MW-like galaxies have rebuilt their disk at z $\leq1$ in a major merger epoch that drastically reshapes their bulges and disks, and is consistent with earlier cumplier evolution.

In summary, there is mounting evidence of the major processes responsible for the assembly and shaping of galaxies at different epochs, and these are complemented with a variety of processes that modify the inside-out formation scenario: stellar migration, bar induced gas inflows, gas-rich minor merger, angular momentum loss due to reorientation of the disk, infall of gas with misaligned angular momentum, etc \citep{aumer14}.

\subsubsection{CALIFA view of the inside-out growth of galaxies}

The results from our studies favor an inside-out growth of spirals.  \citet{perez13} studied the stellar mass growth as a function of the radius and cosmic time in galaxies with $10^{10}\lesssim M_\star\lesssim5\times 10^{11}\, M_\odot$, and showed that the nuclei grow faster than the inner 0.5 HLR, that, in turn, grow faster than the outer 1 HLR. This conclusion is  supported by the stellar age radial profiles presented  in \citet{gonzalezdelgado14a}, and confirmed here in Fig. \ref{fig:ageL_gradient} for most spirals and spheroidals. Further support comes from the ratio HMR/HLR (Fig. \ref{fig:HMR}), a good probe of the spatial variation of the star formation history in  galaxies \citep{gonzalezdelgado14a}. This ratio is lower than 1 (Fig. \ref{fig:HMR}), a fingerprint of the inside-out growth found by \citet{perez13}. 

\begin{figure}
\includegraphics[width=0.48\textwidth]{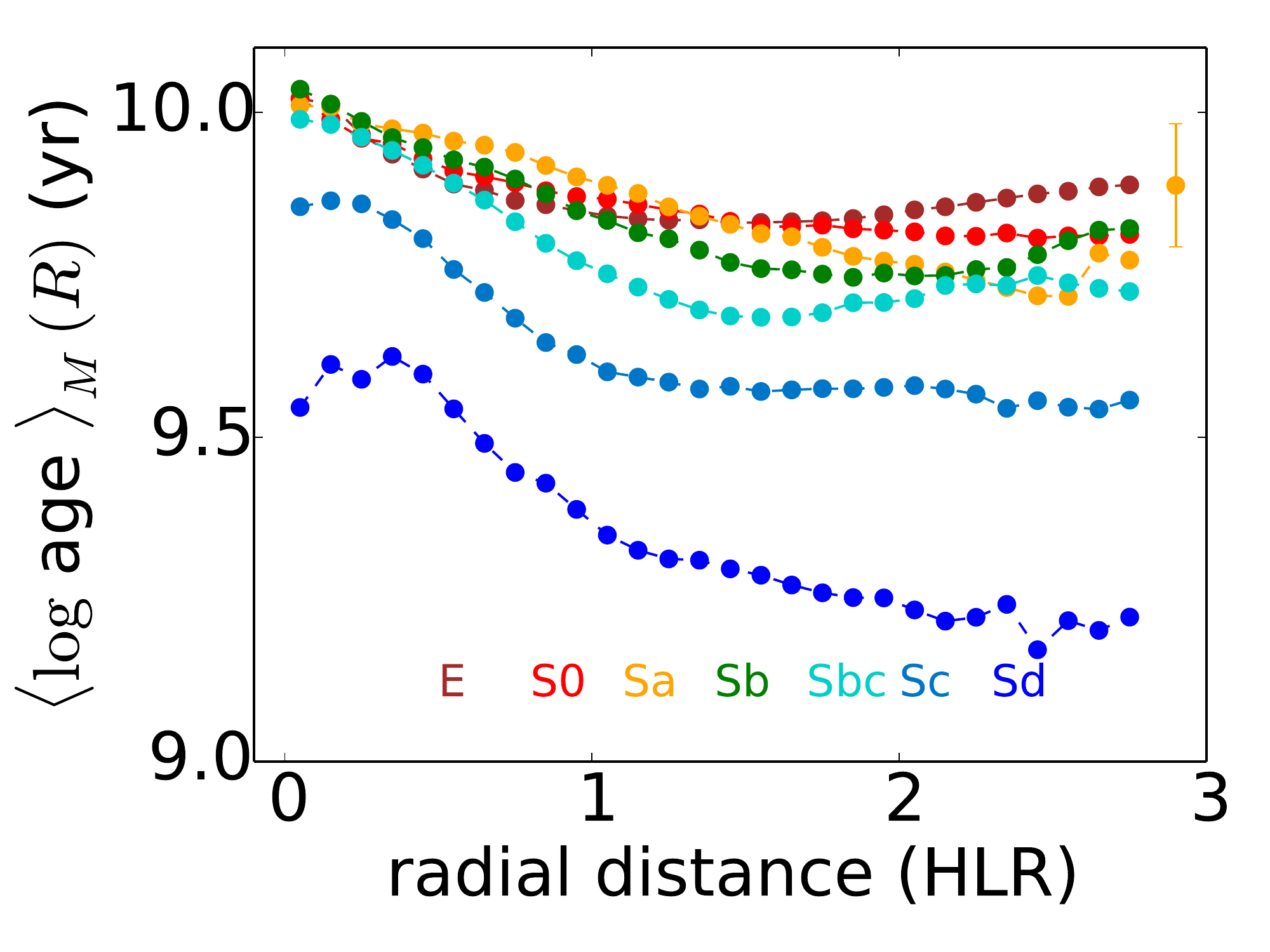}
\caption{ Radial profiles (in units of HLR) of the mass weighted age, \ageM$^{bulge}$, obtained with {\it GMe} base. The results are stacked by morphological type as in Fig.7.} 
\label{fig:ageMradialprofiles}
\end{figure}

Fig. \ref{fig:ageMradialprofiles} shows how the radial profiles of \ageM\ decrease outwards for all the Hubble types. Most of the stellar mass in the center has been formed 10 Gyr ago or  earlier (z$\geq$2). But at 1.5 HLR, \ageM\ ranges from 7 Gyr (z$\sim$1) in E--S0  to 4.5 Gyr (z$\sim$0.4) in Sbc, suggesting that, both early type and MW-like, galaxies have continued accreting or forming in-situ stars in their disks until more recent times, thus supporting the inside-out scenario in these galaxies. 

This trend, however, changes beyond 1.5-2 HLR, where \ageM\ and \ageL\ flatten. This may be interpreted as indicating that the mass was  formed in a more uniformly distributed manner across the outer disk, or that stellar migration shuffles inner born stars to the outer disk, washing out the inside-out formation signs so clearly seen in the inner 1.5 HLR. In the case of E--S0 this may be understood if beyond 2 HLR most of the stellar mass in the galaxies was already accreted at $z\leq1$.


\subsection{Quenching}
\label{sec:subsec_quenching}

\begin{figure*}
\includegraphics[width=\textwidth]{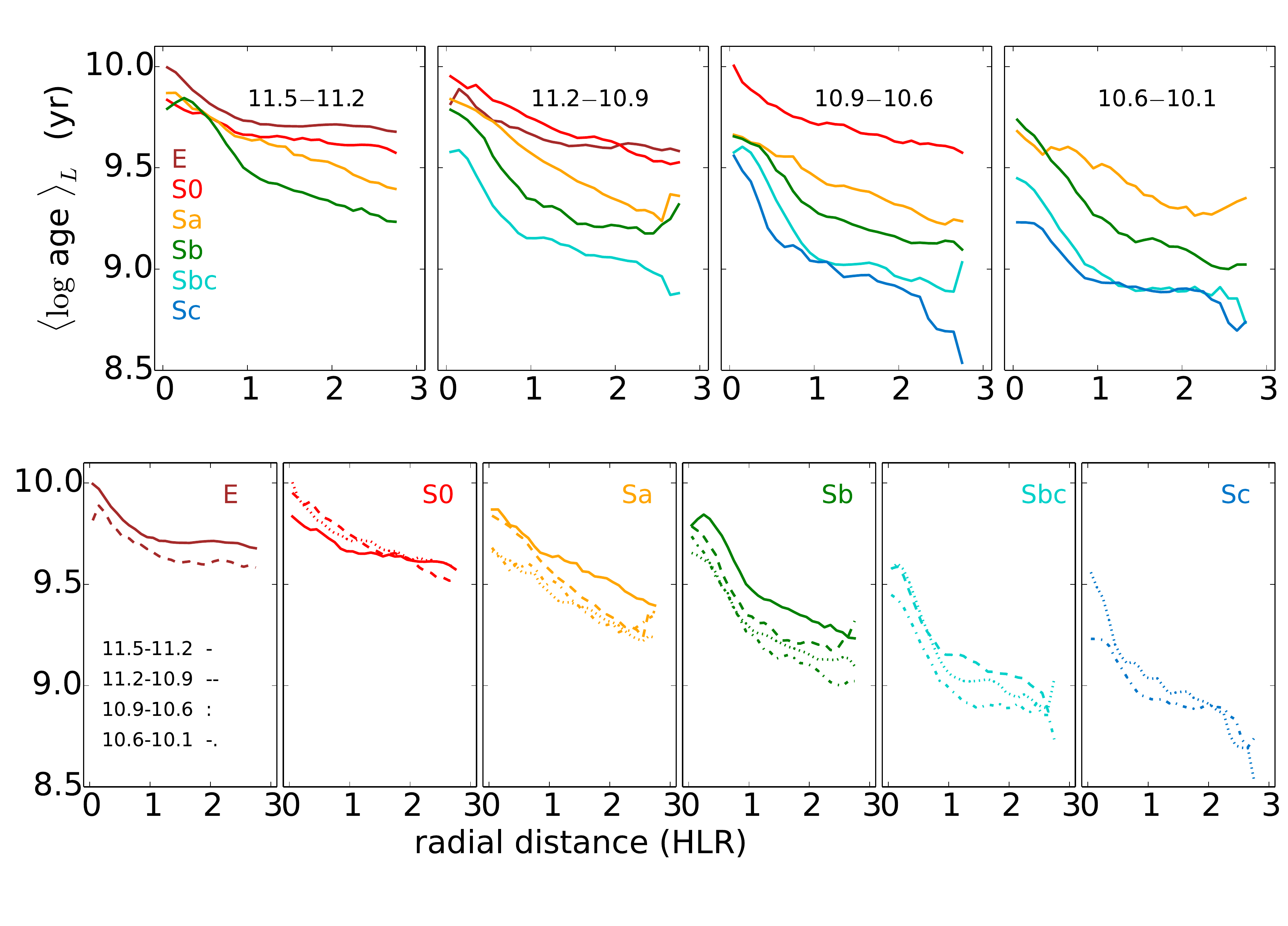}
\caption{Radial profiles of \ageL\ (upper panel)  in four galaxy stellar mass bins. From left to right:  $\log M_\star (M_\odot)$ = 11.2$-$11.5 (continuum line), 10.9$-$11.2 (dashed line), 10.6$-$10.9 (dashed-point line), 10.1$-$10.6 (dotted line). In each panel, the average profile for each Hubble type is plotted if more than four galaxies have galaxy stellar mass in the $\log M_\star$ bin. Bottom: each panel shows the radial profile of each Hubble type averaged in each of the four $\log M_\star (M_\odot)$ bins.
}
\label{fig:at_flux_RadialProfiles_tipo_Mass}
\end{figure*}

Several mechanisms have been proposed to explain the shutdown of star formation in galaxies. Halo mass quenching is one of the most popular ones that explains the bimodal distribution of the properties of galaxies, and it is required to explain the green valley as a pathway towards quenching of star formation in early and late type galaxies \citep[e.g.][]{schawinski14}. In this scheme, galaxies with a halo mass below a critical value (a few $\times$ 10$^{12}$ M$_\odot$) accrete cold gas conducive to star formation. Above this critical mass, the infalling gas reaches the sound speed and cannot form new stars \citep[e.g.][]{cattaneo06, dekel06}. The dependence with environment and clustering strongly supports this quenching mechanism \citep[e.g.][]{weinmann06, peng10}. 

The differential dependence of the stellar mass surface density (Fig. \ref{fig:mu_tipo_mass} and Fig. \ref{fig:mu_radialprofiles}) with the galaxy stellar mass (a proxy of the halo mass) provides further evidence of the halo quenching \citep[e.g.][]{behroozi10}. Estimating of the properties of the stellar populations in SDSS galaxies, \citet{kauffmann03} found that there is a critical mass ($M_\star = 3 \times10^{10}$ M$_\odot$, equivalent to $\sim 6\times10^{10}$ M$_\odot$ for our Salpeter IMF) below which $\log \mu_\star$ scales with $M_\star$, and above which $\log \mu_\star$  is independent of the galaxy stellar mass. Right panels of Fig. \ref{fig:mu_tipo_mass} and Fig. \ref{fig:mu_radialprofiles} support this scenario because the radial profiles of $\log \mu_\star$ scale with $\log M_\star$, and furthermore they do so all along their extent.  Our results also show that $\log \mu_\star$ saturates at high $M_\star$; because the high mass end of the distribution is dominated by early type galaxies (Sa-S0-E), this suggests that the spheroidal component plays a significant role in the quenching of star formation in high mass galaxies. 

The importance of morphology in the quenching of galaxies has also been reported in the literature \citep[e.g.][]{bell08, bell12, barro13, pan14, woo15}.  \citet{martig09}  found that the dependence of  quenching with  morphology is a consequence of the bulge-building mechanism. The steep potential well induced by the formation of a large spheroid component results in the stabilization of the disk, that cuts the supply of the gas, preventing its fragmentation into bound, star forming clumps. Our results support this scenario, as it is explained below, because the dependence of the SFH of galaxies with the morphology.

If the halo mass is the main property responsible for quenching, we should expect that the radial structure of \ageL\  (both, the age values and the gradients) to change more with $M_\star$ than with Hubble type. On the contrary, if quenching is driven by morphology, galaxies of similar stellar mass would have very different \ageL\ structure depending on Hubble type. 
We explore the relevance of morphology versus $M_\star$ in Fig. \ref{fig:at_flux_RadialProfiles_tipo_Mass}:  age radial profiles are shown as a function of $M_\star$ and of morphology, in four  mass bins ($\log M_\star$=11.5$-$11.2, 11.2$-$10.9, 10.9$-$10.6, 10.6$-$10.1). Clearly, morphology is the main driver: it can account for up to 0.75 dex change in age at a given mass (top panels); conversely, at a fixed morphology, mass accounts for less than 0.25 dex (bottom panels). Further, morphology accounts not only for changes in absolute values, but also for changes in  the gradients at a given galaxy mass.

This confirms the similar result obtained above with $\log \mu_\star$, and it implies that galaxies of similar $M_\star$ (equivalent to have similar $M_{halo}$) and with a large spheroid have shutdown their star formation (outside their central core) earlier than galaxies of later morphology. These results
indicate that the SFH and their radial variations are modulated primarily by galaxy morphology, and only secondarily by the galaxy mass, suggesting that the bulge formation has a relevant role in quenching the star formation in galaxies.


\subsection{Radial structure of the stellar population properties in ETG and their relation with galaxy formation models }

\subsubsection{Theoretical predictions from cosmological simulations}

Classical chemical evolution models  of the formation of early type galaxies (ETG) are based in two possible scenarios: 1) dissipative formation, the well known monolithic collapse; and 2) the non-dissipative collapse. 
These scenarios produce very different radial gradients of ages and abundances, being very steep in the first case,
with values of  $\bigtriangledown$[Fe/H] $\sim$ $-$0.5  to $-$1.0 [dex/dex] \citep{larson74,larson75,carlberg84}\footnote{The metallicity gradient measured in spheroids is traditionally calculated as $\bigtriangleup$[Fe/H]/$\bigtriangleup\log r$ and expressed in [dex/dex] units.}, but (almost) flat when there are pure stellar mergers. This second case may even erase a previously existing radial gradient.

The most recent cosmological simulations propose a two phase formation scenario for  ETG's  in which
the central core formed at z $\geq$ 2, and the envelope grows after this through minor mergers \citep[e.g.][]{naab09, oser12, hilz12, navarro-gonzalez13}. Thus: 1) Galaxies assemble their mass through dissipative processes and star formation occurs in-situ. Starbursts formed at the center as a consequence, for example, of large major mergers or monolithic collapse. The star formation is induced by cold flow of accretion or by gas-rich mergers. 2) Galaxies grow in size by mass assembly through the external accretion of satellites; ex-situ star formation formed by dry mergers of galaxies towards  the central most massive one.  

Observationally, there is evidence of a significant size evolution in ETGs. The growth of the galaxy size with M$_\star^{2}$ supports this proposal. A transition region is expected between the in-situ central core of ETG and ex-situ outer regions.  Since the central core of these ETG is enriched very quickly due to major mergers at high redshift ($z \geq$ 2), and the satellites that are accreted are less metal rich than the central massive one, a negative radial gradient of the metallicity is expected, even with a change of the slope in the transition region. Thus, values as  $\bigtriangledown$[Fe/H] = $-0.5$ [dex/dex] \citep{pipino10} or $\bigtriangledown$[Fe/H] = $-0.3$ [dex/dex] \citep{kawata03} are predicted. 

However,  the merger history may change an existing radial gradient: 
while dry major mergers can flatten the pre-existing gradient \citep{kobayashi04, dimatteo09, hopkins09}, dry minor mergers can  steepen the metallicity gradient.   Thus, \citet{kobayashi04} SPH chemodynamical simulations of elliptical galaxies that include radiative cooling, star formation and feedback from SNII-Ia, and chemical enrichment,  (but  do not include kinematic feedback),  found that the steep negative radial metallicity gradient, established during the initial starburst at $z \geq 2$, can flatten significantly  by later major-mergers in the galaxy.  Following these simulations, the average gradient at the present time is $\bigtriangledown$[Fe/H]= $-0.3$ [dex/dex], but it  may decrease to a value of $-0.2$ [dex/dex] when major mergers appear.

Beside the merger history, feedback can change the inner and outer metallicity gradients. Thus, a strong AGN feedback can stop the star formation in the central core of massive galaxies, flattening the inner gradients. Feedback from in-situ star formation can alter the outer metallicity gradient.  Also, the existence of galactic winds may modify the composition of the ISM in a galaxy. \citet{hirschmann15} performed cosmological simulations that include an empirical model for the momentum driven galactic winds, to investigate the dependence of the age and metallicity outer gradients with metal cooling and galactic winds, (in principle required to explain the mass-metallicity relation, MZR). These simulations including winds predict $\bigtriangledown$[Fe/H] = $-0.33$ [dex/dex], steeper than the simulations without winds that predict $\bigtriangledown$[Fe/H] = $-0.11$ [dex/dex]. The main explanation is that in wind models the stars accreted are of lower metallicity than in the simulations with no winds. In both cases, however, they predict a positive {\em age} gradient of $\sim$ 0.03$-$0.04 [dex/dex].

\subsubsection{Implications from this work and comparison with other results from the literature} 

Following our own results in this work,  E and S0 have formed their central core at similar cosmic time since they have similar central ages (see Fig. \ref{fig:agebulgedisk}). Further they must have formed through similar processes since their radial profiles of $\log \mu_\star$, \ageL, and \logZM\ are remarkably similar. They both show small but negative \ageL, and \logZM\ gradients in the central core. In the central 1 HLR, E and S0 in our sample have $\bigtriangledown$\logZM$\sim-0.1$ [dex/dex] (std = 0.15)\footnote{Our gradients, that are measured in a linear scale, are converted here to a logarithmic scale to be compared with predictions from  simulations and other works in the literature.}. (Sligthly steeper,  $\bigtriangledown$\logZM = $-0.2$  [dex/dex], when CBe models are used.) These  are within the range of values found in other studies based on long-slit or IFS data up to one effective radius \citep[e.g.][]{mehlert03,  sanchez-blazquez07, annibali07, rawle08, spolaor10, kuntschner10}. However they are shallow compared with  theoretical expectations if minor mergers are relevant in growing up the central core of E and S0 galaxies. This may indicate that major mergers are more likely the main process building the central regions (up 1 HLR) of ETGs.

Between 1 and 3 HLR, the radial profile of  \logZM\ is of similar slope or slightly shallower than in the inner 1 HLR.  We do not find any evidence of a transition region where the metallicity radial profile steepens to metallicities below solar. If the 1 to 3 HLR envelope of ETG had grown through the accretion of low mass satellites a steepening of metallicity would be expected, as explained before, because the mass-metallicity relation implies that low mass satellites would be of low metallicity. In our results there is no evidence either of an inversion of the age radial profile toward older ages beyond $1-2$ HLR, as expected if these satellites were formed very early on like the core of E and S0 (see Fig. \ref{fig:ageL_radialprofiles} and Fig. \ref{fig:agebulgedisk}).    
These results are in contrast with recent ones  by \citet{greene12, greene13}: for a sample of $\sim$30 early type galaxies they find at 2 R$_{eff}$ an old (10 Gyr) stellar population with [Fe/H]$\sim -0.5$, and  interpret this as the stellar outskirts of these galaxies being built up through minor mergers.  Also \citet{coccato10, labarbera12, montes14} observing a few massive ellipticals have reported a decline of the metallicity to under solar in an old stellar population in their outskirts ($\geq$ 10 R$_{eff}$) suggesting that these galaxies are formed in two phases, the central core through major mergers, and through minor mergers farther out. However, other recent works   show examples of ETGs with an old and metal rich stellar population and a very shallow metallicity gradient up to 3 R$_{eff}$ \citep{trujillo14}, in contrast with the results by \citet{greene12, greene13}.

Our results do not support the minor merger scenario for the size growth of ETGs. Thus, the ages, \ageL$\sim$ 9.7 (yr), and metallicity, \logZM$\sim$ Z$_\odot$, at $1-3$ HLR, and the shallow metallicity gradient, $\bigtriangledown$\logZM$\sim-0.1$ [dex/dex],  that we obtain  are more consistent with the growth of the $1-3$ HLR envelope of ETGs through major mergers. 

Other interesting result reported in the literature that can be compared with ours is the correlation between the metallicity gradient and the galaxy mass (or stellar velocity dispersion, $\sigma_\star$) found for E and S0. \citet{spolaor10} have found that the relation between $\sigma_\star$ and the metallicity gradient shows two regimes of behavior: (i) for galaxies with $\log \sigma_\star \leq$ 2.2 km s$^{-1}$, the metallicity gradient steepens with $\sigma_\star$; (ii) galaxies with $\log \sigma_\star >$ 2.2 km s$^{-1}$ (the most massive ellipticals), have a metallicity gradient that does not correlate with $\sigma_\star$ (or galaxy mass), with a mean value $\sim -0.15$ [dex/dex]. 
On the other hand,  \citet{pastorello14} derive the $[Z/H]$ gradient in the outer $1-2.5$ R$_{eff}$ of a sample of ellipticals, and they find that the gradient covers a wide range of values, from negative very steep ($\sim-2$) to flat or even positive; these values correlate with the galaxy stellar mass and stellar velocity dispersion, with the galaxies of lower $M_\star$ (or $\sigma_\star$)  having the steeper metallicity gradient. However, the most massive galaxies exhibit  the flattest gradients and an average value of $\bigtriangledown$\logZM$\sim-0.2$ [dex/dex] (std = 0.38) for galaxies with $M_\star\geq10^{11}$ $M_\odot$. Both works show a significant scatter in the $\bigtriangledown$\logZM $-$ $M_\star$ relation for $M_\star\geq10^{11}$ $M_\odot$, and reasonable doubts of the existence of the correlation for high mass ellipticals.

Our results (Fig. \ref{fig:logZM_gradient}) indicate that there is no correlation between the metallicity gradient and $M_\star$ for the CALIFA early type galaxies (E and S0). Even so, our results are compatible with \citet{spolaor10} and \citet{pastorello14}, because E and S0 in our sample are all above 10$^{11}$ $M_\odot$ (and $\sigma_\star >$ 100 km s$^{-1}$), for which no correlation is found between the metallicity gradient and $M_\star$ in  \citet{spolaor10} or \citet{pastorello14}. We find that CALIFA ellipticals  have a shallow gradient. This behavior is also in agreement with  \citet{hirschmann13} simulations and their interpretation of the lack of correlation between the metallicity gradient and $M_\star$ in massive ellipticals. Following these authors, massive galaxies accrete higher mass satellites, and because of their deeper potential well they retain their own gas against stellar winds, producing a shallower metallicity gradient in the outer regions of massive ellipticals. 

Because the metallicity at 2$-$3 HLR in E and S0 are similar to the metallicity in the bulge of early spirals, and the stars at these distances are as old as the bulges of Sa-Sb galaxies (see Fig. \ref{fig:ageL_radialprofiles}),  the 1$-$3 HLR envelope of early type galaxies might have built from the centers of early type spirals. In summary, the negative but shallow gradients of the metallicity and ages suggest that massive (M$_\star \geq 10^{11}$ $M_\odot$) early type galaxies  built  their inner 3 HLR through mergers with massive and metal rich spirals.

\subsection{Radial structure of the stellar population properties in spirals and their relation with galaxy formation models}
 	 
New insights on the structure of the Milky Way disk, in particular through the  measurements of chemical abundances of large sample of stars,
are provided by the spectroscopic surveys undertaken in recent times \citep[e.g., SEGUE, RAVE, Gaia-ESO survey, HERMES, APOGEE, LAMOST, etc;][]{yanny09,steinmetz06, gilmore12, zucker12, majewski10, zhao12}.
RAVE (Radial Velocity Experiment) \citep{steinmetz06, boeche14} is studying the radial and vertical chemical gradients using a very large sample of dwarf stars. Close to the Galactic plane, RAVE shows a negative radial gradient of Fe abundance, $-0.054$ dex/kpc\ \footnote{The metallicity gradient in disks is traditionally calculated as $\bigtriangleup$[Fe/H]/$\bigtriangleup$r, and expressed in dex/kpc.}, that becomes flatter or even positive when measured above the disk. So, the [Fe/H] gradient ranges from $-0.039$ to $+0.047$ dex/kpc when measured at heights $0.4-0.8$ kpc, or $1.2-2$ kpc  above the Galactic plane, respectively.

The radial gradient of  abundances in the different regions of a spiral galaxy are important because they are directly related with the formation process. Obviously, not all scenarios of disk/spiral formation are valid, since it is necessary that they produce a radial gradient of abundances in the disk but not in the halo, as observed in the MW and in M31. Thus, the formation of the halo from different fragments or minor mergers  with very short free fall times does not create a radial gradient but a dispersion of abundances, and therefore it was early concluded that the MW halo may be formed from mergers or from the accretion of low mass galaxies (or part of them). 
However, disks are more likely formed from a single cloud falling on and from inside-out.
 
\subsubsection{Theoretical predictions from "classical" chemical evolution models} 

Most classical chemical evolution models claim that  infall of gas with a radial dependence, implying an inside-out scenario for the disk formation,  is essential to reproduce the observed radial gradient of abundances. 
 The key ingredient is the dependence of the disk infall  time scale with the radial distance, that makes  the gas to accumulate faster in the inner disk. Since the SFR depends on the gas density, these assumptions produce a radial dependence of the star formation rate and a negative radial metallicity gradient \citep{ferrini94,molla96,chiappini01,molla05}. Thus \citet{molla96} give a value $-0.08$ for a MW like galaxy, reproducing the value found by \citet{shaver83} and other H{\sc ii} regions studies. \citet{chiappini01}  models predict a gradient of $-0.04$ dex/kpc for a MW-like galaxy. In fact,  as theoretical equations show \citep{gotz92}, the radial gradient of abundances appears in the disk when there is an adequate ratio between star formation rate  to infall rate. It also implies, therefore, that a dependence of the radial gradient  on the morphological type of galaxies may exist.  \cite{molla96} models already predicted radial gradients for galaxies of different morphological types, with values  in the range $-0.025$ (for a M31-like galaxy) to $-0.183$ $\rm dex\, kpc^{-1}$ (for a late type galaxy like NGC~300). More recent works \citep{molla05}  calculate models where the infall rate was a function of the mass distribution (or rotation  curve) of the galaxy, assuming a stronger radial dependence of the infall timescales than in \citet{chiappini01}. Moreover  \citet{molla05} models also depend on an efficiency factor to condense the molecular gas, and to convert the gas reservoir into stars. The metallicity gradients range  $-0.02$  to $-0.15$ $\rm dex\, kpc^{-1}$ with flat gradients for galaxies with the largest efficiency factor, or the most massive ones,  although in the extreme end the low mass and lowest efficiencies models also show flat radial distributions. Thus, the steepest gradients appear in the intermediate mass or intermediate type galaxies. However, there is no dependence on the morphological type when the gradient is normalized to a characteristic value, such as the effective radius as recent results by CALIFA  have found based on HII regions abundances \citep{sanchez13}.

\subsubsection{Theoretical predictions from cosmological simulations} 

Recently, hydrodynamical cosmological simulations have provided evidences in support of the imposed inside-out disk growth scenario adopted within the "classical" chemical evolution models. Like spheroidals, spirals are formed in two phases. In the first phase the bulge formed in a similar way as the core of E-S0. In the second phase, the disk grows by star formation in-situ from the infalling gas \citep{kauffmann93,aumer13}. Metal poor gas  with higher angular momentum at lower redshifts is turned into stars at larger  radii. Negative radial metallicity gradients are expected, as the classical models predict. This assumption is a natural outcome of the mass, momentum, and energy conservation laws, imposed in the simulations of disks in a cosmological context \citep{brook11,brook12,few12,pilkington12a,pilkington12b,gibson13}.  

\citet{pilkington12a} have examined a set of 25 simulations, from several groups, using different codes and initial conditions \citep{stinson10,rahimi11,kobayashi11,few12} to predict the present-day metallicity gradient in MW-like galaxies and its evolution. Although the evolution of the simulated metallicity gradients depends strongly on the choice of the sub-grid physics employed, most of the simulated galaxies tend to a similar present-day  gradient of $\sim -0.05 \rm dex\, kpc^{-1}$, in agreement with the \citet{chiappini01} and \citet{molla05} models for normal galaxies as the MW.

\subsubsection{Implications from this work}

Our findings show that spiral galaxies (excluding Sd) have negative radial gradients as indicative of the inside-out growth of the disk (see Fig. \ref{fig:logZM_radialprofiles}). The average  $\bigtriangledown_{out}$\logZM\  for spirals (excluding Sd and later type) is $-0.08$ dex/HLR or $-0.02$ dex/kpc. These values are compatible with the results obtained by  \citet{sanchez-blazquez14}, that have already derived the metallicity gradients for 62 CALIFA face-on spiral galaxies to study the effect of bars on the properties of the stellar populations. For these galaxies, they find a metallicity gradient of $-0.025$ dex/kpc (std = 0.05), equal ($-0.027$ dex/kpc) to the gradient that we derive for the same group of galaxies.
  
In order to compare with our results, the RaDES simulated galaxies \citep{few12}, 19 galaxies of the \citet{pilkington12a} sample, have been analyzed in a similar way as we have done here, i.e. measuring the gradients in a similar way and using the HLR values of the simulated galaxies (Ruiz-Lara in prep., and Ruiz-Lara et al. private communication). The simulated galaxies analyzed in this work cover a narrow range of morphologies, mainly Sbc-Sc. Therefore, these results cannot be extrapolated to the full work presented here, but they are representative of the state-of-art of cosmological simulations of disk galaxies, and can be used to compare them to similar disks from our observations. Mock B-band images are used to derive the HLR and to perform a bulge-disk decomposition used as a proxy for the morphology. Metallicities are calculated for disk particles using Eq.(2) and the gradient is derived between $0-1$ HLR and also between 1 and 2 HLR.  The stellar metallicity and age gradients of the simulated galaxies are compatible with the results presented here. Keeping in mind that the morphological range covered by these simulations is rather narrow and that they use B/D as a proxy for a morphological classification, the results show a slight dependence of  $\bigtriangledown_{out}$ \logZM\  with B/D ratio, with a steeper slope for  B/D=1 (Sbc galaxies) for which  $\bigtriangledown_{out}$ \logZM\ $\sim-0.1$ dex/HLR. Later type spirals have a flatter gradient of $\bigtriangledown_{out}$ \logZM\ $\sim-0.046$ dex/HLR.  These results go in  line with those found here, namely, that the metallicity radial gradient of spirals shows a dependence on  morphology, with the steepest gradient found in the intermediate Sb-Sbc spirals (see Fig. \ref{fig:logZM_radialprofiles} and Fig. \ref{fig:logZM_gradient}). However, a larger set of cosmological simulations is required covering from early Sa to late Sd, and a large range of galaxy masses (from 10$^9$ to 10$^{11}$ M$_\odot$), in order to confirm the general trend found here. On the other hand, these results indicate that the feedback recipes used in these simulations are able to recover realistic galaxies with small bulges and are fully in agreement with the work presented here. 

Furthermore, our results are also compatible with classical chemical models, and certainly, the CALIFA Sb-Sbc galaxies have stellar metallicity gradients ($-0.025$ dex/kpc) in the range observed in the MW disk, but somewhat shallower than the [Fe/H] gradient measured in the Galactic disk. However, it is necessary to take into account that the gradient usually given in the literature is obtained for young stars or HII regions, while here it is an average value obtained for all stellar populations existing in the studied galaxy or region. Besides that, the number of objects is increased compared with the old studies
and, more important, all of them have been self-consistently analyzed using the same reduction technique and spectral models.

In any case our results favor an inside-out growth of spirals. This conclusion is supported by the stellar age radial profiles presented here: the age decreases outwards for all Hubble types studied\footnote{Sd galaxies, however, show a much flatter age gradient}. Beyond $1.5-2$ HLR the radial distribution of ages flattens, suggesting that the mass forms more uniformly in those regions, or that the stellar mixing brings stars born in the inner disk to the outskirts. This last possibility has been recently investigated by \citet{minchev13,minchev14}, who have performed N-body hydrodynamical models with the chemical evolution implementation \citep{minchev13, minchev14}. They have simulated MW-like galaxies with the aim to investigate whether the Galactic disk can be understood as a single structure with kinematic and chemical features that are continuously distributed, being the thin and thick disks two extreme cases of these structures. Furthermore, they investigate the effect of stellar migration and kinematic heating in the scatter of the age-metallicity relation, and how it changes with the Galactic radius. In fact, an increase of the scatter in the age-metallicity relation and a  flattening of the stellar  metallicity gradient is produced by the stellar radial migration,  that causes a radial mixing in the older stellar population, creating the appearance of a flatter gradient in early times, and leading to a decoupling of the stelar population from their birth interstellar medium \citep{roskar08}. These results also indicate that even though  radial mixing has a significant effect in flattening the metallicity gradient, it can not destroy it.


\section{Summary and conclusions}
\label{sec:Summary}

We have analyzed the stellar population properties of 300 galaxies, observed by CALIFA with the V500 and V1200 gratings and IFU PPak at the 3.5m telescope of Calar Alto, to investigate the trends in the stellar populations properties with radial distance as a function of  Hubble type and  galaxy stellar mass. The sample includes ellipticals, S0 and spirals from early (Sa-Sb) to late types (Sc-Sd). They cover a stellar mass range from 
0.7$\times$10$^9$ to 7$\times$10$^{11}$ $M_\odot$ if Salpeter IMF is assumed,  and a  factor 1.78 (0.25 dex) lower for a Chabrier IMF. A full spectral fitting analysis was performed using the \starlight\ code and a combination of SSP spectra from \citet{gonzalezdelgado05}, \citet{vazdekis10}, or Charlot \& Bruzual (2007, private communication). Our pipeline \pycasso\ is used to process the spectral fitting results to produce present day maps of the spatial distribution of the stellar population properties.  For each galaxy, these maps are azimuthally averaged to produce radial profiles (in units of the half light radius, HLR: $a_{50}^L$) of the stellar mass surface density ($\log \mu_\star$), stellar ages (light weighted, \ageL, and mass weighted, \ageM), metallicity (\logZM), and extinction ($A_V$).  The radial profiles are stacked as a function of Hubble type and of galaxy mass. Radial gradients of these properties measured within the inner 1 HLR and between 1 and 2 HLR are also obtained. 

Our main results are:

\begin{enumerate}

\ls\item{{\it Spatially averaged vs. integrated galaxy properties}: the metallicity, \logZM, galaxy-wide spatially averaged matches the metallicity obtained from the integrated spectrum, and the metallicity at R=1 HLR. This result is equivalent to that obtained for the other stellar population properties, $\log \mu_\star$, \ageL, and $A_V$, as reported by \citet{gonzalezdelgado14a, gonzalezdelgado14c}, proving that effective radii are indeed effective.}

\ls\item{{\it Mass weighted size:} We confirm our earlier finding \citep{gonzalezdelgado14a} that galaxies are more compact in mass than in light by $\sim$20$\%$.  The HMR/HLR ratio shows a dual distribution with  Hubble type, that breaks in the Sb-Sbc, the galaxies with the smaller HMR/HLR. This ratio also shows a dual dependence with $M_\star$: it decreases with increasing mass for disk galaxies, and becomes almost constant in spheroidal galaxies. These results are a signpost of the inside-out growth previously found by \citet{perez13}.}

\ls\item{{\it Stellar mass surface density}: $\log \mu_\star (R)$ shows declining profiles that scale with morphology and with $M_\star$; this behavior is preserved at any given distance. At constant $M_\star$, $\log \mu_\star (R)$ is higher in early type than in late type spirals. E's and S0's show equal $\log \mu_\star (R)$ profiles, independently of $M_\star$. The inner gradient, $\bigtriangledown_{in} \log \mu_\star$, correlates  with Hubble type. 
The negative gradients steepen from late type spirals to spheroids, as well as with galaxy total mass in galaxies with $M_\star \leq$ 10$^{11}$ $M_\odot$. At a constant $M_\star$, $\bigtriangledown_{in} \log \mu_\star$ steepens with  morphology, with E's and S0's having the steepest gradients. These results indicate that morphology, and not only $M_\star$, plays a relevant role in defining $\mu_\star$, and the $\mu_\star$$-$$M_\star$ relation.}

\ls\item{{\it Stellar ages}: \ageL(R)  shows declining profiles that scale with morphology; this behavior is preserved at any given distance.  Early type spirals are always older than late spirals. E's and S0's, although older than spirals, have both similar \ageL(R) profiles, indicating that these galaxies have similar star formation histories. The more massive galaxies are also the older ones; this ``downsizing" behavior is always preserved at any given distance. The negative $\bigtriangledown_{in}$\ageL\  depends on  Hubble type in different ways: steeper from E and S0 to Sbc, and shallower from Sbc to Sd. Thus, Milky Way like galaxies have the steepest age gradient. A $\bigtriangledown_{in}$\ageL $-M_\star$ relation exists, increasing the gradient from the low mass galaxies (which have roughly flat profiles) up to about 10$^{11} M_\odot$, at this point the trend reverses and $\bigtriangledown_{in}$\ageL\ decreases with increasing $M_\star$. However, the dispersion in the $\bigtriangledown_{in}$\ageL $-M_\star$ relation and \ageL$^{HLR}$$-M_\star$ is significant and it is strongly related with the morphology.  Even more, the dispersion of the \ageL(R) profiles of galaxies of equal mass is significant and larger than between the \ageL(R) profiles of galaxies of different $M_\star$ but the same Hubble type. Thus, the SFHs and their radial variations are modulated primarily by the Hubble type, with mass playing a secondary role. }

\ls\item{{\it Stellar metallicity:} \logZM(R) shows mildly decreasing profiles for most  Hubble types, except Sd's that show little, if any, radial dependence. Milky Way like galaxies (Sbc) stand out as the ones with the steepest radial profiles.  \logZM(R) scales with $M_\star$ in a similar way as it does with  morphology. This can be understood as a consequence of the global mass metallicity relation --a primary dependence of the metallicity with $M_\star$. The metallicity gradients are negative but shallow on average, with $\bigtriangledown_{in}$\logZM\  $\sim -0.1$ dex/HLR, and show a small dependence with $M_\star$ up to M$_\star \sim$ 10$^{11}$ $M_\odot$, steepening with increasing mass. Above 10$^{11}\, M_\odot$ (Sa's, S0's and E's) they have similar metallicity gradient. The dispersion in the  $\bigtriangledown_{in}$\logZM\ $-$ $M_\star$ relation is significant and a trend with  morphology is seen, in the sense that, for a given mass, intermediate type spirals are the ones with steeper gradients. }

\ls\item{{\it Stellar extinction:} All the galaxies show A$_V$(R) declining profiles, but  do not have a clear trend with  morphology or with galaxy  mass.  Most  spirals show similar radial variations, and similar average extinction, A$_V\sim0.2$ mag at the disk and up to 0.6 mag in the center, with the inner gradient $\bigtriangledown_{in}$A$_V \sim$ $-0.25$ mag/HLR. However, Sd galaxies show a shallow central gradient. E and S0 also show a negative gradient in the inner 1 HLR (shallower than in early type spirals), but out of the core they are almost dust free. On average, $\bigtriangledown_{in}$A$_V$ gets stronger with increasing $M_\star$ up to 10$^{11} M_\odot$, and weakens towards higher mass. However, the dispersion with respect to the binned mass relation is not related with  Hubble type. A major reason for this is that A$_V$(R) profiles are sensitive to inclination effects, unlike $\mu_\star$, \ageL, or \logZM. Thus, spirals with larger inclination have larger extinction. This is particularly evident in Sb's, that have the largest A$_V^{HLR}$ and the steepest central gradient.}

\end{enumerate}

From these results, we conclude:

\begin{itemize}

\ls\item{
Evidence in favor of the inside-out growth of galaxies is found in the negative radial stellar age gradients. Metallicity gradients and the fact that galaxies are more compact in mass than in light also support this scenario. On the other hand, the flattening of the \ageL(R) profiles beyond 1.5$-$2 HLR may be interpreted as indicative that the mass was formed in a more uniformly distributed manner across the outer disk of spirals. In the case of E's and S0's this may be understood if beyond 2 HLR most of the stellar mass was accreted at z$\leq$ 1. }

\ls\item{The mean stellar ages of disks and bulges are correlated, with disks covering a large range of ages, and late type spirals hosting the younger disks.
The bulges of S0 and early type spirals are old and metal rich as the core of E's.
They formed by similar processes, through mergers. Later type spirals, however,  have younger bulges, and larger contribution from secular evolution are expected. Disks are younger and more metal poor than bulges, as an indicative of the inside-out formation scenario of these galaxies.}

\ls\item{S0's in this sample (all are massive galaxies), act as a transition class between E's and spirals, with $\mu_\star$(R), \ageL(R), and A$_V$(R) between massive E's and Sa's. The gradient in $\mu_\star$ and \ageL\ of S0's is so similar to Sa's galaxies, that they can result from the same formation process. }

\ls\item{It is the Hubble type, not $M_\star$, that drives differences in the  galaxy averaged age, and radial age gradients. These results indicate that the SFH and their radial variations are modulated primarily by galaxy morphology, and only secondarily by the galaxy mass.  This suggest  that galaxies are  morphologically quenched, and that the shutdown of  star formation occurs outwards and earlier in galaxies with a large spheroid than in galaxies of later Hubble type. }

\end{itemize}

From the comparison of the results with the theoretical predictions from cosmological simulations, we conclude:

\begin{itemize}

\ls\item{Major mergers are likely the main process building the central regions of ETGs.
The metallicity gradient within 1 HLR is shallow compared with the theoretical expectation if minor mergers are relevant in the growth of the central core of E's and S0's. In our results there is no evidence either of an inversion of \ageL\ toward older ages beyond 1$-$2 HLR, or of a steepening of the metallicity if these galaxies were growing in size through minor dry mergers. Massive galaxies  probably accreted massive satellites that were able to retain their metal rich gas against winds, producing flatter metallicity gradients \citep{hirschmann15}. Alternatively, the flattening of the metallicity radial profile can result from the quenching of star formation. When this happens, the metal cycle stops and only stars of that last star formation event remain. }

\ls\item{Through the negative metallicity gradients, spirals show evidence of growing inside-out. These gradients are flatter than the predictions by the classical chemical evolution models \citep[e.g]{chiappini01,molla05}, but are similar to those measured above the Galactic disk. 
The largest gradient happens in intermediate types and intermediate galaxy mass, as predicted by the \citet{molla05} models. However, Sbc galaxies have a \logZM\ gradient  similar to the predictions by  RaDES simulations \citep{few12, pilkington12a}. This indicates that the feedback  recipes used in these simulations are able to recover realistic galaxies with small bulges.  However, a larger set of cosmological simulations is required, covering from early type Sa to late Sd, and a large range of galaxy mass (from 10$^9$ to 10$^{11}$ M$_\odot$), in order to confirm the general trend with the Hubble type found in this work.}

\end{itemize}

Thanks to the uniqueness of the CALIFA data in terms of  spatial coverage and resolution,  large sample spanning all morphological types, and homogeneity and quality of the spectral analysis, we are able to characterize the radial structure of the stellar population properties of galaxies in the local universe. The results show that the Hubble sequence is a useful scheme to organize galaxies by their spatially resolved stellar density, ages, and metallicity. However, stellar extinction cannot discriminate so well between the different Hubble types. 
Stellar mass, although responsible for setting the average stellar population properties in galaxies, it is less responsible of the quenching processes. Morphology is, however, more strongly connected with the shut down of the star formation activity in the bulges and disks of galaxies.

\begin{acknowledgements} 
CALIFA is the first legacy survey carried out at Calar Alto. The CALIFA collaboration would like to thank the IAA-CSIC and MPIA-MPG as major partners of the observatory, and CAHA itself, for the unique access to telescope time and support in manpower and infrastructures.  We also thank the CAHA staff for the dedication to this project.
Support from the Spanish Ministerio de Econom\'\i a y Competitividad, through projects AYA2010-15081 (PI RGD), and Junta de Andaluc\'\i a FQ1580 (PI RGD), AYA2010-22111-C03-03, and AYA2010-10904E (SFS).
We also thank the Viabilidad, Dise\~no, Acceso y Mejora funding program, ICTS-2009-10, for funding the data acquisition of this project. 
RCF thanks the hospitality of the IAA and the support of CAPES and CNPq. RGD acknowledges the support of CNPq (Brazil) through Programa Ciencia sem Fronteiras (401452/2012-3).
AG acknowledges support from EU FP7/2007-2013 under grant agreement n.267251 (AstroFIt) and from the EU Marie Curie Integration Grant "SteMaGE" Nr. PCIG12-GA-2012-326466.
CJW acknowledges support through the Marie Curie Career Integration Grant 303912. EP acknowledges support from the Guillermo Haro program at INAOE.
Support for LG is provided by the Ministry of Economy, Development, and Tourism's Millennium Science Initiative through grant IC120009, awarded to The Millennium Institute of Astrophysics, MAS. LG acknowledges support by CONICYT through FONDECYT grant 3140566.
JIP acknowledges financial support from the Spanish MINECO under grant AYA2010-21887-C04-01 and from Junta de Andaluc\'{\i}a Excellence Project PEX2011-FQM7058.
IM, JM and AdO acknowledge the support from the project AYA2013-42227-P.
RAM is funded by the Spanish program of International Campus of Excellence Moncloa (CEI). JMA acknowledges support from the European Research Council Starting Grant (SEDmorph; P.I. V. Wild)


\end{acknowledgements}


\bibliographystyle{aa}
\bibliography{Califa5}

\clearpage

\begin{appendix}

\section{Improvements resulting from the new reduction pipeline}

The quality of the \starlight\ fits to CALIFA version 1.3c spectra was assessed in \citet{cidfernandes14} by averaging $R_\lambda = O_\lambda - M_\lambda$ residual spectra of 107 galaxies ($\sim 10^5$ spectra). Inspection of these residuals revealed low amplitude (a few \%) but systematic features related to unmasked weak emission lines, SSP deficiencies and data calibration imperfections. This exercise needs updating now that the reduction pipeline has changed to version 1.5.

\begin{figure*}
\includegraphics[width=0.85\textwidth]{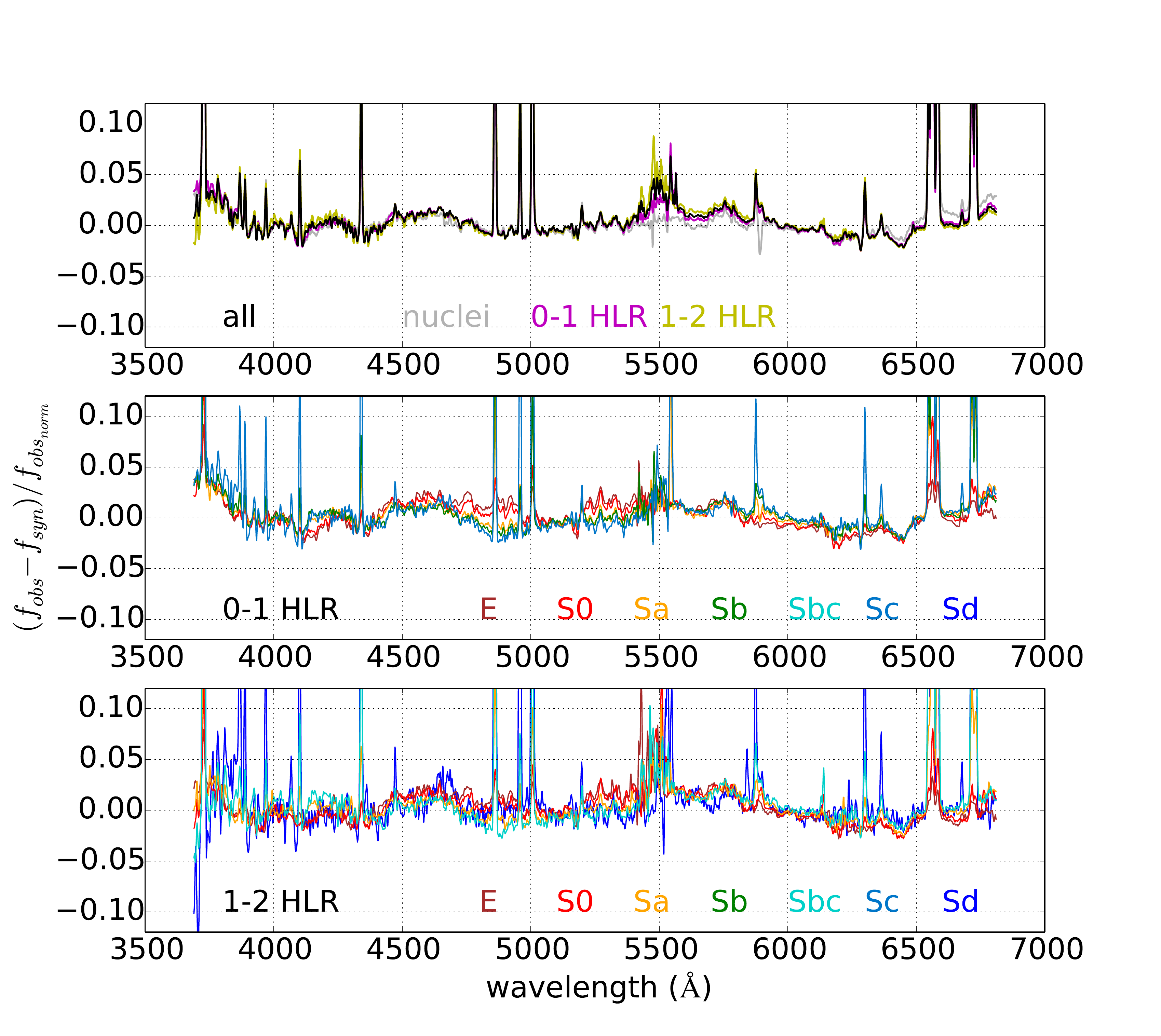}
\caption{Upper panel: Residual spectra averaged for all the spectra (black line), nuclei (grey line) and spectra belonging to zones that are between 0-1 HLR (pink line) and between 1-2 HLR (yellow line)). Middle panel: residual spectra averaged for zones inner to 1 HLR and for Hubble type. Bottom panel: As in the middle panel but for spectra of zones located between 1 and 2 HLR. }
\label{fig:SpecResid}
\end{figure*}

Fig.\ \ref{fig:SpecResid} summarizes the results of this re-evaluation. The plots show stacked $R_\lambda = O_\lambda - M_\lambda$ residual spectra, in units of the median flux in the $5635 \pm 45$ \AA\ window. The top panel shows results for the nuclear extractions, while the middle and bottom ones are built using spectra from zones within radial distances $R = 0 - 1$ and $1 - 2$ Half Light Radius (HLR, computed in the same wavelength range), respectively. Residuals are colored according with the Hubble type of the galaxies. When all galaxies are stacking, the residuals are colored according with the spatial zone extracted. These subdivisions are presented in order to get a sense of how the residuals relate to position within a galaxy and its Hubble type, two central aspects of this paper. 

No matter which panel one looks at, the improvement with respect to version 1.3c is evident to the eye when compared to figure 13 of \citet{cidfernandes14}. The broad trough around H$\beta$ present in the 1.3c spectra, for instance, is much shallower now. In fact, it is confined to late types (compare blue and red lines in the lower panel in Fig.\ \ref{fig:SpecResid}), indicating that its origin is not only related to calibration, but also to the SSP spectra of young stellar populations (as previously reported by \citealt{cidfernandes05} for SDSS data). Residuals are also visibly smaller towards the blue, including the CaII K line, which is now well fitted whereas in version 1.3c a small systematic residual subsisted.\footnote{Because of this improvement, our \starlight\ fits now start at 3700 \AA, whereas in previous articles in this series only the $\lambda > 3800$ \AA\ range was considered.} The humps around 5800 \AA, on the other hand, are still present in version 1.5, particularly noticeable for outer regions, indicating that further refinement of the sky subtraction are warranted.

In short, the spectral fits have improved substantially with the new reduction pipeline. We attribute this to the updated  sensitivity curve used in version 1.5. A more extended discussion of these and other aspects of the data reduction are presented in \citet{garcia-benito14}.

Despite these changes, the stellar population properties derived from the spectral fits did not change much in comparison to those obtained for version 1.3c data. The most noticeable changes were in mean ages, which become 0.1 dex older, and extinction, which is now 0.2 mag smaller on a global average.

\section{Base experiments and uncertainties due to SSP templates}

\label{app:BaseExperiments}
To derive the stellar population properties of these 300 CALIFA galaxies we have fitted $\sim$253000 spectra with the {\it GMe} and {\it CBe} using the cluster Grid-CSIC 
at the Instituto de Astrof\'\i sica de Andaluc\'\i a and the cluster Alphacrucis at IAG-USP Sao Paulo. Examples of the quality of the spectral fits as a function of the Hubble type and radial distance are presented in Fig.\ref{fig:SpecResid}.

\subsection{Mass-metallicity relation}

\begin{figure}
\includegraphics[width=0.5\textwidth]{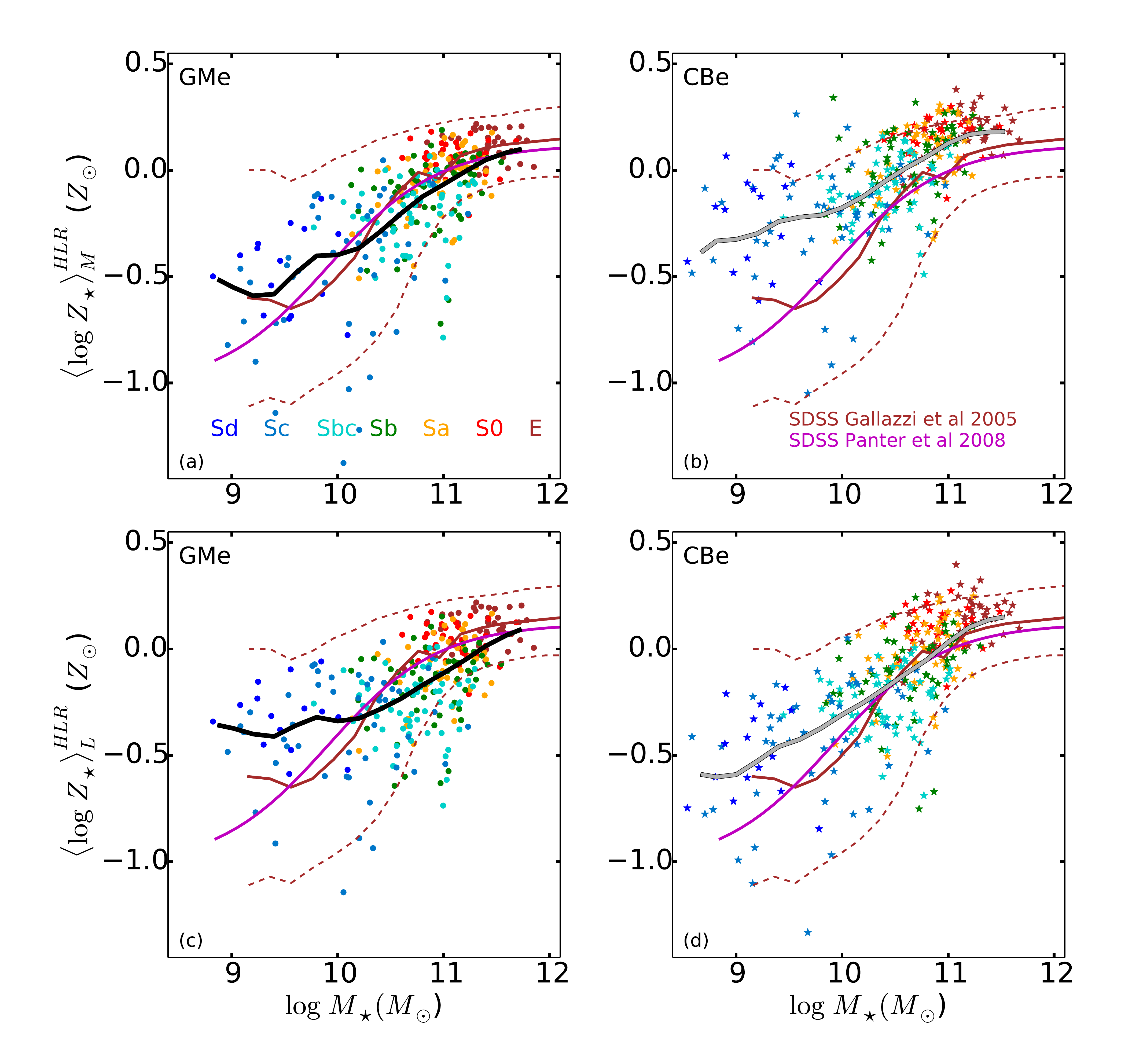}
\caption{The global {\em stellar} MZR for 300 CALIFA galaxies is shown as dots, color coded by the morphological type. It is derived using $\langle \log Z_\star \rangle_M^{HLR}$ (upper panels) and $\langle \log Z_\star \rangle_L^{HLR}$ (lower panels), and the total stellar mass, $M_\star$, obtained with the $GMe$ SSP models  (left panels) and $CBe$ (right panels). 
A mass-binned smooth mean relation is shown as a solid black or grey-black line.
The MZRs obtained for SDSS galaxies by \citet{gallazzi05} and \citet{panter08} are plotted as brown and magenta lines, respectively, with dashed brown indicating the 16 and 84 percentiles of \citet{gallazzi05}. 
}
\label{fig:mass-metallicity-relation}
\end{figure}

Here we want to find out how well our metallicity definitions  follow a MZR that guarantees that galaxies like the MW or Andromeda ($\log M_\star (M_\odot) \sim 11$) have solar metallicity at the disk, while LMC-SMC-like galaxies ($\log M_\star (M_\odot) \sim 9$)  have $\sim 1/4\, Z_\odot$. We do this with mass-weighted and luminosity-weighted definitions of eqs. 2 and 3,  and with the two sets of SSP models ({\it GMe} or {\it CBe}). 

Similarly, the correlation between the galaxy averaged stellar metallicities and the metallicity measured at 1 HLR, and the MZR, guarantee that the metallicity radial profiles  scale with the galaxy stellar mass.
However, the MZR in \citet{gonzalezdelgado14b} was derived using the galaxy averaged stellar metallicity instead of the metallicity measured at 1 HLR, and using the mass weighted definition of the metallicity and only the results with the base {\it GMe}.  
For these reasons, we derive  the MZR that results from using the mass-weighted and also the light-weighted definition of the metallicity, and the {\it GMe} and {\it CBe} bases. 

Fig. \ref{fig:mass-metallicity-relation} shows the correlation of $M_\star$  and $\langle \log Z_\star \rangle_M^{HLR}$  (upper panels), and $\langle \log Z_\star \rangle_L^{HLR}$ (lower panels) for the {\it GMe} (left panels) and {\it CBe} (right panels) SSP models. The mass-metallicity relation found by \citet{panter08} and \citet{gallazzi05} are the magenta and brown lines, respectively\footnote{The \citet{gallazzi05}  and \citet{panter08} relations have been shifted by 0.25 dex to the right to account for the difference in IMF between their results based on models with Chabrier IMF and ours obtained with Salpeter IMF in the two left panels.}.  Note that in the four cases, the metallicities are well in the range of the dispersion given by \citet{gallazzi05} (brown dashed line represent the 16th and 84th percentiles of their distribution). To compare the general trend of these values, we have derived a smoothed mass-binned relation, represented by a black solid  or grey-black line. As  expected from the global MZR derived in \citet{gonzalezdelgado14b}, base {\it GMe} and $\langle \log Z_\star \rangle_M$ predict stellar metallicities for MW and LMC-SMC with the expected values. But $\langle \log Z_\star \rangle_L$ gives a MZR that predict higher metallicities. The opposite happens for the MZR using the base {\it CBe}, that gives mass weighted metallicities higher on average than the SDSS metallicities, but the MZR with  $\langle \log Z_\star \rangle_L$ goes close to the \citet{gallazzi05} relation, and also predicts stellar metallicities for MW-Andromeda-like and LMC-SMC galaxies with the expected values.

In summary, $\langle \log Z_\star \rangle_M$ with {\it GMe} and $\langle \log Z_\star \rangle_L$ with {\it CBe}  provide a mass-metallicity relation  similar to the SDSS MZR, and predict metallicities between -0.7 and -0.4 dex for galaxies with mass between $\sim$10$^9$ and 10$^{10}$ $M_\odot$,  the expected values for LMC and SMC galaxies like, and solar for MW-like galaxies.

\subsection{Variations over bases {\it GMe} and {\it CBe}}

\label{app:Bases_SSP}

\begin{figure*}
\includegraphics[width=\textwidth]{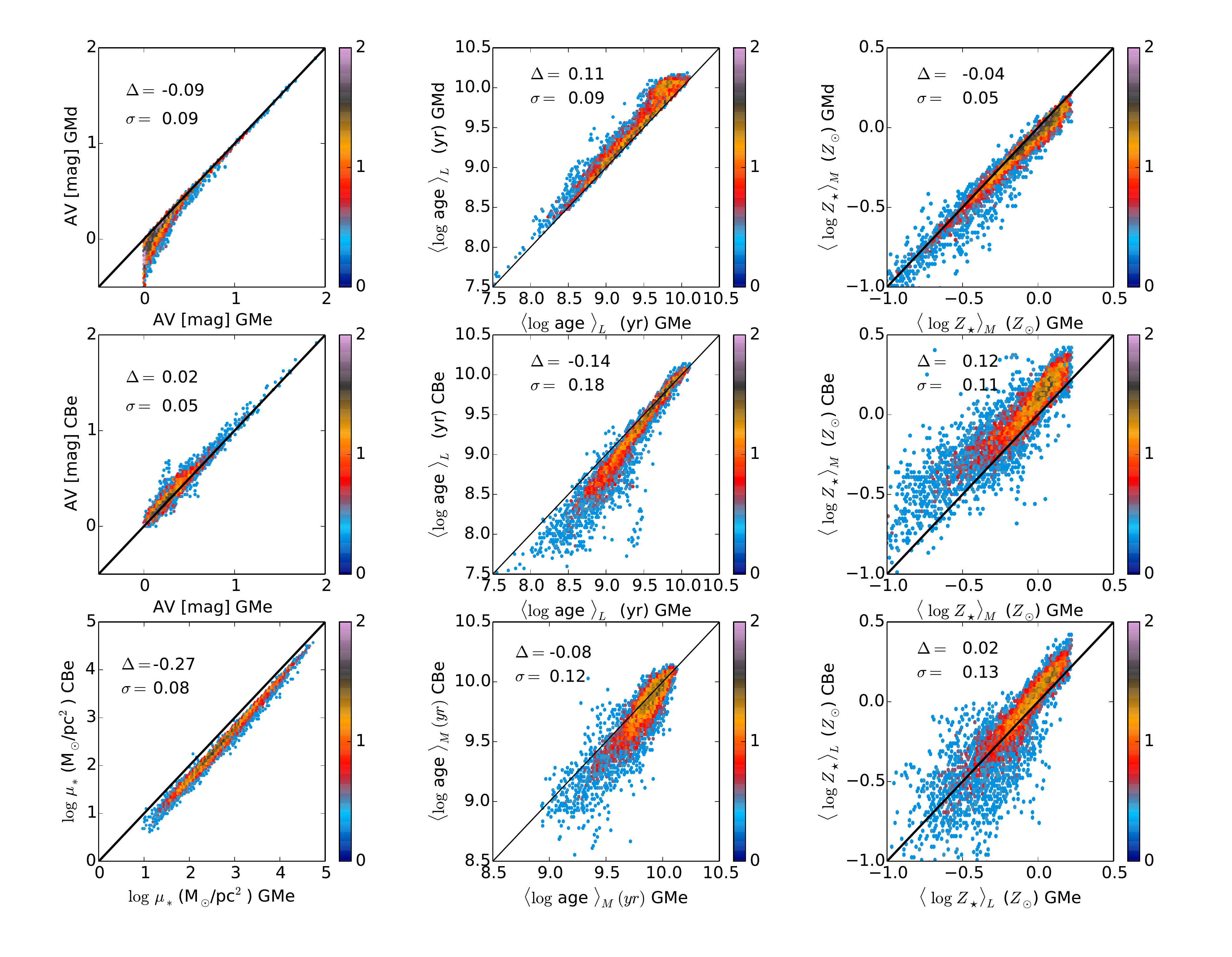}
\caption{Correlation between SP obtained with the {\it GMe} (x-axis) and {\it CBe} (y-axis bottom and middle panels; and {\it GMe} (x-axis) and {\it GMd} (y-axis). The average difference between the property in the y-axis and x-axis is labeled in each panel as $\Delta$, and the dispersion as ($\sigma$). }\label{fig:models_uncertainties}
\end{figure*}

Although the two sets of models are built with the same stellar libraries ($MILES$ and 
 {\sc granada}), base {\it GMe} stops at 1.5$Z_\odot$, while {\it CBe} goes up to 2.5$Z_\odot$. Because $MILES$ is built with stars in the solar neighborhood, it does not contain stars as metal rich as 2.5$Z_\odot$, so {\it CBe} results at over solar metallicity should be interpreted with care. On the other hand, the central parts of galaxies can be as metal rich as 2-3$Z_\odot$, and the base {\it GMe} may be too low to fit spectra of these regions, leading to saturation effects. To avoid these problems, 
 we have also fitted the spectra with another two set of SSPs, that are identical to {\it GMe} and {\it CBe}, but where for each metallicity bin, two SSPs of age 16 and 18 Gyr are added to our "standard" bases. These extra bases, that we name  {\it GMd} and {\it CBd}, allow galaxies older than the age of the universe  if their bulges are very metal rich. Furthermore,  results at very low metallicity also must be  taking with care of. $MILES$ contain only few stars of metallicity below 1/100 $Z_\odot$. For this reason,  \citet{vazdekis10} provide a safe age  range for each metallicity bin, being the models with $\log Z_\star (Z_\odot) \leq -1.7$ only valid between 10 and 18 Gyr. 
 This safety margin is provided to avoid the cases when, due to age-metallicity degeneracy, these old metal poor models fit young metal rich populations, that may happen if the base does not include SSP younger than 100 Myr. Our fits do not suffer this problem because bases {\it GMe} and {\it CBe}  both have spectra of ages as young of 1 Myr.

\subsection{Global results and uncertainties associated to SSP models}

To evaluate to which extent the spectral synthesis results  depend on the choice of SSP models, we now compare the global properties derived with bases {\it GMe} and {\it CBe}. Using our pipeline \pycasso\ we have obtained the radial distribution of the stellar population properties for each galaxy with a spatial sampling of 0.1 HLR. Here we compare the stellar population properties of the 0.1 HLR radially sampled points, instead of comparing the results obtained from the individual 253418  fitted spectra.
Fig \ref{fig:models_uncertainties} shows the results for a total of 6000 points corresponding to a maximum of 20 radial points (from nucleus to 2 HLR) for each of the 300 galaxies analyzed in this work. The figure compares the results for base {\it GMe} in the x-axis, with {\it CBe} in the y-axis, in the bottom and middle panels. The upper panels compare the results of {\it GMe} with  {\it GMd}. Each panel quotes the mean $\Delta$ and its standard deviation, where $\Delta =$ property({\it CBe}) $-$ property({\it GMe}) or  $\Delta =$ property({\it GMd}) $-$ property({\it GMe}).

{\it GMe}-based $\mu_\star$-values are higher than {\it CBe} by 0.27 dex on average, reflecting the different IMF used. Apart from this offset, the two stellar mass surface density agree to within 0.08 dex. Mean extinction is also in good agreement with a dispersion of 0.05 mag. Ages are higher in  {\it GMe} than {\it CBe} by 0.14 dex for $\langle \log\,age\rangle_L$ and 0.08 dex for $\langle \log\,age\rangle_M$, with dispersion 0.18 dex and 0.12 dex, respectively. In someway, this result is expected since base {\it GMe} also differs from {\it CBe} in IMF and isochrones.  The differences in opacities in the equation of state between Padova 2000 ({\it GMe}) and 1994 ({\it CBe}) tracks produce somewhat warmer stars in the red giant branch in the former. Thus, older ages are expected with {\it GMe} than with {\it CBe}. However,  the metallicities are lower in {\it GMe} than in {\it CBe} by 0.13 dex for $\langle \log\,(Z_\star/Z_\odot)\rangle_M$ and very similar (on average) for  $\langle \log\,(Z/Z_{\odot})\rangle_L$. In both cases, the dispersion is similar, 0.11 and 0.13 dex, respectively. Note that for $Z_\star \geq Z_\odot$, the metallicities (weighted in light or in mass) are always higher with {\it CBe} than {\it GMe}, reflecting the saturation effects in the base  {\it GMe} due its limitation to $Z\leq 1.5 Z_\odot$. The shift at under-solar metallicities may be reflecting the age-metallicity degeneracy, {\it CBe} giving higher metallicity and younger the ages.

The upper panels compare the results of {\it GMe} with the {\it GMd}. Here, we see two relevant effects. The results of {\it GMd} also differ from {\it GMe} in the the range of extinctions allowed to \starlight. While with {\it GMe} and {\it CBe} \starlight\ always assumes A$_V\geq$ 0, with {\it GMd}, \starlight\ can bluer the SSP spectra by up to $A_V$ = -0.5 mag. This is allowed to avoid the effect of saturation at $A_V$ = 0. The global effect is that ages can be 0.11 dex younger with {\it GMe} than {\it GMd}. Metallicity is not affected  by this choice of $A_V$. 
But the extension to ages  older than the age of the Universe in  {\it GMd} has some effect on the metallicity above Z$_\odot$, so metallicities are slightly lower and ages slightly older.

\section{2D maps and tables\label{ap:2dmap}}

\subsection{2D maps of $\log \mu_\star$}

\begin{figure*}
\begin{center}
\includegraphics[width=0.85\textwidth]{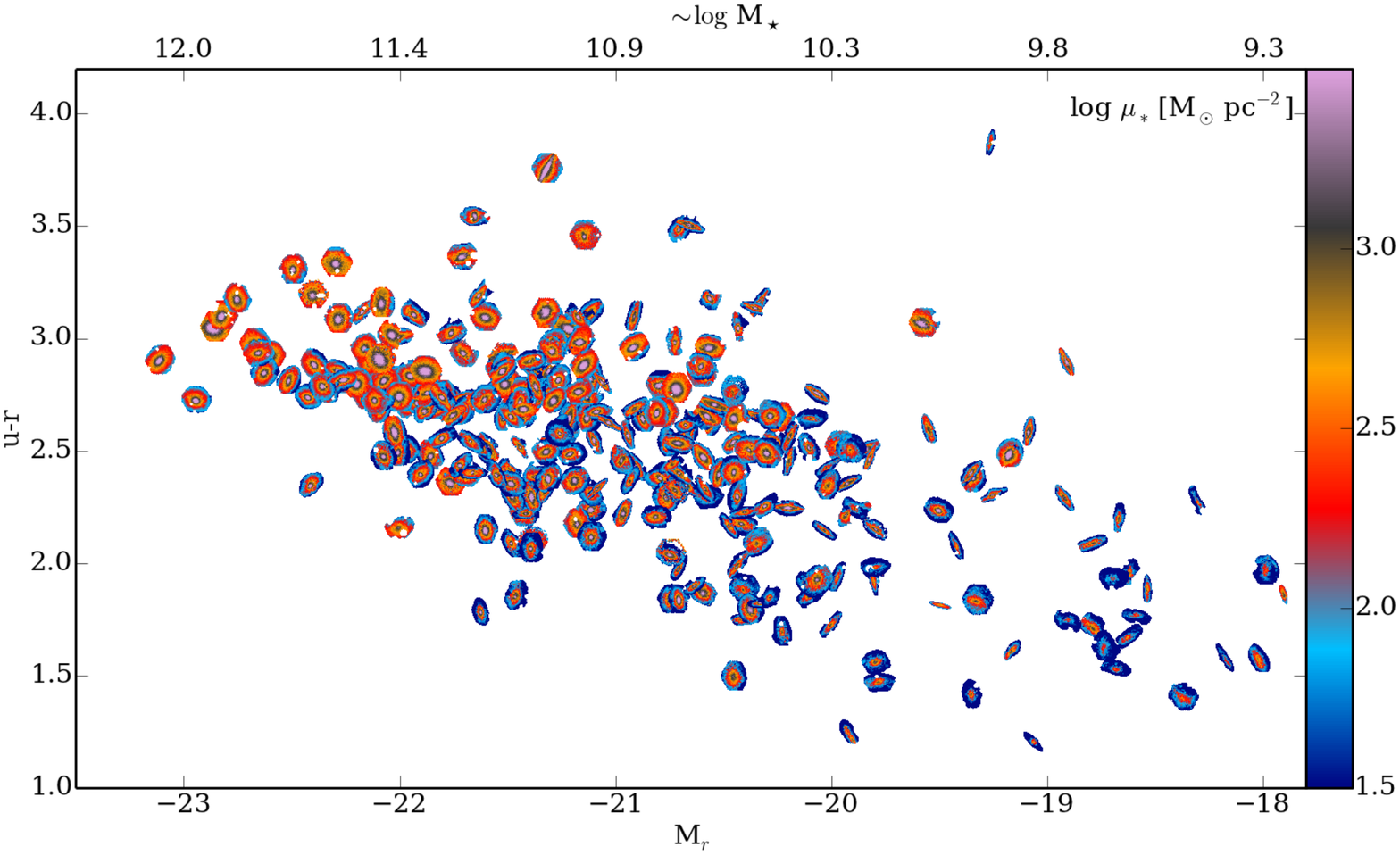}
\caption{2D maps of stellar mass surface density, $\mu_\star$. Each galaxy is placed in its location in the $u-r$ vs. $M_r$ diagram, where color and magnitude correspond to its global values.  The 2D maps are shown with North up and East  to the left. }
\label{fig:cmd_mu}
\end{center}
\end{figure*}

Fig. \ref{fig:cmd_mu} shows the $M_r$ vs. $u-r$ CMD for the 300 CALIFA galaxies of our sample. Each galaxy is represented by its 2D map of the $\log \mu_\star$  located at the position of its integrated $M_r$ and $u-r$ values. In this plot the SFH is compressed into the present-day stellar mass surface density, that measures the end product of the SFH. Because our analysis  accounts for extinction, these $\log \mu_\star$ values and their radial variations are free from extinction effects. Fig. \ref{fig:cmd_mu} shows\footnote{These 2D maps are the results from the {\it GMe} SSP base.}  clearly that $\log \mu_\star$ correlates with $M_r$, and spheroids are significantly denser than late type galaxies by one to two orders of magnitude at the center, and by one order of magnitude at distances 1-2 HLR. At the center, 2.0 $\leq$ $\log \mu_\star$ ($M_\odot$ pc$^{-2}$) $\leq$ 4.7, while $\leq$ $\log \mu_\star$ ($M_\odot$ pc$^{-2}$) $\leq$ 3.4 at 1 HLR,  and 1.0 $\leq$ $\log \mu_\star$ ($M_\odot$ pc$^{-2}$) $\leq$ 2.9 at 2 HLR.

\subsection{2D maps of \ageL}

\begin{figure*}
\begin{center}
\includegraphics[width=0.85\textwidth]{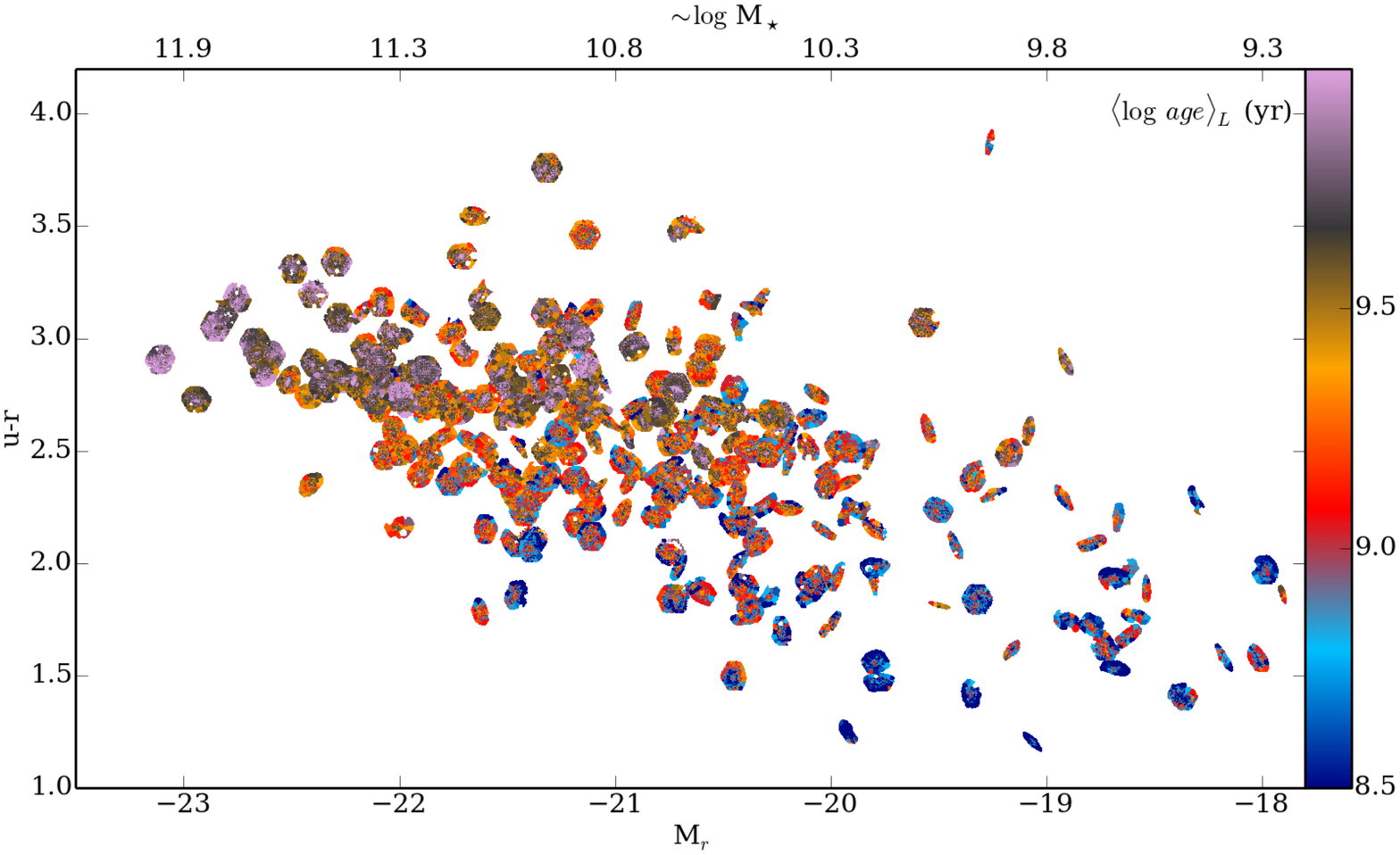}
\caption{As Fig. \ref{fig:cmd_mu}, but for images of the luminosity weighted mean age,  \ageL. }
\label{fig:cmd_ageL}
\end{center}
\end{figure*}

Similarly to Fig. \ref{fig:cmd_mu}, Fig. \ref{fig:cmd_ageL} shows the 2D maps of \ageL.
It portrays the correlation between the average age of the stellar populations and $M_r$ and colors, with the most luminous  and red galaxies being older, while the bluest galaxies are the youngest. Gradients of the stellar population ages are also clearly detected within each galaxy in these 2D maps, and more remarkably in galaxies located in the green valley. At the center, 7.3 $\leq$ \ageL\ (yr) $\leq$ 10.1, while  8.3 $\leq$ \ageL\ (yr) $\leq$ 10.1 at 1 HLR, and  7.5 $\leq$ \ageL\ $\leq$ 9.9 at 2 HLR.

\subsection{2D maps of \logZM}

\begin{figure*}
\begin{center}
\includegraphics[width=0.85\textwidth]{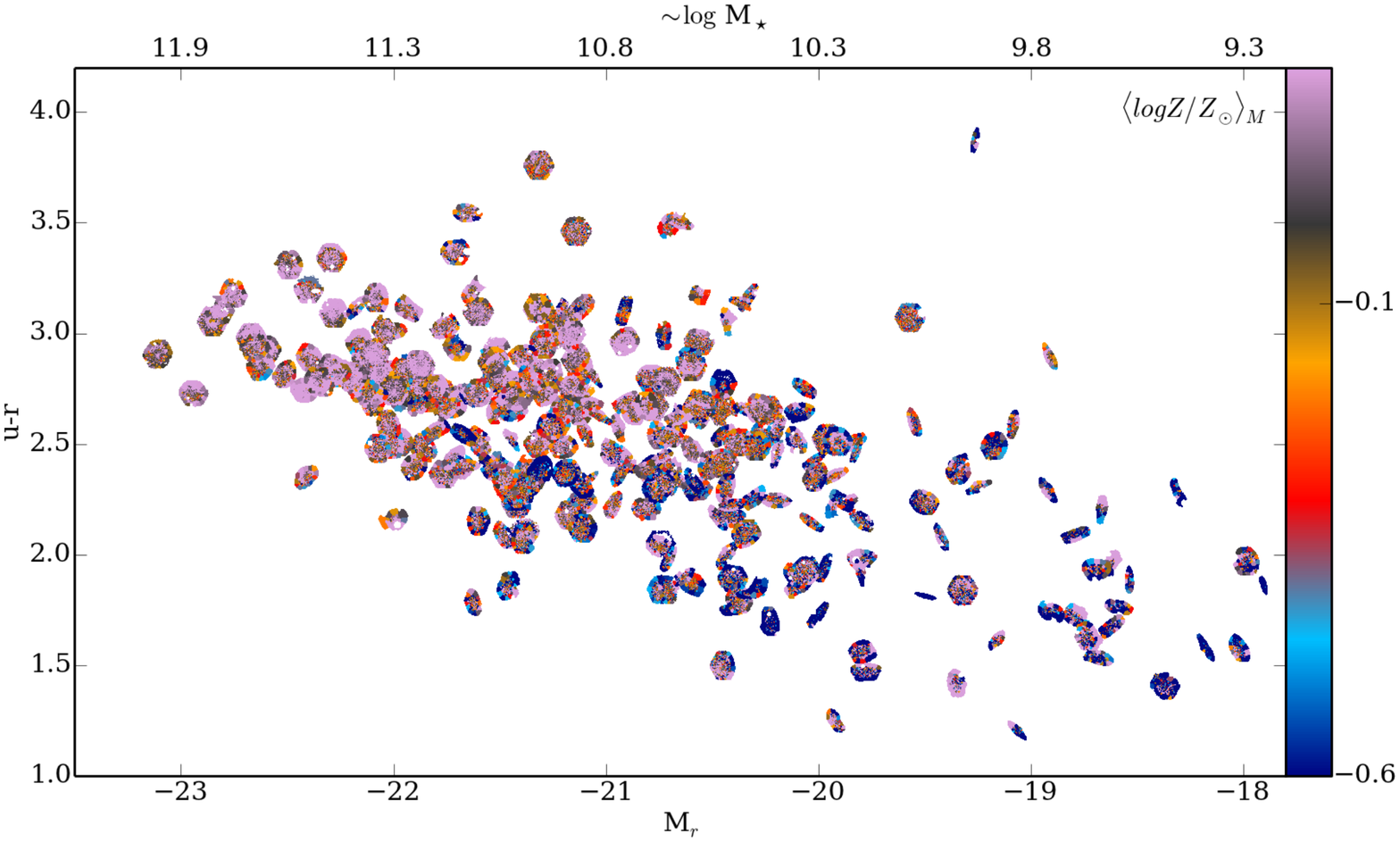}
\caption{As Fig. \ref{fig:cmd_mu}, but for images of the mass weighted mean metallicity,  \logZM. }
\label{fig:cmd_logZM}
\end{center}
\end{figure*}

Similarly to Fig. \ref{fig:cmd_mu}, Fig. \ref{fig:cmd_logZM} presents the 2D maps of \logZM. It clearly shows the correlation between the stellar metallicity and galaxy luminosity,  equivalent to the mass metallicity relation. Gradients of the stellar metallicities are more clearly seen in these 2D maps in galaxies with intermediate luminosity ($-22 \leq M_r \leq -20$). The most luminous galaxies have solar or over solar metallicity producing a visual saturation in the 2D maps. The stellar metallicities range from \logZM = -1.4 to 0.22.

\subsection{2D maps of A$_V$}

\begin{figure*}
\begin{center}
\includegraphics[width=0.85\textwidth]{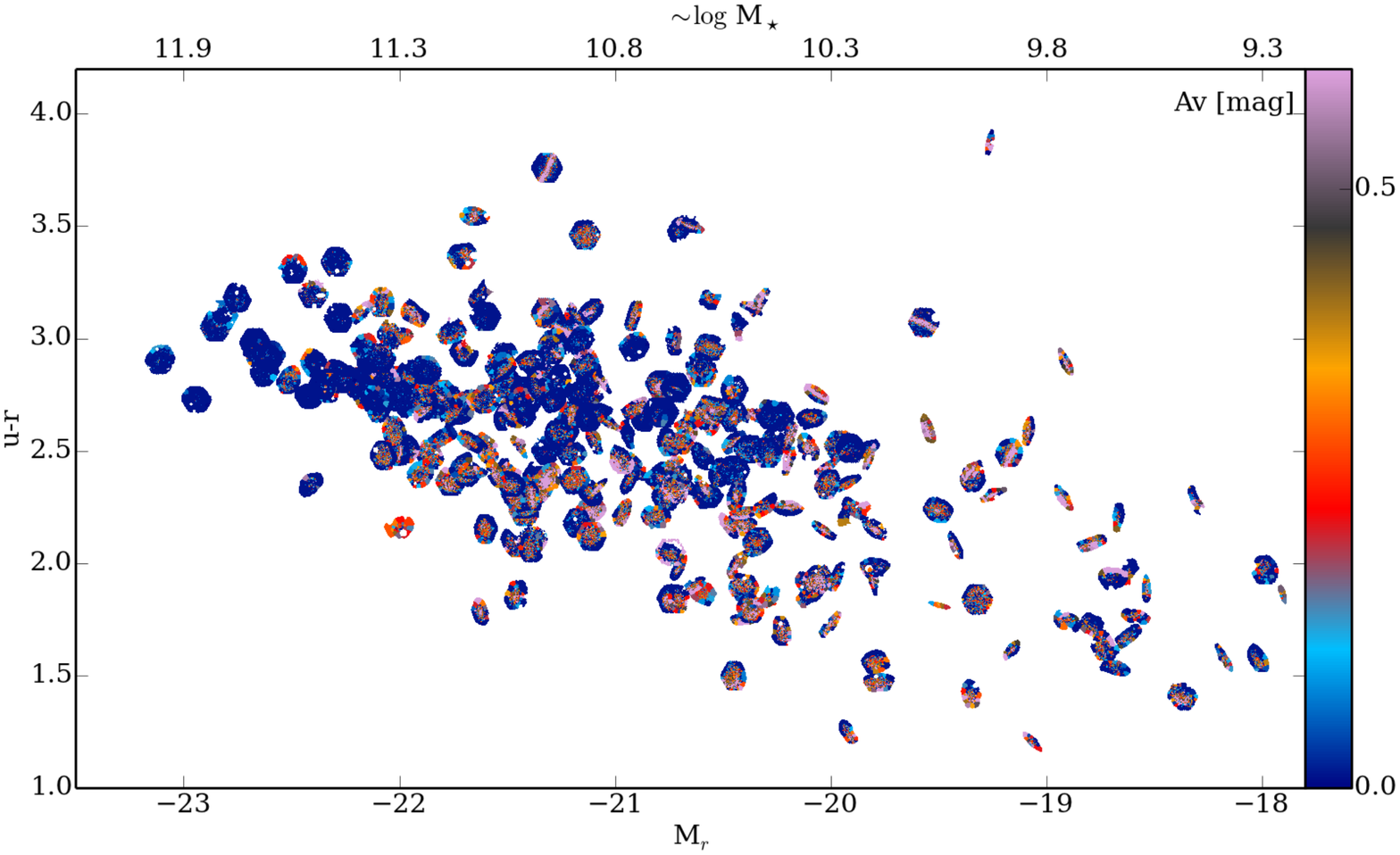}
\caption{As Fig. \ref{fig:cmd_mu}, but for images of the stellar extinction, $A_V$ [mag] }
\label{fig:cmd_AV}
\end{center}
\end{figure*}

Similarly to Fig. \ref{fig:cmd_mu}, Fig. \ref{fig:cmd_AV} presents 2D maps of  $A_V$. 
Effects of spatial binning are visible in the $A_V$ maps, where all the pixels within a Voronoi zone have the same value. These effects are not noticeable in the $\mu_\star$ images (Fig. \ref{fig:cmd_mu}) because $\mu_\star$ is an extensive property, and the zoning effect was softened by scaling the value at each pixel by its fractional contribution to the total flux in the zone; this is not possible for $A_v$ (Fig. \ref{fig:cmd_AV}), \ageL\ (Fig. \ref{fig:cmd_ageL}), or \logZM\ (Fig. \ref{fig:cmd_logZM}), because these are intensive properties. Fig. \ref{fig:cmd_AV}  shows how $A_V$ changes across the CMD: the most luminous galaxies are  little affected by extinction, while $A_V$ is higher in spirals of intermediate type and with blue colors.
The mean (dispersion) $A_V$ values at the nuclei, 1 HLR, and 2 HLR are 0.47 (0.37), 0.19 (0.16), and 0.13 (0.13), respectively. 2D maps and the difference in the mean values at different distance indicate that stellar extinction shows radial gradients.

\newpage
\onecolumn

\tiny
\begin{landscape}
\begin{longtable}{lcccccccccccccc}
\caption{Stellar population properties: $CBe$}\\
\hline\hline
CALIFA  &   Name & type &  $\log\,M_\star$  & $\log\mu_\star$[0]  & $\log\mu_\star$[HLR]  &   $A_V$[0]  & $A_V$[HLR]  & $\langle\log age$[0]$\rangle_L$  & $\langle\log age$[HLR]$\rangle_L$ & $\langle\log Z$[0]$\rangle_M$  & $\langle\log Z$[HLR]$\rangle_M$ &  HLR  & HMR & C \\ 
     ID\# & NED   & morph. & $M_\odot$   & $M_\odot$/pc$^2$ &  $M_\odot$/pc$^2$ & mag & mag & (yr)  &  (yr)   & [$Z_\odot$]   & [$Z_\odot$]           &  pc     &  pc &  \\      
\hline
\endfirsthead
\caption{continued.}\\
\hline\hline
CALIFA  &   Name & type & $\log\,M_\star$  & $\log\mu_\star$[0]  & $\log\mu_\star$[HLR]  &   $A_V$[0]  & $A_V$[HLR]  & $\langle\log age$[0]$\rangle_L$  & $\langle\log age$[HLR]$\rangle_L$ & $\langle\log Z$[0]$\rangle_M$  & $\langle\log Z$[HLR]$\rangle_M$ &  HLR  & HMR & C \\ 
     ID\# & NED   & morph. & $M_\odot$   & $M_\odot$/pc$^2$ &  $M_\odot$/pc$^2$ & mag & mag & (yr)  &  (yr)   & [Z$_\odot$]   & [Z$_\odot$]  &  pc     &  pc &  \\      
\hline
\endhead
\hline
\endfoot
001 & IC5376 & 2  A & 10.60 & 3.87 & 2.41 & 0.42 & 0.53 & 9.97 & 9.38 & 0.25 & -0.17 & 3883 & 2076 & 3.3 \\
002 & UGC00005 & 3  A & 10.84 & 3.29 & 2.23 & 0.63 & 0.41 & 9.51 & 8.88 & -0.03 & -0.04 & 6842 & 5146 & 2.1 \\
003 & NGC7819 & 4  A & 10.04 & 2.80 & 1.59 & 0.63 & 0.05 & 8.46 & 8.86 & 0.13 & -0.27 & 5455 & 4057 & 2.1 \\
004 & UGC00029 & -1  A & 11.16 & 3.67 & 2.49 & 0.44 & 0.03 & 9.89 & 9.75 & 0.37 & 0.14 & 5400 & 4372 & 3.1 \\
005 & IC1528 & 3  AB & 10.09 & 3.03 & 1.95 & 0.80 & 0.35 & 8.97 & 8.64 & -0.22 & -0.29 & 4348 & 3417 & 2.1 \\
007 & UGC00036 & 1  AB & 10.91 & 3.89 & 2.70 & 0.73 & 0.19 & 9.66 & 9.51 & 0.39 & 0.32 & 4104 & 3142 & 2.7 \\
008 & NGC0001 & 3  A & 10.68 & 3.85 & 2.53 & 0.43 & 0.63 & 9.55 & 8.95 & 0.07 & -0.06 & 2781 & 1790 & 3.0 \\
010 & NGC0036 & 2  B & 11.00 & 3.71 & 2.35 & 0.33 & 0.34 & 9.97 & 9.31 & 0.05 & 0.09 & 5717 & 4152 & 2.5 \\
\end{longtable}
\end{landscape}

\newpage
\onecolumn

\begin{landscape}
\begin{longtable}{lcccccccccccccc}

\caption{Stellar population properties: $GMe$}\\
\hline\hline
\hline\hline
CALIFA  &   Name & type &  $\log\,M_\star$  & $\log\mu_\star$[0]  & $\log\mu_\star$[HLR]  &   $A_V$[0]  & $A_V$[HLR]  & $\langle\log age$[0]$\rangle_L$  & $\langle\log age$[HLR]$\rangle_L$ & $\langle\log Z$[0]$\rangle_M$  & $\langle\log Z$[HLR]$\rangle_M$ &  HLR  & HMR & C \\ 
     ID\# & NED   & morph. & $M_\odot$   & $M_\odot$/pc$^2$ &  $M_\odot$/pc$^2$ & mag & mag & (yr)  &  (yr)   & (Z$_\odot$)   & (Z$_\odot$)           &  pc     &  pc &  \\      
\hline
\endfirsthead
\caption{continued.}\\
\hline\hline
CALIFA  &   Name & type &  $\log\,M_\star$  & $\log\mu_\star$[0]  & $\log\mu_\star$[HLR]  &   $A_V$[0]  & $A_V$[HLR]  & $\langle\log age$[0]$\rangle_L$  & $\langle\log age$[HLR]$\rangle_L$ & $\langle\log Z$[0]$\rangle_M$  & $\langle\log Z$[HLR]$\rangle_M$ &  HLR  & HMR & C \\ 
     ID\# & NED   & morph. & $M_\odot$   & $M_\odot$/pc$^2$ &  $M_\odot$/pc$^2$ & mag & mag & (yr)  &  (yr)   & (Z$_\odot$)   & (Z$_\odot$)  &  pc     &  pc &  \\      
\hline
\endhead
\hline
\endfoot
001 & IC5376 & 2  A & 10.82 & 4.05 & 2.64 & 0.41 & 0.49 & 10.00 & 9.50 & 0.09 & -0.26 & 3883 & 2324 & 3.3  \\
002 & UGC00005 & 3  A & 11.16 & 3.60 & 2.54 & 0.66 & 0.36 & 9.58 & 9.00 & 0.00 & -0.05 & 6842 & 5175 & 2.1  \\
003 & NGC7819 & 4  A & 10.39 & 3.37 & 1.91 & 0.62 & 0.06 & 8.59 & 9.12 & 0.06 & -0.33 & 5455 & 3561 & 2.1  \\
004 & UGC00029 & -1  A & 11.41 & 3.90 & 2.73 & 0.43 & 0.05 & 9.96 & 9.82 & 0.17 & -0.01 & 5400 & 4542 & 3.1  \\
005 & IC1528 & 3  AB & 10.42 & 3.49 & 2.29 & 0.79 & 0.27 & 9.20 & 8.91 & 0.01 & -0.37 & 4348 & 3204 & 2.1  \\
007 & UGC00036 & 1  AB & 11.13 & 4.12 & 2.92 & 0.77 & 0.21 & 9.74 & 9.59 & 0.16 & 0.10 & 4104 & 3355 & 2.7  \\
008 & NGC0001 & 3  A & 10.96 & 4.13 & 2.80 & 0.47 & 0.61 & 9.57 & 9.09 & -0.00 & -0.18 & 2781 & 1871 & 3.0  \\
010 & NGC0036 & 2  B & 11.21 & 3.88 & 2.62 & 0.38 & 0.29 & 9.90 & 9.46 & -0.08 & -0.10 & 5717 & 4621 & 2.5  \\
\end{longtable}
\end{landscape}

\end{appendix}

\end{document}